\documentclass{aa}
\usepackage[varg]{txfonts}

\usepackage{graphicx}
\usepackage{amssymb}
\usepackage[english]{babel}
\usepackage[varg]{txfonts}
\usepackage{chngpage}
\usepackage{natbib}
\usepackage{threeparttable}
\usepackage{pgffor}
\usepackage{tikz}
\usepackage{lipsum}
\usepackage{rotating}
\usepackage{geometry}
\usepackage{lscape}
\usepackage{longtable}
\usepackage{appendix}
\usepackage{comment}
\usepackage{tablefootnote}
\usepackage{hyperref}
\usepackage{xspace}
\usepackage{subfigure}
\usepackage{caption}
\usepackage{placeins}
\usepackage{needspace}

\def\twco{$^{12}$CO\xspace}

\def\kms{km\,s$^{-1}$\xspace}
\def\msunyr{$M_{\odot}$\,yr$^{-1}$\xspace}
\def\msun{$M_{\odot}$\xspace}
\def\lsun{$L_{\odot}$\xspace}
\def\lstar{$L_{\star}$\xspace}
\def\teff{$T_{\mathrm{eff}}$\xspace}
\def\vexp{$v_{\mathrm{exp}}$\xspace}
\def\vLSR{$v_{\mathrm{LSR}}$\xspace}
\def\mdot{$\dot{M}$\xspace}

\def\rstar{$R_{\star}$\xspace}
\def\rhalf{R_{1/2}}

\def\aap{A\&A}

\def\apjs{ApJS}
\def\apjl{ApJL}

\def\nodata{}

\begin{document}

%\title{A millimeter perspective on mass loss among RSGs in galactic open clusters}
\title{The molecular circumstellar environments of red supergiants in galactic open clusters}

\author{M. A. Siebert\inst{1}
  \and E. De Beck\inst{1}
  \and G. Quintana-Lacaci\inst{2}
  \and T. Khouri\inst{1}
  \and M. Saberi\inst{3,4}
  \and M. Maercker\inst{1}
  \and W.~H.~T.~Vlemmings\inst{1}
  }

\institute{Department of Physics and Astronomy, Chalmers University of Technology, Gothenburg,  Sweden 
 		 \and Department of Molecular Astrophysics, Instituto de F\'isica Fundamental, IFF-CSIC, Serrano 123, 28006, Madrid, Spain
         \and Institute of Theoretical Astrophysics, University of Oslo, P.O. Box 1029 Blindern, NO-0315 Oslo, Norway
         \and Rosseland Centre for Solar Physics, University of Oslo, P.O. Box 1029 Blindern, NO-0315 Oslo, Norway
         }

\date{Received xxx / Accepted yyy}

\abstract
{The mass loss of red supergiants (RSGs) is a critical parameter both in their evolution and to the observed characteristics of Type~II-P supernovae (SNe). Empirical constraints on the relationship between RSG mass-loss rate ($\dot{M}$) and stellar parameters ($L$, $T_{\rm eff}$, $M_{\rm init}$), on which stellar evolution codes rely, largely derive from mid-infrared studies of circumstellar dust emission in nearby RSG populations. However, dust-based mass-loss determination depends heavily on uncertain gas-to-dust-ratios, and supporting constraints on the circumstellar gas remain available for only a limited number of very close RSGs.}
{We aim to expand the sample of RSGs with direct measurements of gas in their circumstellar environments with millimeter observations toward two of the largest coeval RSG populations in the Galaxy: the open clusters RSGC1 and RSGC2.}
{We present an interferometric molecular line study of 28 RSGs in RSGC1 and RSGC2 using observations from two programs carried out on the Atacama Large Millimeter/submillimeter Array (ALMA). The primary targeted emission is CO $J=2$--$1$ and continuum, with additional thermal and maser SiO lines observed for RSGC2 targets. We measure terminal expansion velocities, systemic velocities, emission-region sizes, and provide updated dust models and luminosities using the new long-wavelength constraints. 
Finally, we perform 1D radiative transfer modeling of CO rotational-line emission to estimate mass-loss rates. }
{We detect \twco toward six RSGs in RSGC2 and five in RSGC1. Except for the extreme object DFK\,52, the spatial distribution of all
\twco emission is relatively compact or unresolved by the ALMA beam ($r\lesssim6000$\,au). The obtained mass-loss rates range from $\log(\dot{M}/M_\odot\,\mathrm{yr}^{-1})=-5.5$ to $-3$, but are highly uncertain due to a lack of stringent constraints on the envelope sizes and temperatures. The lower-limit $\dot{M}$-values we obtain are high compared to standard empirical prescriptions and favor dust models adopting the radiatively driven wind assumption for the circumstellar shell. We also highlight DFK\,49 as a clear outlier in the sample, as its revised luminosity is the lowest in both clusters, yet it harbors a strong wind with $\dot{M}>10^{-5}$\,\msunyr.}
{The spatial distribution of CO around RSGC1 and RSGC2 members is considerably smaller than expectations set by closer individual objects and theoretical models, and suggests a unique phase structure for the CSM around RSGs in cluster environments. This work demonstrates the mutual utility of molecular line and dust observations for constraining RSG winds, while highlighting the need for future spatially resolved, multi-tracer observations of RSGC targets to establish the physical structure of their circumstellar environments and improve mass-loss determinations.}

\maketitle 
\nolinenumbers

\section{Introduction}\label{sect:intro}

Massive stars with initial masses of 8--40\,\msun play a key role in the chemical and dynamical evolution of galaxies through their synthesis of heavy elements and the energetic supernovae (SNe) they produce. During all phases of their lives, massive stars experience mass loss which impacts the trajectory of their evolution on the HR-diagram \citep{vanLoon2025_RSG_MdotReview,levesque2010_physicsofRSGs}. Mass loss becomes especially important as they leave the main sequence and become red supergiants (RSGs), and material in their cool extended atmospheres is expelled in the form of dusty circumstellar winds with mass-loss rates $\dot{M}=10^{-7}{\sim}10^{-4}$\,\msunyr \citep{vanLoon2025_RSG_MdotReview}. The efficiency of these winds and its dependence on stellar properties are key factors in determining the post-RSG evolutionary stages a star will experience% (e.g. YSG, BSG, or WR)
, as they regulate how much of the hydrogen envelope is removed prior to core collapse \citep{Zapartas2025_RSG_mdot}.

As RSGs are the direct progenitors of Type II-P supernovae, the amount and composition of circumstellar medium (CSM) built up by mass loss during the RSG phase also strongly influence observable properties of individual transients. In recent years, it has become apparent that the interpretation of early light curves, as well as flash spectroscopy of events like the extremely nearby SN2023ixf \citep{Jencson2023_SN2023ixf,Kilpatrick2023a_SN2023ixf,Pledger2023_SN2023ixf} requires significant amounts of close-in ($10^{14}-10^{15}$\,cm) CSM to be photo-ionized by the early shock breakout \citep{Gal-Yam2014_SN2013cu_WR,JacobsonGalan2024_SN2024ggi,kilpatrick2025_SN2025pht_CrichRSG,Moriya2017_RSGsuperwinds}, with estimated densities up to $\rho{\sim}10^{-12}$\,g\,cm$^{-3}$ and implied $\dot{M}>10^{-2}$\,\msunyr. These signatures persist and evolve with long-term monitoring of these SNe: changes in nebular line profiles and flattening of light curves after ${\sim}200$\,days indicate interaction with material also at $r>10^{15}$\,cm and imply $\dot{M}\approx10^{-6}$\,\msunyr \citep{Folatelli_2023ixf_late,Weil2020_2017eaw_late,Dessart2022_lateSNeInteraction}. Finally, observations of, e.g., the Cas A supernova remnant show evidence of strong interactions with asymmetric mass loss from the progenitor RSG \citep{DeLooze2024_CasA}.

The impact of RSG mass loss on stellar evolution in turn affects the overall demographics of progenitor properties (mass, temperature, composition) for core-collapse transient events \citep{Georgy2013_stellarmodels}. Archival SN progenitor searches gave rise to the ``red supergiant problem'', referring to the apparent absence of high-mass ($\gtrsim18\,M_\odot$) RSG progenitors compared to the distribution expected from stellar evolution models and the initial mass function \citep{Smartt2009_RSGproblem}. However, \citet{Beasor2025_RSGproblem} recently showed that the discrepancy can be attributed to an uncertain luminosity determination, relying on single-band photometry and affected by assumptions on spectral types, progenitor variability, and a lack of knowledge about extinction by circumstellar dust.

Despite the widespread impact of RSG mass loss on massive-star evolution and SN measurements, its underlying physical mechanisms are poorly understood and stellar-evolution codes rely on empirical prescriptions for the mass-loss rate. However, existing prescriptions vary by several orders of magnitude and produce very different evolutionary tracks \citep{Zapartas2025_RSG_mdot,Merritt2026_Mdot_prescrips}. The most common method of constraining the $\dot{M}-L$ relation is through radiative transfer modeling of circumstellar dust emission, which appears as excess infrared (IR) flux in a star's spectral energy distribution (SED). This allows one to obtain dust-mass-loss rates for large samples of nearby coeval RSG populations \citep[e.g.][]{humphreys2020,beasor2020,Antoniadis2024_LMC_RSG_precr}. This method is widely applicable, but notably requires assumptions about dust composition, outflow velocity, and, critically, the dust-to-gas ratio which varies by an order of magnitude across Galactic RSGs \citep[e.g.][]{justtanont1999_alphaOri_alphaSco,matsuura2014_vycma_spire_pacs}. Because of this, it is important to supplement these studies, when possible, with additional constraints from millimeter/sub-millimeter observations of rotational line emission \citep{debeck2010_comdot,debeck2025_nmlcyg,andrews2022,singh2023, decin2024_rsgc1}. These observations offer insights into important physical characteristics of the wind, including its kinematics and morphology, and allow one to derive the bulk mass loss of RSGs through radiative transfer modeling of dense gas phase tracers like \twco. Such models require their own set of assumptions, most notably regarding the CO abundance and temperature structure of the envelope; however, the resulting uncertainties are smaller than those associated with the gas-to-dust ratio, with uncertainties of order a factor of $\sim3$ \citep{ramstedt2008}.

Due to the weak intrinsic brightness of rotational CO lines, existing interferometric studies of RSG winds are focused on individual, very nearby ($<$\,2\,kpc) objects like VY CMa \citep{singh2023,kaminski_vycma_2019}, NML Cyg \citep{debeck2025_nmlcyg}, and Betelgeuse \citep{ogorman2012_alphOri_CO}. However, closer to the Galactic center, several large associations of evolved massive stars offer many additional targets well within sensitivity capabilities of current facilities. The ``RSGC'' clusters \citep{Figer2006_RSGC1,davies2007_rsgc2,Clark2009_RSGC3} are a collection of young massive open clusters located near the base of the Scutum-Crux spiral arm. Together, they contain upwards of 30\% of all known RSGs in the galaxy and are individually estimated to have initial masses of $>10^4$\,\msun \citep{Messineo2016_RSGs_innerMW}. These clusters likely represent large-scale multi-seeded starburst activity that took place near the tip of the galactic bar 10--20\,Myr ago \citep{negueruela2012}. 

RSGC1 and RSGC2 (also called Stephenson 2 or Ste 2) host the largest numbers of RSGs among the RSGC clusters and have recently served as useful test beds for testing \mdot-prescriptions. \citet{beasor2020} and \citet{humphreys2020} both analyzed the SEDs and calculated dust-derived mass-loss rates for members of these clusters, but obtain significantly different results. \citet{decin2024_rsgc1} quantified \mdot for five RSGs in RSGC1 based on ALMA observations of \twco $J=2-1$ emission, reporting systematically lower values than those derived from SED-modeling. 

RSGC2 is a more complex region than RSGC1, with a larger angular distribution of RSGs, a potential evolutionarily-distinct sub-cluster \citep[``Ste 2 SW'';][]{deguchi2010}, and a wider range of measured luminosities indicative of a spread in age or initial mass for stars currently in the RSG phase \citep[14--20\,Myr, 12--25\,\msun;][]{humphreys2020,Eldridge2020_RSGbinary_populations}. RSGC2 is also host to several unique or peculiar RSGs, including ``Stephenson 2-18'' (referred to as DFK\,1 in this paper), one of the largest purported stars if taken at the cluster distance \citep{fok12,negueruela2012}. Additionally, DFK\,49 is a member that has been proposed as having recently left the RSG \citep{humphreys2020}, and DFK\,52 was shown to harbor a massive detached outflow despite its relatively low luminosity \citep{Siebert2025_DFK52}.

In this work, we present an ALMA study of the circumstellar environments of 14 RSGs in RSGC2 targeting molecular line and continuum emission. We also re-analyze the archival observations of RSGC1 first presented by \citet{decin2024_rsgc1} to enable a self-consistent cross-cluster comparison of the mass-loss properties in both samples. With this approach, we aim to provide a holistic view of the current body of millimeter observations toward young massive RSG clusters, highlighting where these long-wavelength constraints agree with or diverge from previous studies of RSGs to guide future modeling and observational efforts. 

The paper is structured as follows: in Sect.~\ref{sect:obs}, we summarize the observations and data reduction methods, in Sect.~\ref{sect:analysis_results}, we list the detections, characterize the spatial and spectral appearance of circumstellar emission, and perform non-local thermodynamic equilibrium (non-LTE) radiative transfer calculations to determine CO mass-loss rates. In Sect.~\ref{sect:discussion} we place these results in the context of previous SED studies of the RSGC clusters and compare with popular \mdot-prescriptions for RSGs.  We present our conclusions in Sect.~\ref{sect:conclusion}.

\section{Observations}\label{sect:obs}

\begin{table*}[!t]
\caption{Summary of ALMA line and continuum detections toward RSGC2 and RSGC1.}
\centering
\begin{tabular}{p{1.2cm}l|c|c c c|c|c c|p{4cm}}
\hline\hline
\rule{0pt}{2.5ex}
&ID & $\dot{M}_{\mathrm{SED}}$ & \twco\ & SiO$_{\nu=0}$\ & SiO$_{\nu=1}$\ & $F_{1.3\,\mathrm{mm}}$ & Line$^{(a)}$ & Cont. & Additional Lines\\
 &&  & $J=2{-}1$ & $J=5{-}4$ & $J=5{-}4$ &  & sens. & sens. & \\
 && [$10^{-6}$\,M$_\odot$\,yr$^{-1}$] & \multicolumn{3}{c|}{Peak flux (mJy)} & (mJy) & (mJy) & ($\mu$Jy) & \\
\hline\\[-2ex]
\textbf{RSGC2} & 1$^{(b)}$ & 13.5$^{(c)}$ & 75.5 & 260 & 58.2 & 0.82 & 3.0 & 68 & SO $5_5-4_4$ \\
&2 & 13 & 17.3 & 57.3 & 96.4 & 0.52 & 2.5 & 69 & -- \\
&3 & 0.51 & -- & -- & 5.95 & 0.11 & 0.7 & 21 & -- \\
&5 & 5.8 & 7.26 & 7.37 & -- & -- & 2.3 & 72 & -- \\
&6 & 0.2 & -- & -- & -- & 0.12 & 0.6 & 19 & -- \\
&8 & 0.26 & 17.6 & 2.24$^{(d)}$ & 4.24 & -- & 0.8 & 16 & -- \\
&49 & 770 & 30.4 & 152$^{(d)}$ & 24.8 & 0.25 & 2.1 & 74 & SO $5_5-4_4$, H$_2$O$_{\nu_2=1}$ $5_{5,0}-6_{4,3}$ \\
&52$^{(e)}$ & 1.4 & 348 & 11.2 & 6.09 & 17.2 & 0.8 & 21 & H$_2$O$_{\nu_2=1}$ $5_{5,0}-6_{4,3}$ \\
\hline\\[-2ex]
\textbf{RSGC1} & 1 & 17 & 9.19 & -- & -- & 0.64 & 2.2 & 56 & -- \\
&2 & 9 & 9.01 & -- & -- & 0.65 & 2.4 & 47 & -- \\
&3 & 15 & 13.5 & -- & -- & 0.27 & 2.2 & 49 & -- \\
&4 & 9.7 & 11.2 & -- & -- & -- & 3.2 & 67 & -- \\
&13 & 27 & 26.2 & -- & -- & 0.36 & 2.6 & 50 & -- \\
\hline\hline
\end{tabular}
\tablefoot{SED mass-loss rates are adopted from \citet{humphreys2020} unless noted otherwise. CO and SiO peak fluxes are obtained using a shell profile model fit to ISM-free channels. No SiO observations were obtained for RSGC1 sources. Sources with no detections in any observed tracer are excluded from the table (RSGC2: 68, 10, 11, 14, 13, 23; RSGC1: 5--12, 14).  (a) Line sensitivity at the native spectral resolution (1.27\,\kms). (b)  Stephenson\,2-18. Possibly a foreground giant; membership discussed by \citet{negueruela2013} and \citet{humphreys2020}. (c) Mass-loss rate from \citet{fok12}. (d)  Irregular line profile. (e) Extreme morphology, evidence of prior massive ejection; see \citet{Siebert2025_DFK52}. }
\label{tab:detections}
\end{table*}

Our study primarily utilizes ALMA data acquired through Program ID 2023.1.01519.S, which covered a sample of 14 RSGs in RSGC2. Observations were performed with two configurations of the main 12m array using the Band 6 receiver \citep{Ediss2004_Band6}, with noise-level goals based on SED-derived mass-loss rates, $\dot{M}_{\rm SED}$, from \citet{humphreys2020}. RSGC2 sources with no observed IR-excess were not included in this sample, and all targets have $\dot{M}_{\rm SED}>10^{-7}$\,\msunyr (total, assuming a gas-to-dust ratio of 200).

All observations were reduced with the standard ALMA calibration pipeline \citep{Hunter2023_ALMApipeline}, and visibilities from different configurations were combined in the \textit{uv}-plane with antenna-specific weighting. Continuum subtraction was performed in the visibility plane, and imaging was done with the \texttt{CASA} \citep{casa2022} task \texttt{tclean} using Briggs weighting and a robust parameter of 0.5 for molecular lines. For continuum images, natural weighting was utilized to optimize the detection limit. After array combination and imaging, final brightness sensitivities range from $0.7-3$\,mJy for spectral lines and $16-70$\,$\mu$Jy/beam for continuum.
% , depending on the source 
(Table~\ref{tab:detections}). The achieved spatial resolution was 0.35\arcsec\/ at 230\,GHz (${\sim}3\times10^{16}$\,cm at 5.8\,kpc) with a maximum recoverable scale (MRS) of 12.5\arcsec. Four spectral windows were positioned in the $214-233$\,GHz range targeting $^{12}$CO $J=2-1$ at 230.538\,GHz and SiO $J=5-4$ at 215.596\,GHz. The spectral resolution was 1.27\,\kms, and the total bandwidth was 1.8\,GHz. 

Flux loss due to spatial filtering is estimated to be small given the distance of RSGC2 \citep[5.8\,kpc;][]{davies2007_rsgc2} and the physical scales probed. 
Because the full sample of RSGs studied in this work have spatial extents that are more compact, by a factor of ${\sim}10$, than DFK\,52, which was discussed in detail by \citet{Siebert2025_DFK52}, we expect spatial filtering to be negligible. The flux calibration error was taken as the standard 5\%$-$10\% for ALMA observations \citep{Cortes2020_ALMAtechnical}.

A common feature across all our observations was a significant amount of contamination from the interstellar medium (ISM) in CO. 
In our treatment of the molecular emission, we inspected image cubes channel-by-channel to determine which velocity bins are affected, and removed those from further analysis. On average, we found that in the observed \twco $J=2-1$ lines, ${\sim}60$\% of the velocity extent is free of this contamination for our targets. For continuum maps, line-free channels were used in the imaging process, and the $^{12}$CO spectral window was omitted entirely to ensure the ISM features were excluded. 

We also utilized the archival ALMA data for RSGC1 (ID 2013.1.01200.S) presented also by \citet{decin2024_rsgc1} to enable a self-consistent comparison of molecular emission between the two clusters. These observations reach a comparable angular resolution (0.42\arcsec) to that achieved for RSGC2 and also target the \twco $J=2-1$ line with the same spectral resolution. The main difference in these data is the significantly smaller MRS of 4.4\arcsec\/ ($\sim$$4\times10^{17}$\,cm at 6.6\,kpc), since only one array configuration, with fewer short baselines, was used. Additionally, the SiO $J=5-4$ transition was not included in the spectral tuning. We reduce and image these data with the same methods detailed above, and restrict our analysis for RSGC1 to the \twco and continuum images. The line sensitivity reached in these observations is comparable to that of the brightest RSGC2 targets ($\sigma{\sim}2.5$\,mJy/beam).

\begin{figure*}[!t]
    % --------- Row 1 ---------
    \centering
    \includegraphics[width=0.3\linewidth]{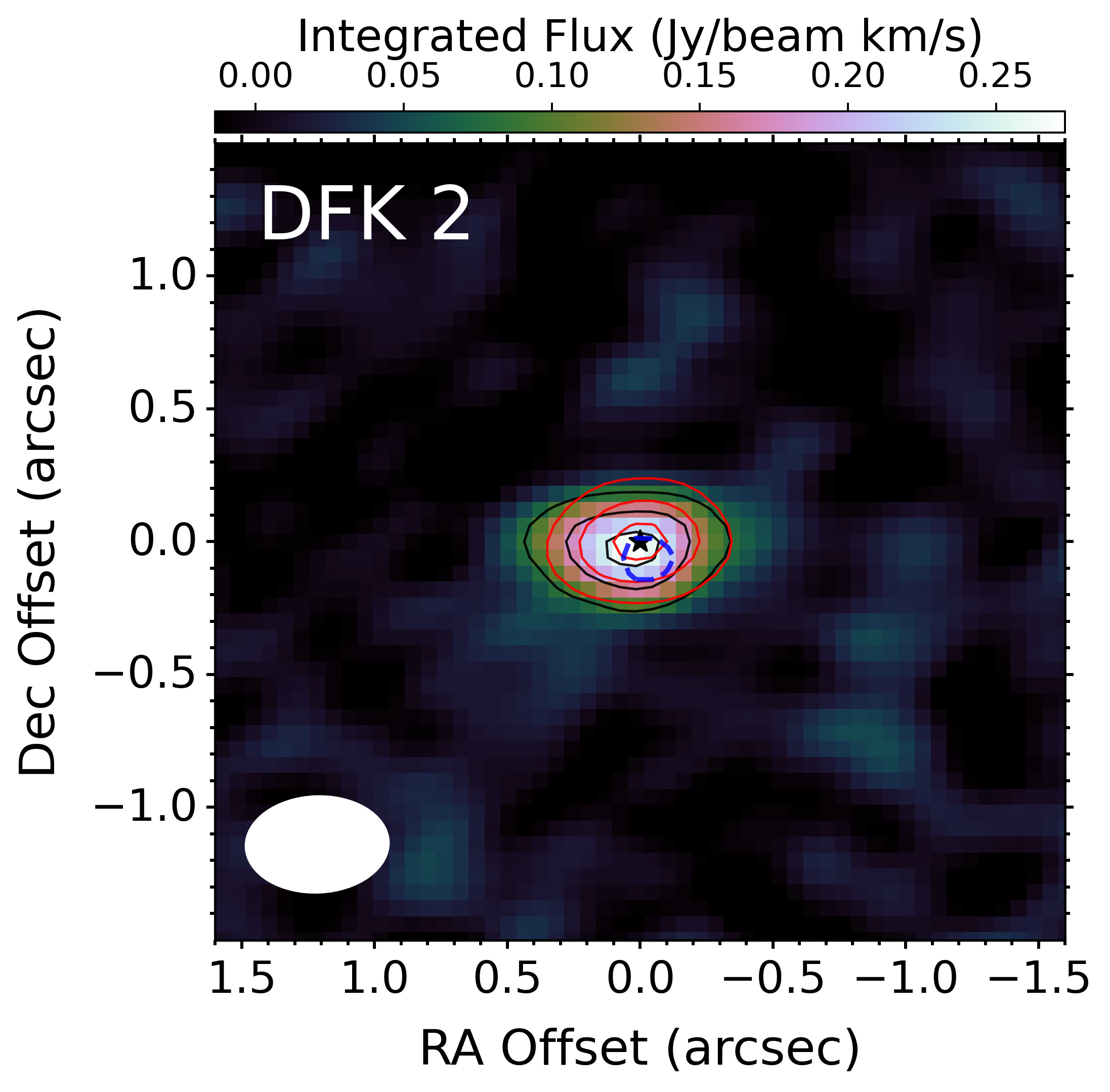} \label{fig:DFK_2_summA}
    \includegraphics[width=0.3\linewidth]{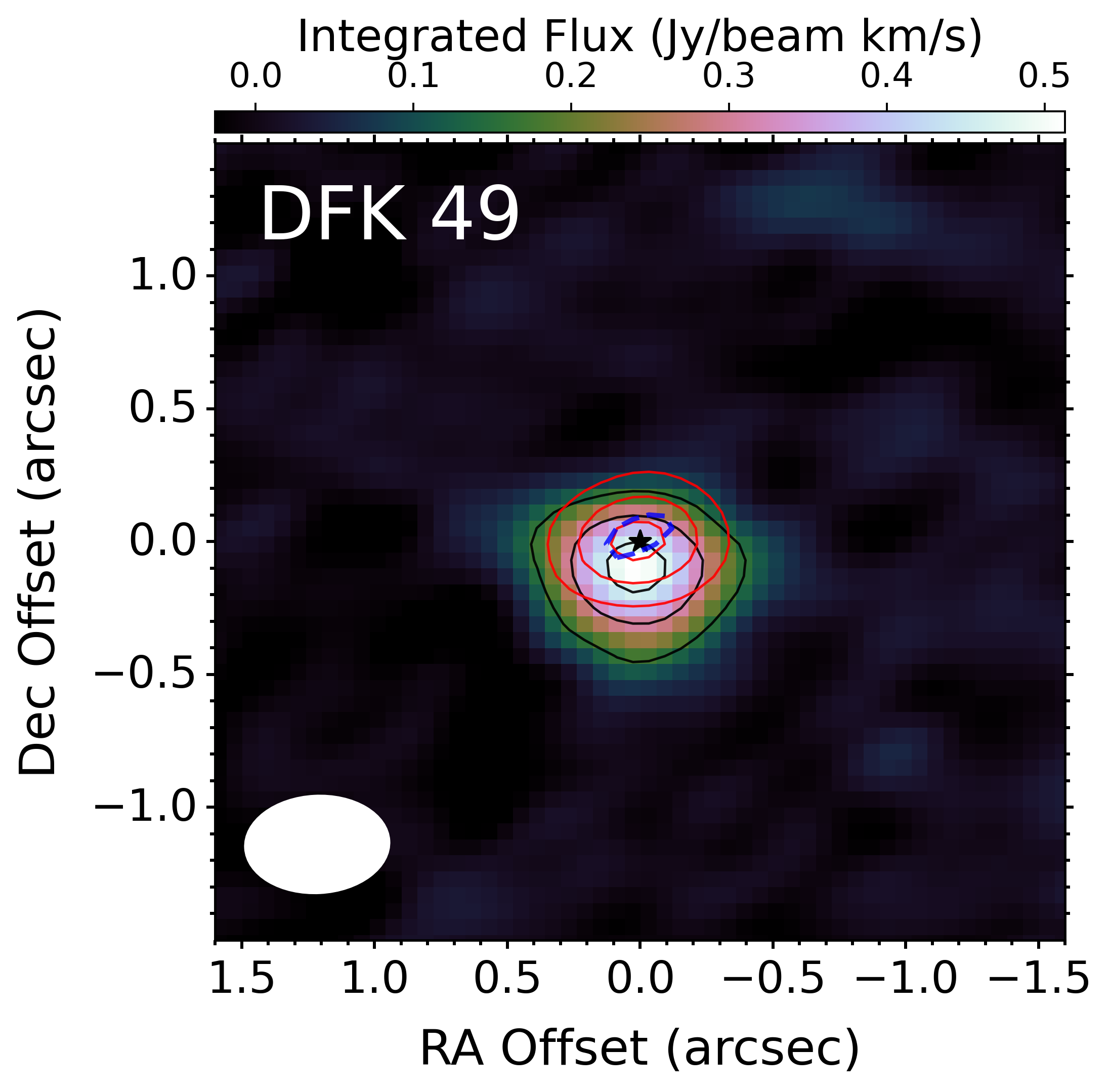} \label{fig:DFK_49_summA}
    \includegraphics[width=0.3\linewidth]{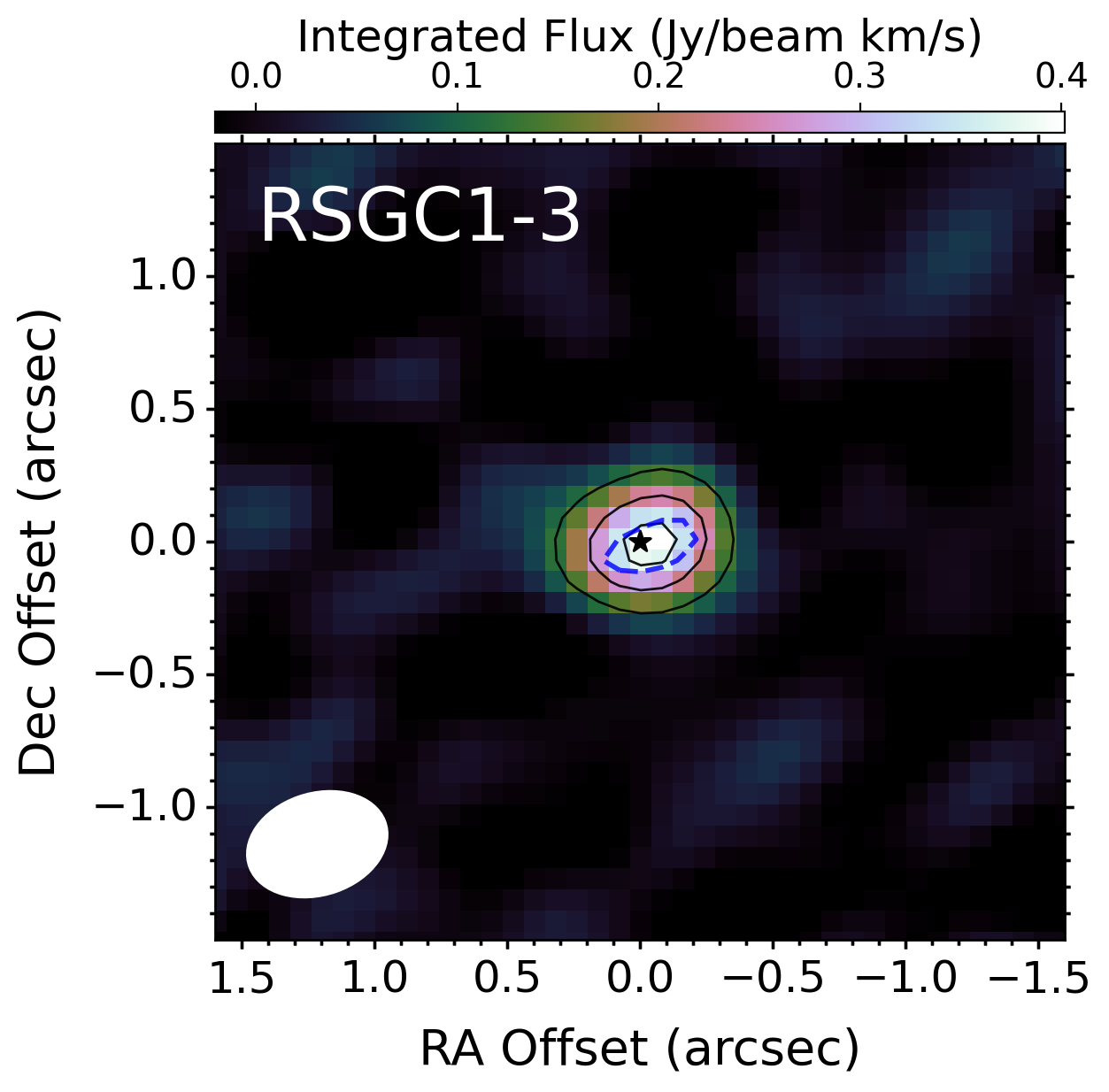} \label{fig:DFK_1_summA}
 
    % --------- Row 3 ---------
    \includegraphics[width=0.3\linewidth]{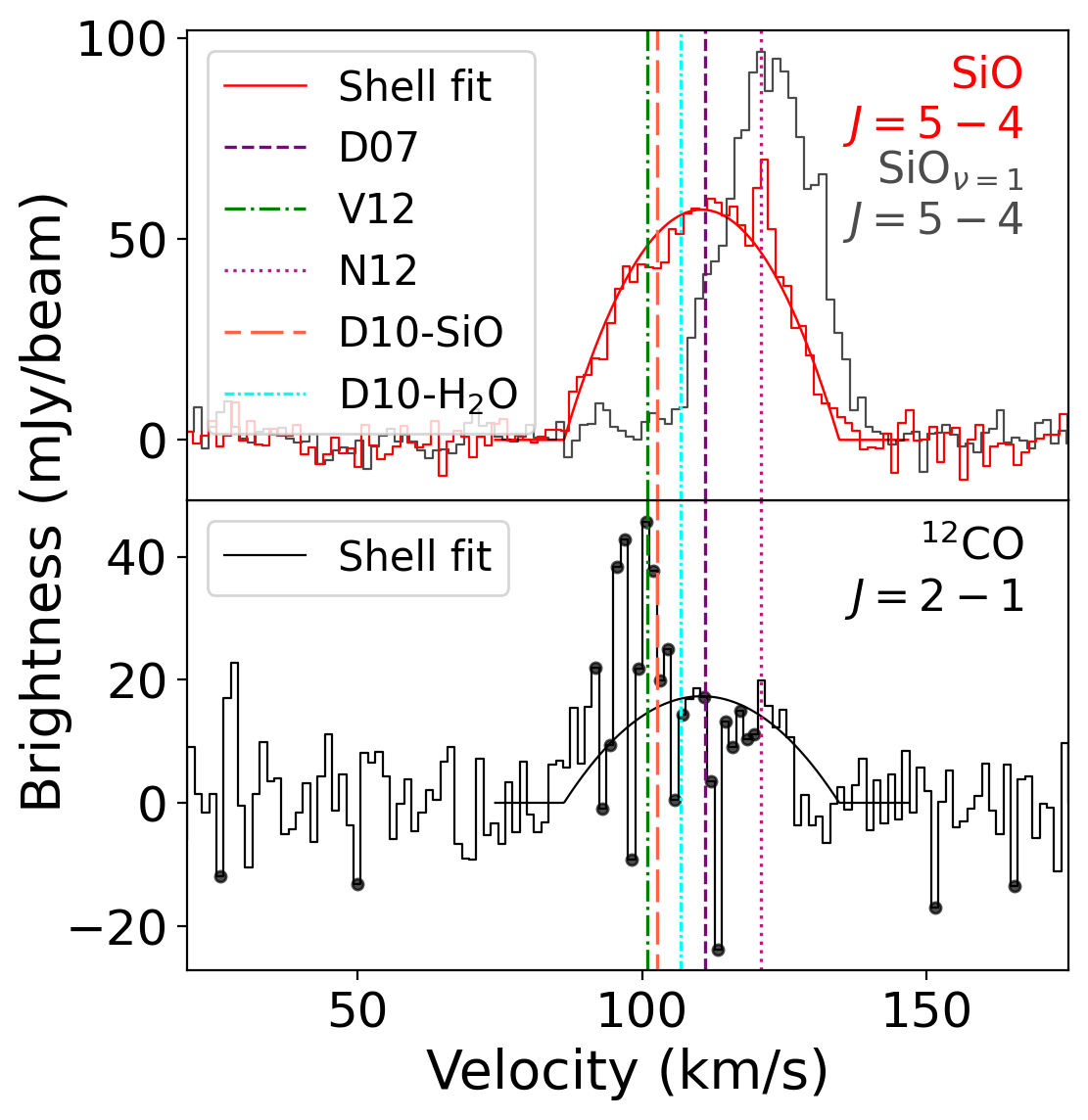} \label{fig:DFK_2_summB}
    \includegraphics[width=0.3\linewidth]{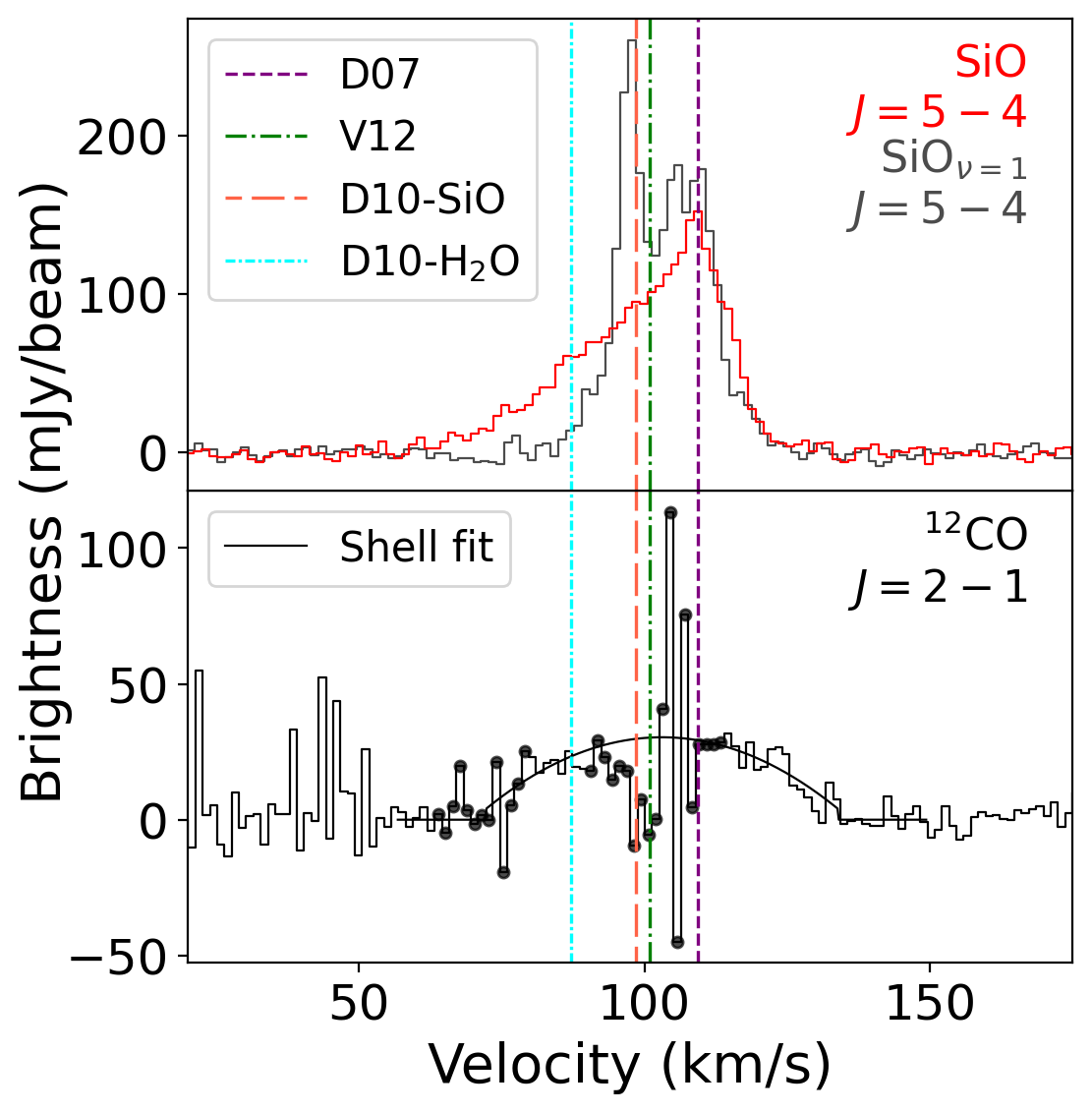} \label{fig:DFK_49_summB}
    \includegraphics[width=0.3\linewidth]{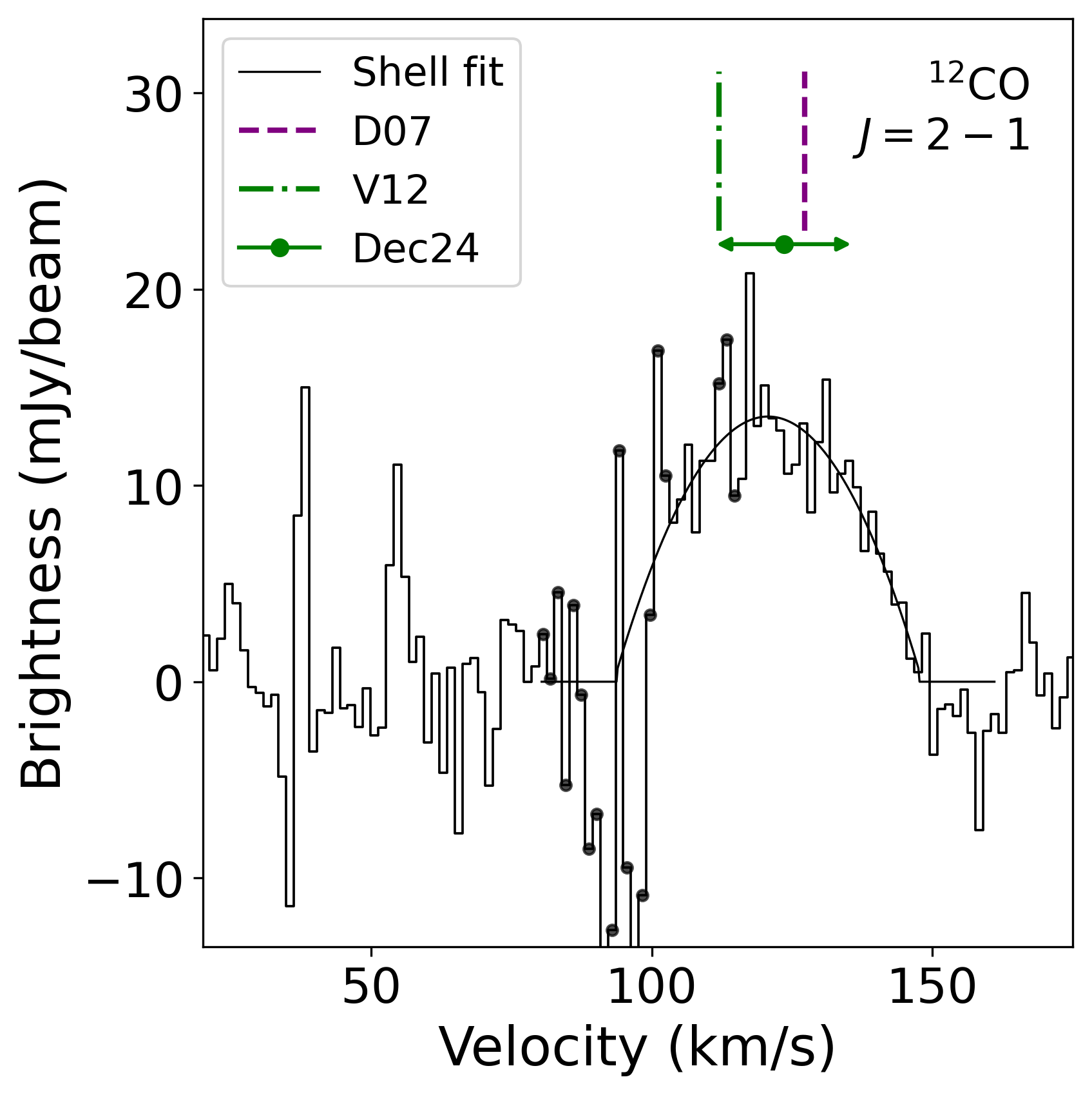} \label{fig:DFK_1_summB}

    \caption{
    ALMA observations for DFK\,2 (\textit{left}), DFK\,49 (\textit{middle}), and RSGC1-3 (\textit{right}). \textit{Top:} Images show the fit stellar positions (\textit{black star}), contours at 0.3, 0.6, and 0.9 times the peak intensity (when above $3\sigma$) for integrated CO (\textit{black}; \textit{color map}; ISM-free) and SiO (\textit{red}) and 1.3\,mm continuum (\textit{blue dashed}) emission, and the ALMA synthesized beam (\textit{white filled ellipse}). 
    \textit{Bottom:} Spectra for \twco $J=2-1$ (\textit{black histogram}), SiO$_{\nu=0}$ $J=5-4$ (\textit{red histogram}), and SiO$_{\nu=1}$ $J=5-4$ (\textit{gray histogram}) extracted at the stellar position from a 0.6\arcsec\ beam. A shell-profile fit (Sect.~\ref{sect:linefits}) is overlaid on symmetric CO and SiO$_{\nu=0}$ lines (\textit{black} and \textit{red}, respectively). Source velocities from the literature are marked as vertical lines: D07 \citep{davies2007_rsgc2}, V12 \citep{verheyen2012}, N12 \citep{negueruela2012}, and D10 \citep{deguchi2010}. ISM-contaminated channels are indicated with circles. For RSGC1-3, the CO line center and width (FWZM) derived by \citet{decin2024_rsgc1} are also indicated (\textit{green point and arrow}).
    }\vspace{-1em}
    \label{fig:obs_summary}
\end{figure*}

\section{Results}\label{sect:analysis_results}
\subsection{Summary of detections}

A list of molecular-line and 1.3\,mm continuum detections for RSGs in both clusters, as well as the survey sensitivities, is provided in Table~\ref{tab:detections}. In RSGC2, we detect continuum emission toward 6 out of 14 targets with fluxes on the order of 0.1--1\,mJy. The exception to this is the unique source DFK\,52 which has a complex extended dust component producing a much higher total flux \citep{Siebert2025_DFK52}. We also detect SiO$_{\nu=1}\,J=5-4$ maser emission from 6 RSGs. The strongest three (DFK\,1, DFK\,2, and DFK\,49) have other known masers found by previous maser surveys \citep{deguchi2010,verheyen2012} of the RSGC clusters. For the other three (DFK\,52, DFK\,3, and DFK\,8), these are the first reported maser detections.  SiO$_{\nu=0}\,J=5-4$ emission is detected in all objects with a \twco $J=2-1$ detection and typically exhibits equal or higher peak brightness.

\twco emission is detected toward 6 out of 14 RSGC2 sources, with only four sources detected in both CO and continuum, while two CO detections lack continuum counterparts and two continuum detections lack CO counterparts (Table~\ref{tab:detections}). In general, CO detections are found for sources with highest $\dot{M}_{\rm SED}$; however, surprisingly, some sources with bright CO emission (DFK\,8, most notably) have low SED mass-loss rates and conversely some targets with substantial $\dot{M}_{\rm SED}$ are only marginally or not at all detected in CO (e.g.\ DFK\,5, DFK\,3). Without any additional modeling one would naively assume the CO brightness to correlate with $\dot{M}_{\rm SED}$, so these mismatches suggest that there could be shortcomings in the SED models and/or the role of CO as a mass-loss tracer.

Toward RSGC1, five targets are detected in \twco, four of which have accompanying continuum detections. These are the same detections that are discussed in detail by \citet{decin2024_rsgc1}. The range of peak continuum and line intensities are similar between the two clusters, with the exception of DFK\,52 in RSGC2. However, the detections toward RSGC2 span a wider range in $\dot{M}_{\mathrm{SED}}$, reflecting the larger range in luminosity and implied initial mass noted by previous comparisons of the clusters \citep{Eldridge2020_RSGbinary_populations,humphreys2020}. The non-detected RSGC1 members have $\dot{M}_{\rm SED}$ on the order of a few $10^{-6}$\,\msunyr, whereas those in the RSGC2 survey have a few $10^{-7}$\,\msunyr \citep{beasor2020,humphreys2020}. 

\subsection{Spatial distribution of millimeter emission}\label{sect:spatial}
Figure~\ref{fig:obs_summary} shows a summary of the mm-continuum and circumstellar rotational-line emission for selected RSGs; others can be found in App.~\ref{app:linefits}. Toward RSGC2 members, we note that SiO$_{\nu=0}\,J=5-4$ and mm-continuum distributions are almost all unresolved and spatially coincident. Because the SiO line is generally detected at higher signal-to-noise than the continuum, we use it to measure the stellar positions using the \texttt{CASA} task \texttt{imfit}, and report these values in Table~\ref{tab:linefits}. For RSGC1 targets, the continuum peak is used for the stellar position instead, as SiO$_{\nu=0}\,J=5-4$ was not observed.

\begin{figure*} [!t]
    \centering
    \includegraphics[width=0.85\linewidth]{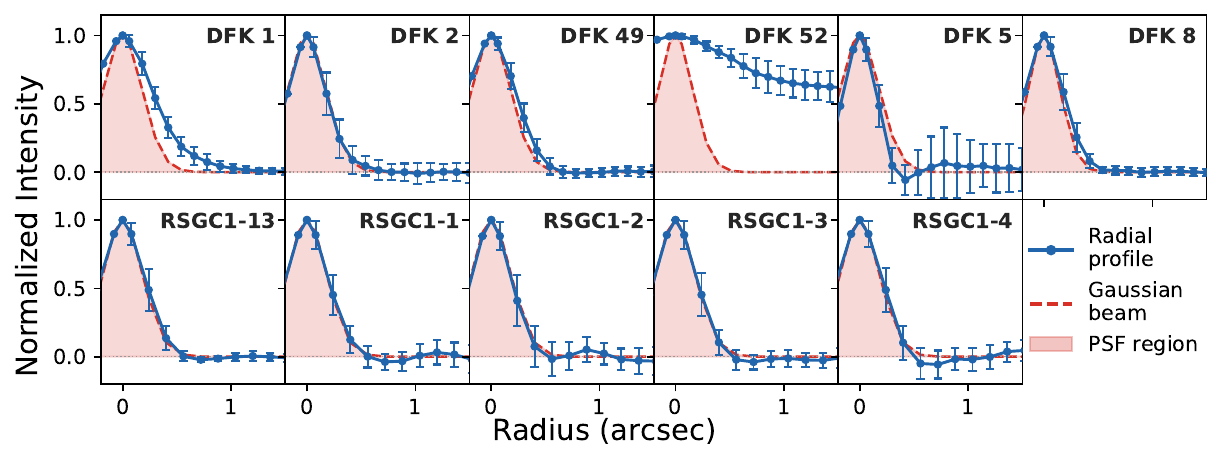}
    \vspace{-0.2cm}
    \caption{Radial profiles of \twco emission for all detected RSGC1 and RSGC2 members, 
    based on the integrated-intensity maps excluding ISM-affected channels. Error bars represent the 1$\sigma$ standard deviation 
    in each radial bin. The brightness profile of the point spread function (PSF) is plotted as reference for each source, normalized to the peak intensity at the stellar position.}
    \label{fig:radial_profs}
\end{figure*}

The integrated \twco $J=2-1$ maps\footnote{Channel maps of all detected RSGs in this study are also available at Zenodo: \url{https://doi.org/10.5281/zenodo.22794283}}
toward RSGC2 members show only marginally or entirely unresolved spatial extents ($\lesssim0.^{\prime\prime}6\approx3500$\,au), apart from the extreme source DFK\,52 \citep{Siebert2025_DFK52}. Figure~\ref{fig:radial_profs} explores this further, showing the radial distribution of integrated \twco emission (excluding ISM-contaminated channels), averaged over bins with width equal to half the beam major axis. When compared with the characteristic radial profile for an unresolved point source (the clean beam), DFK\,2, DFK\,5, and DFK\,8 exhibit no statistically significant emission that would suggest a spatially resolved CO envelope. The two high-$\dot{M}_{\mathrm{SED}}$ sources DFK\,49 and DFK\,1 exhibit some slightly extended circumstellar CO emission out to 0.6\arcsec--1.0\arcsec. This is apparent in their moment-zero maps (Fig.~\ref{fig:obs_summary}). DFK\,49 also shows a slight offset (0.1\arcsec) in its CO emission peak from the position of SiO and the continuum.

For RSGC1 members, Fig.\ \ref{fig:radial_profs} shows virtually no deviation from the Gaussian restoring beam in the detected \twco emission. Given the larger distance to RSGC1 \citep[6.6\,kpc;][]{davies2008} and the slightly larger beam, the upper limit on the physical size of the CO-emitting region is higher for RSGC1 stars (${\sim}2800$\,au) than it is for unresolved RSGC2 sources (${\sim}2000$\,au). As noted by \citet{decin2024_rsgc1}, it is possible that the \twco emitting region is larger than the ALMA beam but below the brightness-sensitivity limit given that the images are noise limited (for both clusters). However, we argue that if this were the case, it would produce a noticeable impact on the radial profiles in Fig.~\ref{fig:radial_profs} as well as on the emerging line shapes which we discuss in the following section. 

\subsection{Line profiles}\label{sect:linefits}

The spectra extracted from the stellar position (Figs.\ \ref{fig:obs_summary} and \ref{fig:obs_summary_full}, lower panels) show a variety of emission profiles and demonstrate the impact of ISM contamination on CO lines in both clusters. Contaminated channels were identified in channel maps, appearing as large-scale bands of emission and absorption varying greatly from channel-to-channel. Toward all RSGC2 sources (including non-detections) a prominent grouping of ISM channels appears around $v_{\mathrm{LSR}}\approx104$\,\kms, which is quite close to the cluster velocity of 109\,\kms \citep{davies2007_rsgc2}, making determination of systemic velocities difficult with \twco alone. Toward RSGC1, interference from line-of-sight ISM is also present in the range 80\,--\,115\,\kms. In contrast to RSGC2, this only cuts into the blue-shifted side of the \twco line, so channels from the line center and red-shifted wing are without contamination.

The SiO$_{\nu=0}\,J=5-4$ lines observed for RSGC2 sources generally show parabolic profiles, consistent with optically thick, spatially unresolved emission in a circumstellar envelope. We find a highly asymmetric line profile for DFK\,49, indicating either non-thermal excitation mechanisms or non-uniform kinematics in this wind. Toward DFK\,2 and DFK\,8 we also note narrow spikes in the line profile, which may be associated with ISM-affected velocity ranges or non-thermal excitation. SiO emission is not expected under typical ISM conditions, but it is known to trace interstellar shocks \citep{Rybarczyk2023_SiOshocks_MCs} which might not be uncommon given the turbulence and stellar feedback inherent to a massive-cluster environment. Unlike \twco, however, channel maps of this irregular SiO emission show no spatial components located away from the stellar position, so we consider such a cluster origin unlikely.

To determine systemic velocities, terminal wind velocities, and integrated line strengths in the presence of non-uniform ISM contamination, we fit empirical line profiles to the SiO and CO lines, as defined in the \texttt{CLASS/GILDAS} software package\footnote{\href{https://www.iram.fr/IRAMFR/GILDAS/doc/html/class-html/node38.html}{https://www.iram.fr/IRAMFR/GILDAS/doc/html/class-html/node38.html}} as ``shell'' profiles:
\begin{equation}
    S(v) = \frac{A}{2v_{\mathrm{exp}}\left(1+H/3\right)}\left(1+4H\left(\frac{v-v_\mathrm{LSR}}{2v_{\mathrm{exp}}}\right)^2\right),
\label{eq:shell}
\end{equation}
where $A$ is the integrated line flux, $v_{\mathrm{exp}}$ is the expansion velocity, and $H$ is a shaping parameter ranging from $-1$ (parabolic) to 0 (flat-topped) and $>0$ (double-horn). To measure the width and position of spectral-line emission for each detected star in RSGC2, we fit the thermal SiO profile on a grid of $v_\mathrm{LSR}$ and $v_{\mathrm{exp}}$, with $H$ fixed to $-1$ and $A$ left as a free parameter. Uncertainties and covariances are calculated from the resulting $\chi^2$-surface, as well as bootstrap resampling of the input spectra (App.~\ref{app:linefits}). These uncertainties are non-uniform across the data set, primarily due to the variable impact of ISM on the observed line.
For \twco, $v_{\mathrm{exp}}$ and $v_{\mathrm{LSR}}$ were fixed to the SiO best-fit values (due to the strong ISM contamination) and $H$ was left as a free parameter, as these lines do not strictly appear parabolic in these data. In the case of DFK\,49, the asymmetric SiO line prevents us from constraining the $v_{\mathrm{LSR}}$ and $v_{\mathrm{exp}}$ in this way, so we instead adopt values consistent with the approximated center and total velocity extent of both CO and SiO.

The resulting line-profile fits for RSGC2 targets are summarised in Table \ref{tab:linefits} and Figs.\ \ref{fig:obs_summary} and \ref{fig:obs_summary_full}.
The $v_{\rm LSR}$ obtained here is compared with the radial velocities from \citet{davies2007_rsgc2} measured using the edge of the CO bandhead at 2.3\,$\mu$m, SiO and H$_2$O maser measurements from \citet{verheyen2012} and \citet{deguchi2010}, and optical spectroscopy of \citet{negueruela2012}. We find that our reported velocities are typically in statistical agreement with the measurements of \citet{davies2007_rsgc2} given the nominal error (2\,\kms) in that work and the uncertainties in our line-profile fits. 
For objects with previously measured SiO masers, their peak velocities fall within ${\sim}5$ \kms of our derived \vLSR values, and the SiO$_{\nu=1}$ $J=5-4$ transition observed in this work traces similar velocity components. Toward DFK\,2 we observe vibrationally-excited SiO emission centered at \vLSR+10\,\kms which matches the position of the narrow (4 channels $\approx$\,5\,\kms) component seen on top of the ground state SiO line (Fig.~\ref{fig:obs_summary}), suggesting a region of non-thermal excitation in this source. We note that this velocity is also coincident with the atmospheric modeling result of \citet{negueruela2012}, who fit the Doppler shift using absorption lines in the 8370--8900\,Å range where O$_2$ is the most prominent feature \citet{Negueruela2011_RSGC3}, so the narrow SiO component could be tracing a similar location.

For DFK\,49, we find a low $v_{\rm LSR}$ value of 103.1\,kms, which is in line with previous maser studies, but deviates from the results of \citet{davies2007_rsgc2} by $-6$\,km/s. This downward correction brings the $v_{\rm LSR}$ of DFK\,49 closer to that of DFK\,1. The latter has previously been proposed as foreground giant owing to its apparent brightness and $v_{\mathrm{LSR}}$ (93\,\kms). But now we find that both of these stars, located together in the ``Ste 2 SW" region, might have more similar $v_{\rm LSR}$ values than previously reported, suggesting a velocity (and potentially distance) gradient along the elongated NE-SW direction of RSGC2. However, it is important to note that DFK\,5, the only other RSG detected by ALMA in the Ste 2 SW sub-region, has a velocity consistent with the cluster mean.

We perform the same profile fitting for the RSGC1 sources and report the measured intensities and velocity constraints in Table \ref{tab:linefits} and Fig.~\ref{fig:obs_summary}. However, in this case, the thermal SiO line was not observed, so \vLSR and \vexp are determined from the $^{12}$CO emission alone. We again flag and ignore ISM-affected channels, and perform a $\chi^2$-analysis to examine how this contamination impacts our ability to constrain these parameters (App.~\ref{app:linefits}). Under the assumption of a shell line profile, measured line widths and positions from these data are uncertain due to a combination of low signal-to-noise, and the consistent impact of ISM which appears at $-20$\,\kms relative to the cluster velocity \citep[123\,\kms][]{davies2008}. The lack of these channels introduces a degeneracy between $v_{\mathrm{exp}}$ and $v_{\mathrm{LSR}}$, and errors in the fit are consequently of the order of $1-3$\,\kms. For RSGC1-1 and RSGC1-4, two local minima are present in the $\chi^2$-surface, so we list both corresponding fits in Table \ref{tab:linefits}.

The wind expansion velocities we derive for RSGC1 members differ substantially from those reported by \citet{decin2024_rsgc1} using the same data set, and the values we obtain are systematically larger by a factor of 1.5--3. The reason for this is that the authors in that work assume the CO lines are somewhat spatially resolved and exhibit low-to-medium optical depth ($\tau<2$), producing double-peaked emission profiles which cover much smaller velocity extents (see App.~\ref{app:linefits}). However, as we note in the previous section, we find that there is no significant evidence of extended emission for these objects, and the observed line shapes are more parabolic based on our empirical fits, indicating optically thick unresolved emission. We also make sure to allow the line fit to extend into the heavily contaminated blue-shifted channels for these sources (e.g. $<105$\,\kms for RSGC1-3 in Fig.\ \ref{fig:obs_summary}), as it is important to ensure that the ISM absorption edge is not treated as an intrinsic property of the RSG wind. After taking this into account, our best-fit expansion velocities for RSGC1 targets are in the 20\,--\,35\,\kms range instead of the 8\,--\,15\,\kms obtained by \citet{decin2024_rsgc1}.

\begin{figure}[!t]
    \centering
    \includegraphics[width=\linewidth]{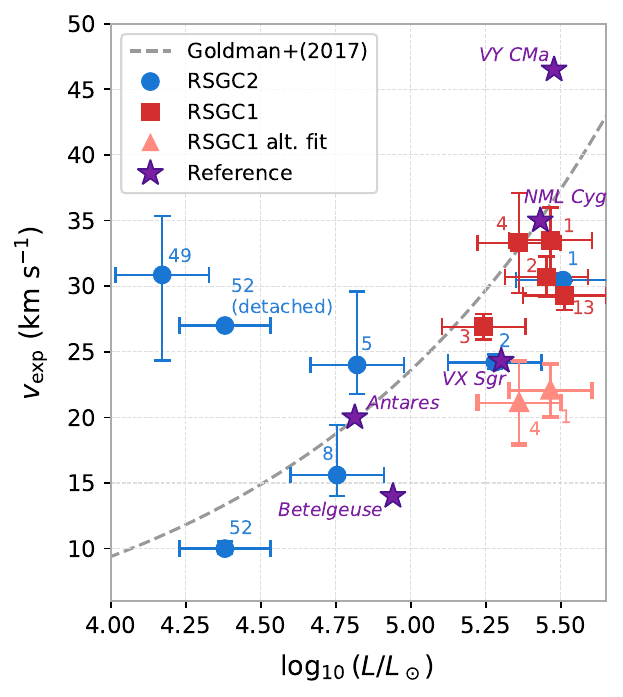}
    \caption{RSG wind-expansion velocities ($v_{\mathrm{exp}}$) as a function of stellar luminosity. $v_{\mathrm{exp}}$ values and uncertainties for RSGC1 (\textit{red squares}) and RSGC2 members (\textit{blue circles}) are determined with the empirical line fits described in Sect.~\ref{sect:linefits}, using the \twco line for RSGC1 sources and DFK\,49, and SiO$_{\nu=0}\,J=5-4$ for all others. For RSGC1 objects with two acceptable fits, an alternate expansion velocity is shown (\textit{pink triangles}). Cluster RSG luminosities are determined from SED fits in App.~\ref{app:SEDs}. The $v_{\rm exp}\propto ZL^{0.4}$ relation from \citet{Goldman2017_windspeeds} is shown as a dashed line, taking their value of $Z=0.236$ for stars with solar metallicity. Nearby RSGs are plotted for comparison (\textit{purple stars}), with $v_{\rm exp}$ measurements from \citet{Braun2012_alphaSco_hydro}, \citet{ogorman2012_alphOri_CO}, \citet{debeck2010_comdot}, and \citet{debeck2025_nmlcyg}. }
    \label{fig:vexp_lum}
\end{figure}

The measured $v_{\rm exp}$ values for both clusters are plotted as a function of luminosity in Fig.~\ref{fig:vexp_lum}, along with a comparison to nearby well-studied RSGs. Excluding the obvious outliers DFK\,49 and the fast, large-scale, detached component of DFK\,52, we observe that expansion velocities form a positive trend with luminosity consistent with observations of nearby RSGs. The RSGC1 members are grouped near the upper range of velocities, whereas the wider spread in RSGC2 wind speeds likely reflects the larger range of luminosities in this cluster. This trend is in agreement with the $v_{\rm exp}\propto L^{0.4}$ relation derived by \citet{Goldman2017_windspeeds} for galactic AGBs and RSGs using OH-1612 MHz maser emission. We emphasize the effect this may have on current mass-loss prescriptions, as SED constraints on RSG-$\dot{M}$ typically adopt a fixed $v_{\rm exp}$ across samples. Because dust-derived mass-loss rates scale linearly with $v_{\rm exp}$, the results of this work and \citet{Goldman2017_windspeeds} suggest that this assumption could introduce a luminosity-dependent bias in the empirically derived $\dot{M}-L$ relations.

Figure \ref{fig:vexp_lum} demonstrates that DFK\,49 and (the extended component of) DFK\,52 exhibit much higher wind speeds than any RSGs in their luminosity class, suggesting a unique evolutionary state for both sources. DFK\,52 has a complex, detached circumstellar component described in detail by \citet{Siebert2025_DFK52}. However, its present-day compact wind traced by SiO emission suggests an expansion velocity in agreement with the trend seen for other RSGs. This would suggest a recent change in the stellar luminosity, or the wind launching mechanism.

DFK\,49 was previously classified as a post-RSG for its large infrared excess, earlier spectral type (K4), and high amount of near-IR extinction \citep{davies2007_rsgc2,humphreys2020}, but in Appendix~\ref{app:SEDs}, we revisit its full SED with new long-wavelength constraints and find that its bolometric luminosity has previously been overestimated by an order of magnitude. This correction now makes it the least luminous target across both samples, which is striking given its strong and asymmetric \twco and SiO emission observed here with $v_{\rm exp}>30$\,\kms, which is more typical of ``extreme'' RSGs like VY CMa or the prototypical YHG IRC+10420 \citep{QL2016_IRC10420}. Unlike DFK\,52, DFK\,49 shows no current evidence of a recent change in its wind expansion velocity, as no narrow line component has been detected.

\begin{table*}[!t]
\caption{Derived positions, and spectral line measurements for ALMA-detected RSGs in RSGC2 and RSGC1.} 
\label{tab:linefits}
\centering
\begin{tabular}{p{1cm}l|c c|c c|c c|c c c}
\hline\hline
\rule{0pt}{2.5ex}
\rule{{0pt}}{{2.5ex}}
&ID & \multicolumn{2}{c|}{Position$^{a}$} & SiO$_{5-4}$$\int S\mathrm{d}v$ & $^{12}$CO$_{2-1}$$\int S\mathrm{d}v$ & $v_{\mathrm{exp}}$ & $v_{\mathrm{LSR}}$ & $v_{\mathrm{LSR}}^{b}$ & $v_{\mathrm{LSR}}^{c}$ & $v_{\mathrm{LSR}}^{d}$ \\
\rule{{0pt}}{{2.5ex}} 
&& $\alpha$(J2000) & $\delta$(J2000) & (Jy \kms) & (Jy \kms) & (\kms) & (\kms) & \multicolumn{3}{c}{(\kms)} \\
\hline\\[-2ex]
% \multicolumn{10}{l}{\rule{0pt}{2.5ex}  \\
% \hline
\rule{0pt}{2.5ex}\textbf{RSGC2} &1 & 18:39:02.370 & -06:05:10.64 & $10.54_{ \pm 0.1}$ & $9.16_{ \pm 0.4}$ & $30.5_{ \pm 0.3}$ & $93.4_{ \pm 0.2}$ &  94.0 & 89.0 & 92.7 \\
\rule{0pt}{2.5ex} &2 & 18:39:19.607 & -06:00:40.86 & $1.85_{ \pm 0.03}$ & $0.56_{ \pm 0.08}$ & $24.2_{ \pm 0.4}$ & $110.5_{ \pm 0.3}$ &  111.1 & 101.0 & 102.7 \\
\rule{0pt}{2.5ex} &5 & 18:39:08.040 & -06:05:24.38 & $0.24_{ \pm 0.04}$ & $0.28_{ \pm 0.08}$ & $24.0_{ \pm 3}$ & $108.4_{\pm 2}$ &  113.3 & -- & -- \\
\rule{0pt}{2.5ex} &8 & 18:39:19.882 & -06:01:48.21 & $0.05_{ \pm 0.006}$ & $0.46_{ \pm 0.06}$ & $15.6_{ \pm 2}$ & $102.0_{ \pm 1}$ &  104.1 & -- & -- \\
\rule{0pt}{2.5ex} &49 & 18:39:05.566 & -06:04:26.65 & $3.70_{ \pm 0.3}$$^{{e}}$ & $1.33_{ \pm 0.07}$ & $30.9_{ \pm 2}$$^{{f}}$ & $103.1_{ \pm 1}$$^{{f}}$ &  109.4 & 101.0 & 98.6 \\
\rule{0pt}{2.5ex} &52 & 18:39:23.407 & -06:02:16.12 & $0.16_{ \pm 0.006}$ & -- & $10.0_{ \pm 0.4}$ & $108.4_{ \pm 0.3}$ &  111.2 & -- & -- \\
\hline\hline\\[-2ex]
\rule{0pt}{2.5ex}\textbf{RSGC1}&1 & 18:37:56.306 & -06:52:32.32 & -- & $0.44_{ \pm 0.04}$ & $33.5_{ \pm 2}$ & $112.5_{ \pm 2}$ &  129.5 & 116.0 & -- \\
\rule{0pt}{2.5ex}&  & '' & '' & -- & $0.29 \pm 0.03$ & $22.1_{ \pm 2}$ & $121.5_{ \pm 2}$ &  '' & '' & '' \\
\rule{0pt}{2.5ex}&2 & 18:37:55.292 & -06:52:48.49 & -- & $0.41_{ \pm 0.03}$ & $30.7_{ \pm 2}$ & $114.1_{ \pm 1}$ &  114.2 & -- & -- \\
\rule{0pt}{2.5ex}&3 & 18:37:59.744 & -06:53:49.51 & -- & $0.50_{ \pm 0.03}$ & $26.9 _{\pm 1}$ & $120.7_{ \pm 1}$ &  127.2 & 112.0 & -- \\
\rule{0pt}{2.5ex}&4 & 18:37:50.888 & -06:53:38.49 & -- & $0.52_{ \pm 0.07}$ & $33.3_{ \pm 4}$ & $119.3_{ \pm 3}$ &  121.2 & 121.0 & -- \\
\rule{0pt}{2.5ex}&  & '' & '' & -- & $0.33 \pm 0.06$ & $21.1_{ \pm 3}$ & $128.1_{ \pm 3}$ &  '' & '' & '' \\
\rule{0pt}{2.5ex}&13 & 18:37:58.898 & -06:52:32.15 & -- & $1.20_{ \pm 0.08}$ & $29.3_{ \pm 1}$ & $115.9_{ \pm 2}$ &  125.4 & 113.0 & -- \\
\hline\hline
\end{tabular}
\tablefoot{Stellar positions were determined from a 2D gaussian fit to the integrated SiO map if available, or continuum if not. Systemic and expansion velocities are based on shell–profile fits to the observed SiO line for RSGC2 and $^{12}$CO for RSGC1, unless otherwise noted. Errors derived from bootstrap resampling of the spectral-line fit are shown in subscripts, in units of the last quoted digit. Literature velocities from previous studies are included in the last three columns for comparison. (a) The positional accuracy is 80\,mas for a $5\sigma$ detection, and scales inversely with the S/N \citep{Cortes2020_ALMAtechnical}. (b) $2.3\mu$m CO band-head measurement from \citet{davies2007_rsgc2,davies2008}, (reported error $\pm2$\kms). (c) SiO$_{\nu=1}$ $J=2-1$ maser velocity from \citet{verheyen2012}. (d) SiO$_{\nu=1}$ $J=1-0$ maser velocity from \citet{deguchi2010}. (e) No fit performed due to asymmetric line profile; total flux obtained by direct integration of the spectrum. (f) $^{12}$CO used for velocity determination. }
\end{table*}

\subsection{Dust models}\label{sect:DUSTmodeling}
Previous SED-fitting of RSGC1 and RSGC2 members was performed by \citet{humphreys2020} and \citet{beasor2020}. Both works use the \texttt{DUSTY} radiative transfer code, but with different assumptions on the physical structure of the dust shell. \citet{humphreys2020} vary the density power-law index $n$, allowing for a non-steady mass-loss history, while \citet{beasor2020} fix an $r^{-2}$ profile and instead vary the temperature at the inner boundary of the dust shell. Additionally, the two works assume different optical constants, to which the grain temperature and resulting optical depths are sensitive. As a result, the retrieved dust mass-loss rates differ by ${\sim}1$ order of magnitude, with \citet{humphreys2020} obtaining the higher values. Furthermore, neither study covers all sources in our sample: \citet{humphreys2020} treat both clusters but classify DFK\,1 as a non-member and therefore do not model it, while \citet{beasor2020} do not include RSGC1-4, RSGC1-13, or any RSGC2 targets in their analysis.

To obtain a consistent treatment of the dust mass-loss rates and stellar luminosities across all RSGs studied in this work, we model their SEDs again here using the Monte Carlo radiative transfer code \texttt{RADMC-3D} \citep{Dullemond2012_RADMC3D}. This step is essential for our analysis, as we require a homogeneous set of stellar and circumstellar parameters to serve as input to the CO modeling described in Sect.~\ref{sect:COmodeling}. Re-deriving these parameters within a single, self-consistent framework ensures that any trends identified in the CO emission reflect physical differences between the sources rather than inconsistencies in their SED treatments.

The dust modeling procedure and its results are described in detail in App.~\ref{app:SEDs}; however, in brief, we run models for steady-state winds with a range of $\dot{M}_{\rm SED}$, using a fixed inner radius dust temperature of 1000\,K and a $\rho\propto r^{-2}$ density profile. We also investigate the impact of assumed optical constants on the retrieved wind parameters by computing models for ``traditional'' astronomical silicates \citep[e.g.][]{Draine2003_dustopacs,Oseenkopf1992_silicates}, as well as the laboratory-measured opacities of \citet{Demyk2022_lab_silicates}. The best-fitting models are found via $\chi^2$-minimization from the SED shape after scaling to the observed photometry, and adjusting the luminosities, mass-loss rates, and inner radii accordingly. In general, the dust shell optical depths and luminosities we obtain are consistent with previous studies; however, we note substantially smaller $\dot{M}_{\rm SED}$ measurements in comparison to \citet{humphreys2020} due to our assumption of a non-variable mass-loss rate. This is discussed further, along with a comparison of all dust models with constraints from CO, in Section \ref{sect:dustgas_comp}.

\subsection{CO models}\label{sect:COmodeling}
\subsubsection{Radiative-transfer modeling approach}
We use a one-dimensional non-LTE Monte Carlo radiative transfer code \citep{schoeier2001} to model the CO $J=2-1$ line emission. We describe the CSEs with a standard model: they are formed by radially expanding, spherically symmetric outflows with a constant mass-loss rate \mdot. The stellar temperatures and luminosities assumed in the models are $T_{\star}$ and $L_{\mathrm{fit}}$ from Table~\ref{tab:sed_results}. The inner radius of the CO envelope in the model, $R_{\mathrm{in}}$, is defined as the inner radius of the dust envelope (from the SED models in Section \ref{sect:DUSTmodeling}), corresponding to a temperature $T(R_{\mathrm{in}})=1000$\,K, varying within the range $7-20\,R_{\star}$ across the entire modeled sample. The gas kinetic-temperature profile is described as
\begin{equation}
    T(r) = T(R_{\mathrm{in}}) \times \left(R_{\mathrm{in}}/r\right)^{0.8},
\end{equation}
with a minimum of 10\,K, due to photoelectric heating, in line with typical temperature profiles found by \citet{schoeier2002}. 
The velocity profile across the outflow is assumed to follow
\begin{equation}
    v(r) = v_0 +(v_{\mathrm{exp}}-v_0)\left(1-R_{\mathrm{in}}/r\right)^{\beta},
\end{equation}
with \vexp the expansion velocity derived in Sect.~\ref{sect:linefits}, $v_0=3$\,\kms, and an assumed $\beta=0.5$. The modeling procedure also includes a turbulent velocity of 1\,\kms throughout the CSE. Neither $v_0$, nor the turbulent velocity are currently constrained for the modeled RSGs, but this velocity profile is commonly assumed for the CSEs of evolved stars where the gas is assumed to be accelerated by momentum exchange between dust grains and gas \citep[e.g.][]{decin2010}. We additionally note that tests with constant expansion velocity lead to insignificant differences to the retrieved best-fitting part of the sampled parameter space, providing support for the current analysis as a first step. 
The description of the CO excitation, including CO-H$_2$ collisional rates is taken from \citet{yang2010_CO} and includes the rotational levels $J=0\ldots40$ in the vibrational states $\nu=0,1$. The CO abundance $f_{\mathrm{CO}}=$\,CO/H$_2$ varies radially according to 
\begin{equation}
    f_{\mathrm{CO}}(r) = f_{\mathrm{CO},0} \exp(-\ln(2) \left( r/\rhalf \right)^2),
\end{equation}
with $\rhalf$ the radius at which the abundance has fallen to half of $f_{\mathrm{CO},0}$, the peak value. We calculate $f_{\mathrm{CO},0}$ from the C and O abundances reported by \citet{davies2009} and the assumption of maximum association, i.e. consuming all C into CO. The variation of these abundances among individual RSGs is ${\sim}$0.2 dex, and the reported uncertainties are ${\sim}20\%$ \citep{davies2009}. For consistency in our treatment within the samples, we adopt $f_{\mathrm{CO},0}=1.8\times10^{-4}$ for RSGC1 targets and $f_{\mathrm{CO},0}=1.6\times10^{-4}$ for RSGC2.

Each model is calculated for $10^3$ photon packages injected at the inner boundary across a radial grid of $N_{\mathrm{shells}}=200$ shells spaced from $R_{\mathrm{in}}$ to a maximum radius of $3\rhalf$, according to the function
\begin{equation}
    r_{i} = (R_{\mathrm{in}}^p + (\frac{1-R_{\mathrm{in}}^p}{N_{\mathrm{shells}} -2})(i-1))^{1/p},
\end{equation}
and with $p=-0.2$, ensuring tight enough sampling of the inner region, without under-sampling the outer region of the CSE.
%evenly in logarithmic space 
Per source, we calculate a model grid over two free parameters, varying $\log(\dot{M}/(M_{\odot}/\mathrm{yr}))$ across the range $-7.0, -6.5, \ldots, -3.0$ and $\log(R_{1/2}/\mathrm{cm})$ across the range $15.00, 15.25, \ldots, 17.00$. These grids are calculated per source, given the differences in \teff, \lstar, $R_{\mathrm{in}}$, and \vexp across the sample. 

The evaluation of the models is based on the \twco $J=2-1$ line intensities and line profiles. We compare the synthetic spectra to the spectral lines extracted for different beam sizes to sample spatial information (0.6\arcsec, 1.1\arcsec, 1.6\arcsec, 2.1\arcsec, 2.6\arcsec). A failure to reproduce the observed trend of flux across spatial scales disqualifies a model. Since the CO emission for most sources appears compact (Fig.~\ref{fig:radial_profs} and Sect.~\ref{sect:spatial}), integrated flux and line shape assessment is primarily based on the smallest beam extraction, which has the highest S/N. Larger beams are then used to ensure accepted models do not overpredict the amount of spatially resolved emission. The line-profile assessment is partially based on how well the synthetic profiles reproduce the shell fit (see Sect.~\ref{sect:linefits}), but is ultimately done by eye to account for the noise and ISM contamination across the spectra, which varies considerably across the modeled sources.

\begin{figure}
 \centering
    \includegraphics[width=\linewidth]{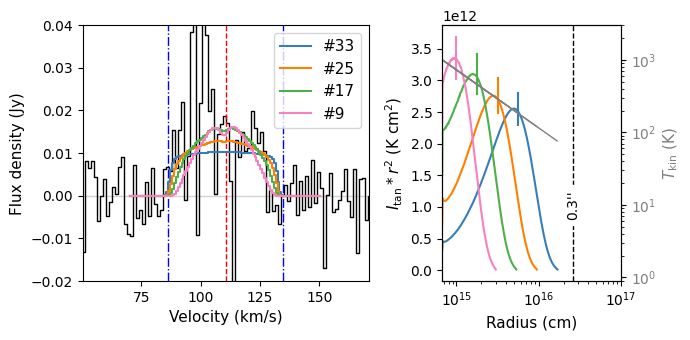}
    \includegraphics[width=\linewidth]{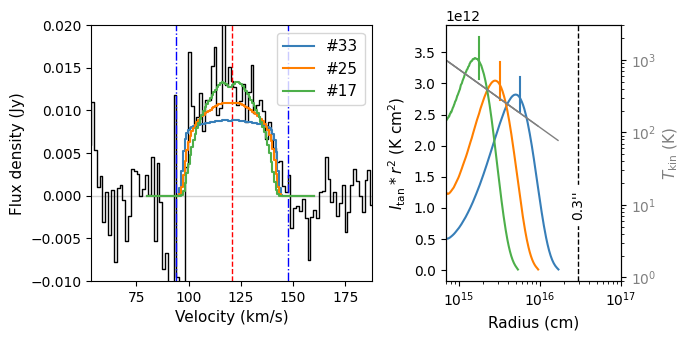}
   \caption{CO radiative-transfer results. For each of the stars (\textit{top:} DFK\,2, \textit{bottom}: RSGC1-3), the left-hand panel shows the spectrum extracted for a beam size of 0.6\arcsec\/ (\textit{black histogram}), 
   % the shell fit according to the description in Sect.~\ref{sect:linefits} (\textit{red shaded curve}), 
   and the accepted simulated line profiles (\textit{colored according to models listed in Table~\ref{tab:CO_RT_inputs}}). The curves in each right-hand panel show a measure for the region contributing to the $J=2-1$ emission, where the vertical dashes indicate the location of each model's $\rhalf$. The kinetic temperature profile of the models is shown in gray and corresponds to the right-hand side vertical axis. The results for the other cluster stars are shown in Fig.~\ref{fig:comodels-rsgc-appendix}. % and \ref{fig:comodels-rsgc1-appendix}.
    }
    \label{fig:CO-RT-results-selected}
\end{figure}

\subsubsection{CO results}

\begin{figure*}
 \sidecaption
\includegraphics[height=6.25cm]{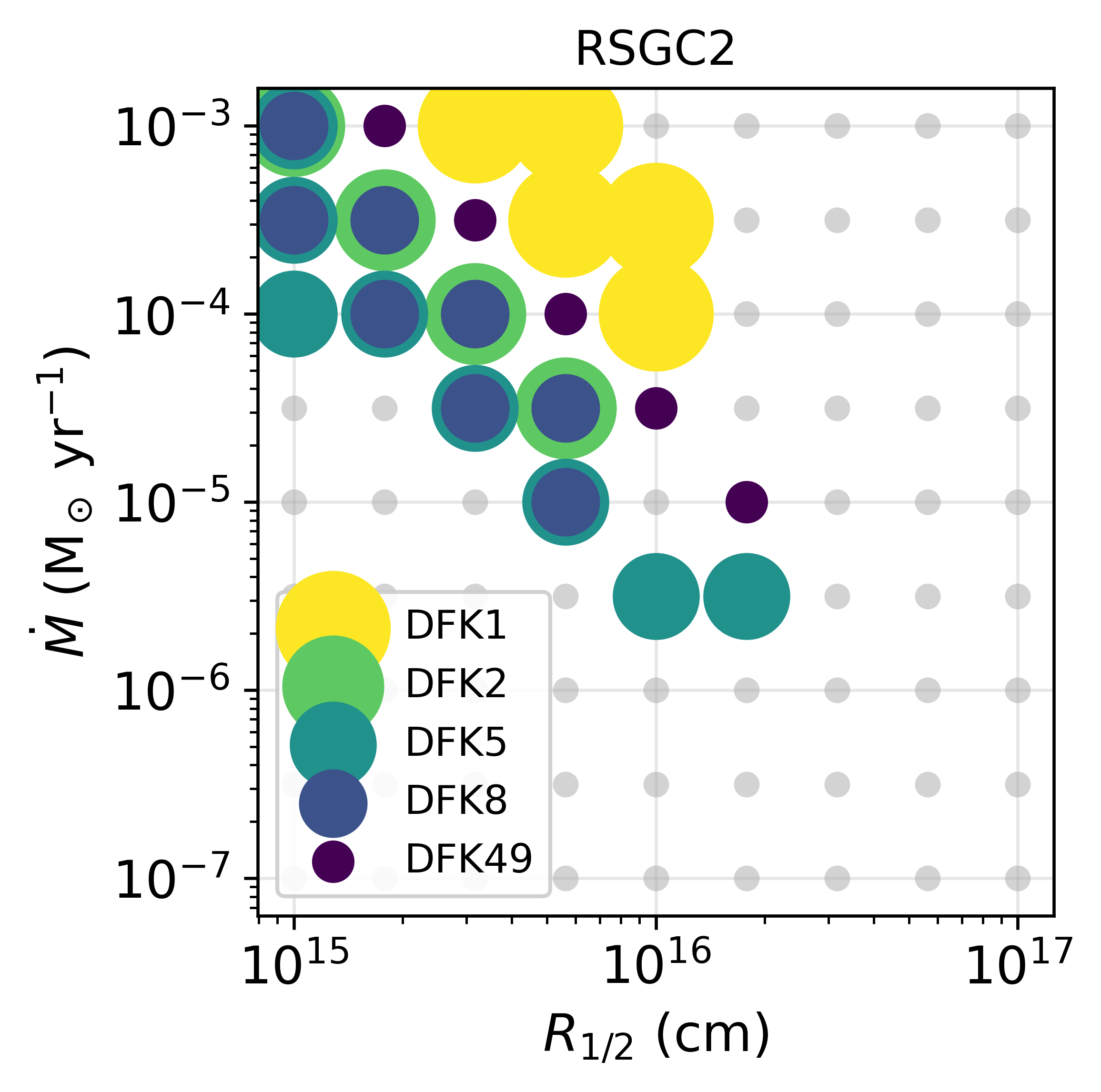} 
\includegraphics[trim=2.1cm 0cm 0cm 0,clip,height=6.25cm]{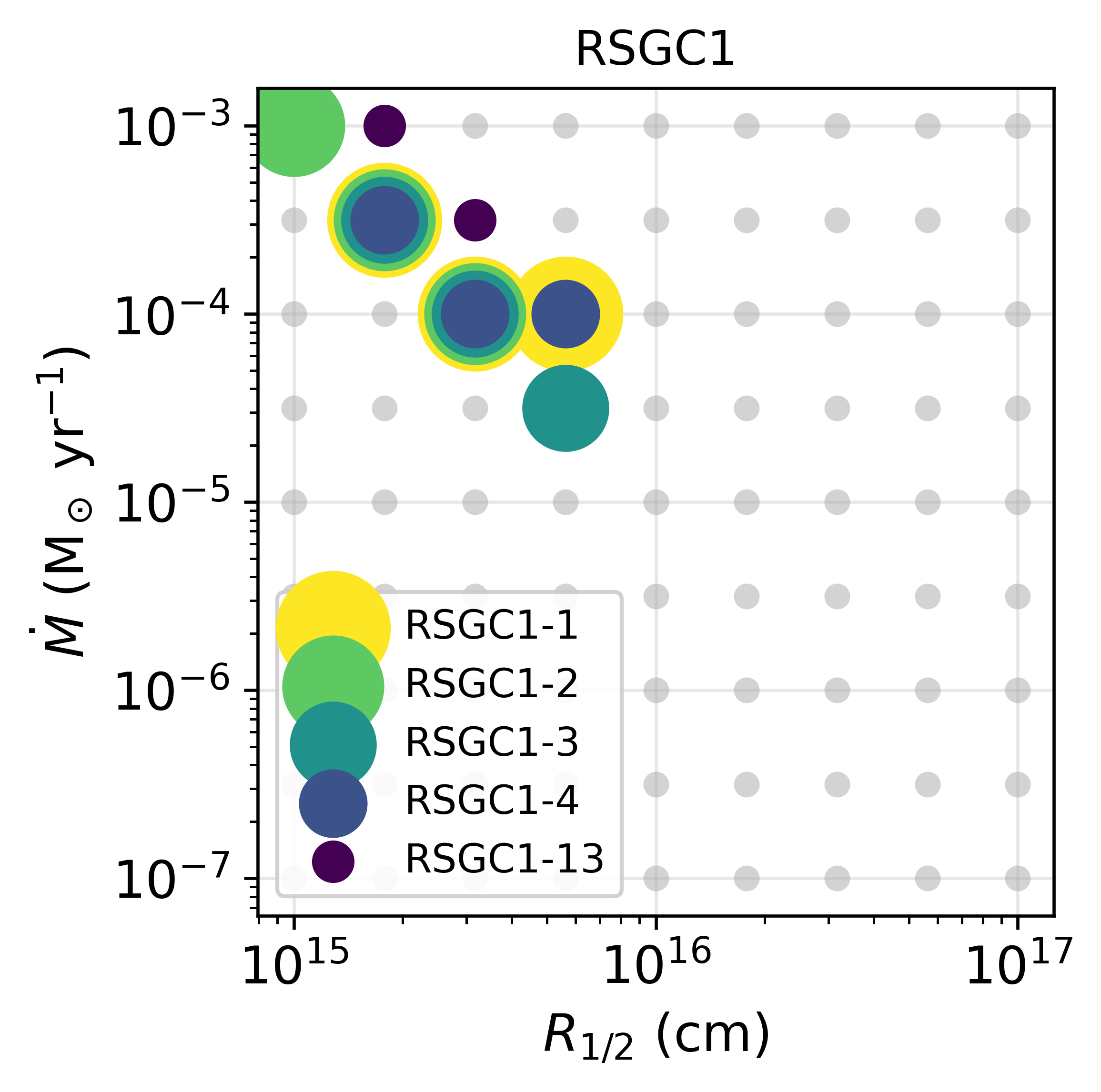}
\caption{Parameter space coverage and accepted models for modeled members of RSGC2 (\textit{left}) and RSGC1 (\textit{right}). All sampled points in the ($\rhalf$,\mdot)-grid are indicated as gray filled circles. The ($\rhalf$,\mdot)-combinations of the accepted models, from Figs.~\ref{fig:CO-RT-results-selected}, \ref{fig:comodels-rsgc-appendix}, 
% \ref{fig:comodels-rsgc1-appendix},
and Table~\ref{tab:CO_RT_inputs} are color- and size-coded per source.}
\label{fig:parameterspace}
\end{figure*}

A summary of the radiative-transfer model results can be found in Fig.~\ref{fig:CO-RT-results-selected} for DFK\,2 and RSGC1-3; similar plots for all additional sources with CO detections are shown in Fig.~\ref{fig:comodels-rsgc-appendix}.
Table~\ref{tab:CO_RT_inputs} and Fig.~\ref{fig:parameterspace} summarize the \mdot and $\rhalf$-values for the accepted RT models for all modeled targets.

\paragraph{RSGC2} Our accepted models have mass-loss rates in the range $(0.03-10)\times10^{-4}$\,\msunyr and $\rhalf$ in the range $0.1-1.8\times10^{16}$\,cm, equivalent to $\approx10-1000\,R_{\star}$; however, for the large majority of sources we require $\rhalf<10^{16}$\,cm ($500\,R_{\star}$). As can be seen from Fig.~\ref{fig:CO-RT-results-selected} for the specific case of DFK\,2 and Table~\ref{tab:CO_RT_inputs}, all accepted models reasonably reproduce the line flux and line profile, even for a high-quality spectrum, highlighting a significant modeling degeneracy given the current observational constraints. Figure~\ref{fig:parameterspace} demonstrates this degeneracy in detail, showing the accepted models in the $\rhalf-\dot{M}$ grid for each source.

In the cases of DFK\,5 (Fig.~\ref{fig:comodels-dfk5}) and DFK\,8 (Fig.~\ref{fig:comodels-dfk8}), the low signal-to-noise ratio in the observed spectra allows for an even larger range in accepted model parameters. DFK\,49 is the RSGC2 member for which we find the largest CO envelope, in line with the partial spatial extension reported in Sect.~\ref{sect:spatial}. Therefore, we took into consideration the spectra extracted for three different beam sizes (Fig.~\ref{fig:comodels-dfk49}) when judging the model results. 

We note that none of our models can reproduce the spectra toward DFK\,1 across multiple apertures in a satisfactory way (Fig.~\ref{fig:comodels-dfk1}). This mismatch, where we are unable to simulate the increase of the observed flux densities with increasing apertures, could have its origin in the processing (cleaning) of the data set, or perhaps indicate that this star is not a cluster member and needs a revised distance. We note that, of all the stars here modeled as members of RSGC2, DFK\,1 is the star with the most deviating systemic velocity at 93.4\,\kms. However, radiative-transfer models for alternative distances (4.0\,kpc, 4.5\,kpc, and 5.0\,kpc) did not alleviate this problem sufficiently to make a strong case.  We, therefore, treated DFK\,1 as a cluster member and focused on the smallest-beam (0.6\arcsec) spectrum alone. 

Since many models favor high \mdot, we explored an even higher-\mdot regime, including values up to $10^{-2}$\,\msunyr. Those models produce synthetic line profiles which no longer reproduce the observed line profiles, as they become more triangular owing to the higher optical depth. Even though an increase of \vexp could possibly alleviate some of this issue, the observed SiO emission does not support such an assumption. 

We did not model DFK\,52 here, given the outflow complexity. \citet{Siebert2025_DFK52} modeled a high-\mdot, fast detached component ($\dot{M}>10^{-4}$\,\msunyr, $v_{\mathrm{exp}}\approx27$\,\kms) and a low-\mdot slow close-in component ($\dot{M}\approx3\times10^{-6}$\,\msunyr, $v_{\mathrm{exp}}\approx10$\,\kms) to reproduce the CO emission of this star.  

\paragraph{RSGC1} For all five RSGC1 sources we recover a few accepted models, with \mdot-values in the range $(0.3-10) \times 10^{-4}$\,\msunyr. As was the case for the RSGC2 sources, we find extremely compact CO envelopes, with $\rhalf$ in the range $10-75$\,\rstar ($1.0-5.6\times 10^{15}$\,cm). 

Gas-radiative-transfer models using the model inputs reported by \citet{decin2024_rsgc1}, including larger envelope sizes than our best-fit models, produce increasing fluxes for increasing apertures. This is reflected in the line shapes shown by \citet{decin2024_rsgc1}, where the central depression that appears for multiple of their sources in their Fig.~1 reflects partial spatial resolvedness. This behavior across different scales is not supported by the observations (see Fig.~\ref{fig:radial_profs} and Sect.~\ref{sect:spatial}). 

Another difference between our models and those of \citet{decin2024_rsgc1} is that our \vexp-values used as model input are notably larger, see Sect.~\ref{sect:linefits}. We explored the (\mdot, $\rhalf$)-grid for the alternative combinations of \vLSR and \vexp for RSGC1-1 and RSGC1-4 presented in Table~\ref{tab:linefits} and retrieved the same (\mdot, $\rhalf$)-combinations as the best-fitting models.

\begin{figure*}[t]
\includegraphics[width=0.47\linewidth]{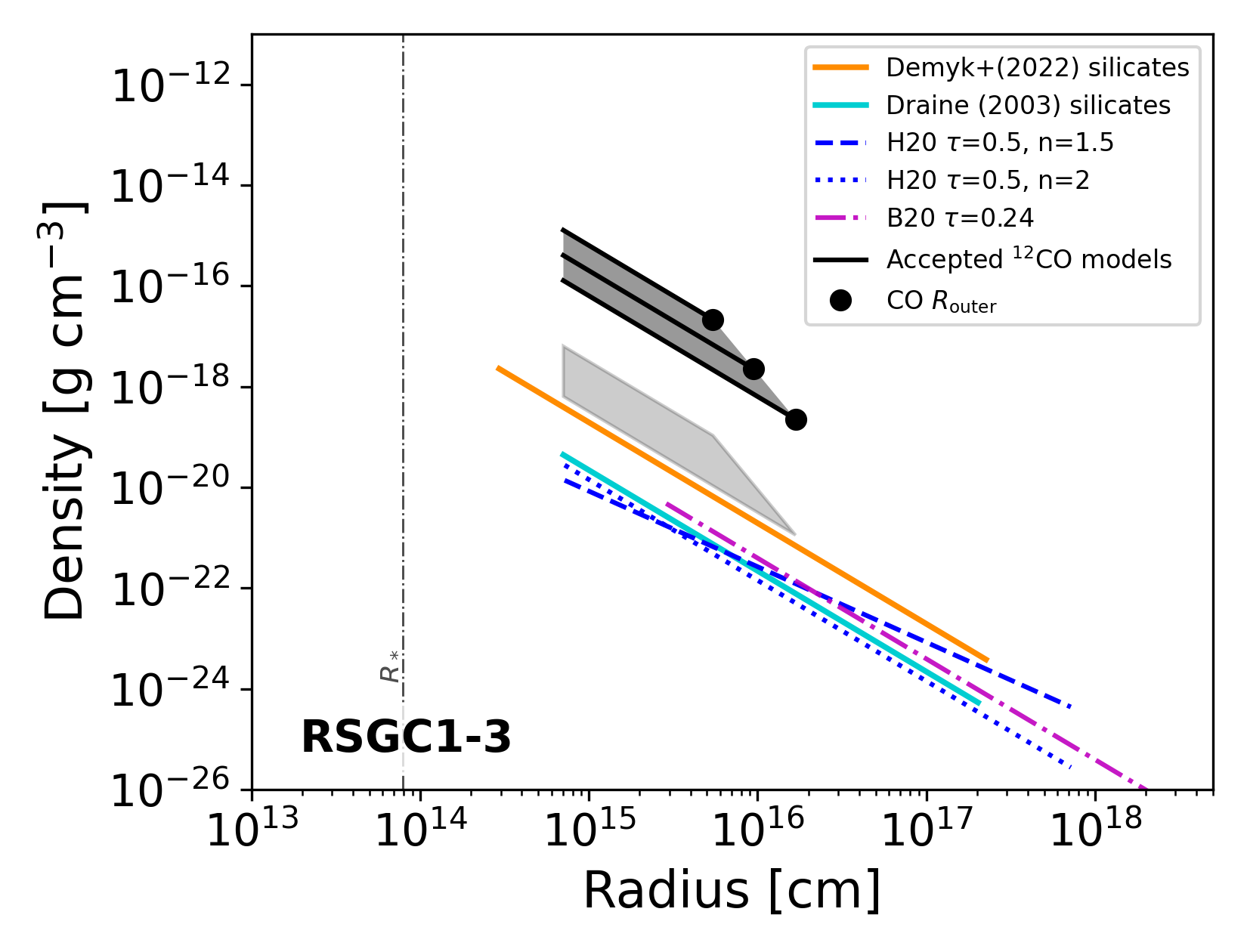}
\hfill
\includegraphics[width=0.47\linewidth]{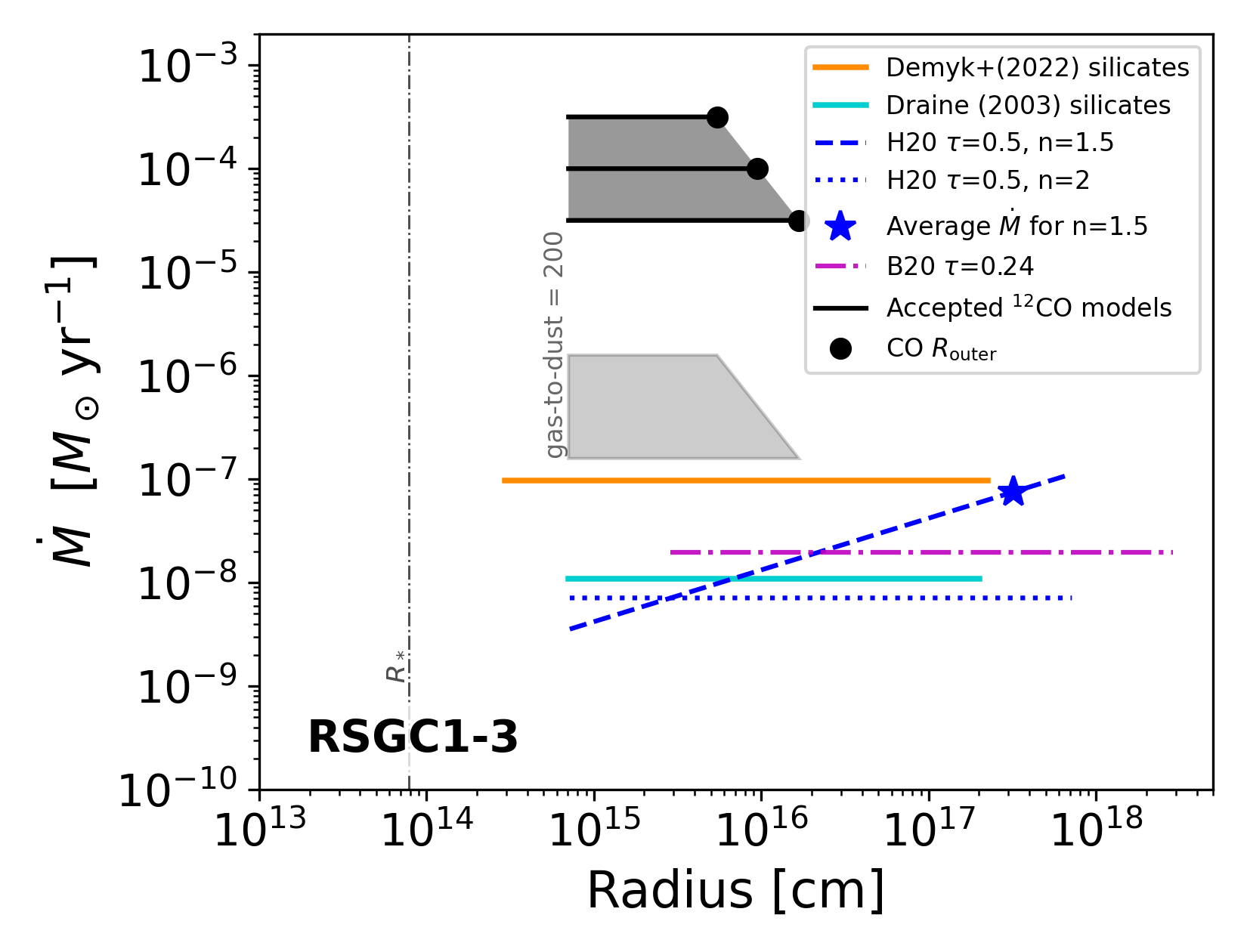}
\hfill
\caption{
Radial density (left) and mass-loss rate (right) profiles for the CSM of RSGC1-3. Black lines and shaded regions denote the gas densities and mass-loss rates obtained from RT-modeling of \twco $J=2-1$, where the lower polygon represents the dust densities if a gas-to-dust ratio of 200 is assumed. Colored lines indicate dust profiles derived from the SED fits in this work (\textit{solid orange and cyan}), \citet[][H20, \textit{blue}]{humphreys2020}, and \citet[][B20, \textit{magenta}]{beasor2020}. The inner boundary of the dust profiles is $R_{\rm in}$ (Table \ref{tab:sed_results}; where $T_{\rm dust}=1000$\,K). Two profiles are shown on each panel for the H20 fit: the first (dashed) corresponds to their variable-$\dot{M}_{\rm dust}$ model with $\rho(r)\propto r^{-n}$ and $n\neq2$, while the second (dotted) is the constant-$\dot{M}_{\rm dust}$ which yields equivalent $\tau$. For the former, a marker indicates the average $\dot{M}_{\rm dust}$ over the variable-\mdot model (the reported $\dot{M}_{\rm SED}$ from that study). For the B20 dust model, the inner radius boundary is calculated based on the reported $T_{\rm in}$. Since an outer radius of that model is not specified, it is displayed as $1000\,R_{\rm in}$ here. Similar figures for the full detected sample are available at Zenodo: \url{https://doi.org/10.5281/zenodo.22794283}
}
\label{fig:RSGC1_3_Mdot_dens_profiles}
\end{figure*}

\section{Discussion}\label{sect:discussion}
\subsection{Envelope sizes}\label{sect:COsizes}
The most unexpected result of this study is the very small spatial scales of \twco $J=2-1$ emission toward RSGC1 and RSGC2 RSGs. For a purely observational context, the physical diameter of this line observed toward the RSG VX Sgr is $8\times10^{16}$\,cm, as measured by \citet{Gottlieb2022_ATOMIUM}. Scaling this to the cluster distances would imply an angular scale of $0.8-1.0$\arcsec, which would be cleanly resolved by $2-3$ beams in the observations presented here. Likewise, oxygen-rich AGB stars with $\dot{M}\gtrsim10^{-6}$\,\msunyr are seen with typical envelope sizes larger than ${\sim}3\times10^{16}$\,cm \citep{ramstedt2020_deathstar}, where we would still expect some marginally resolved emission. Considering there are numerous RSGs in both RSGC1 and RSGC2 with comparable $\dot{M}_{\rm SED}$ to these examples, their compact observed envelopes point toward some unexpected truncation of CO. % at large radii.

Likewise, the spatial CO distribution constraints resulting from our RT-models also present puzzling discrepancies with similar objects and model predictions for circumstellar environments. Nearly all RSGs in our sample have half-abundance radii $\rhalf$ in the range $0.1-1\times10^{16}$\,cm ($10-500$\,\rstar), significantly smaller than what is typical for the outflows of cool evolved stars in the solar neighborhood \citep[$\rhalf>1000$\,\rstar][]{ramstedt2020_deathstar}, especially given the rather high retrieved \mdot-values. The winds of both RSGs and AGB stars are assumed to have CO envelope sizes constrained by photodissociation by the interstellar radiation field (ISRF). Adopting the CO photodissociation models of \citet{saberi2019} to investigate the potential impact on our results of the radiation field near the RSGC clusters, we find that, even with an ISRF scaling factor of 10 or 20 relative to the solar neighborhood, a $10^{-5}$\,\msunyr wind with $v_{\rm exp}=25$\,\kms is expected to have $R_{1/2}$ of ${\sim}3-4\times10^{16}$\,cm for an initial CO abundance of $2\times10^{-4}$ (see Fig.~\ref{fig:co_efold}).

This is still significantly larger than our maximum allowed envelope sizes and would yield spatially resolved lines, especially considering that the CO $J=2-1$ emission region extends beyond $\rhalf$ (see e.g. Fig.~\ref{fig:CO-RT-results-selected}). At a galactocentric radius of 3.5\,kpc, the ambient UV flux is predicted to be a factor of ${\sim}8$ higher than in the solar neighborhood \citep[applying Eq.\ 22 of][]{Vernetto2016_gammaabs_ISRF}, so the radiation field near both RSGC1 and RSGC2 would need to be dramatically higher than expected in order for it to be the sole contributor to our small CO envelope sizes. Additionally, a much higher flux of dissociating radiation would make the presence of DFK\,52 in RSGC2 even more puzzling, as this object demonstrates that CO \textit{can} exist very far ($r=3\times10^{17}$\,cm) from the central star in this cluster environment \citep{Siebert2025_DFK52}.

The small CO envelopes and the measured \vexp imply that our observations only trace about one hundred years of mass loss. 
Since the column of shielding material, and consequently $R_{1/2}$, is set by the preceding mass loss, a recent increase in $\dot{M}$ would cause the CO envelope sizes to appear ``too small'' for their derived densities, which correspond to current mass loss. However, this interpretation stands in contrast to the SED models for these stars, which require a substantial amount of cold dust grains and even has been interpreted as evidence for a recent \textit{decrease} in mass-loss rate (discussed further in Section \ref{sect:dustgas_comp}). Furthermore, we find it statistically unlikely that all of the most luminous RSGs in both clusters have simultaneously (on stellar evolution timescales) experienced a very recent increase in their mass-loss. 
In any case, the ALMA observations suggest that 
neither the SEDs, nor the CO emission can be used as sole tracers of the 
stars' mass-loss histories and provide stringent input to RSG mass-loss prescriptions. Considering the high likelihood of an increased ISRF, observations of photodissociation products, e.g. atomic carbon, will be instrumental in tracing the gas-phase component of the outflows from cluster RSGs. 

A possible alternative interpretation is that we are not detecting CO emission from a freely expanding wind around the RSGs, but rather from a gravitationally bound layer. In that case, the CO emission could provide a measure of the total mass in this layer, rather than a mass-loss rate. In order to explain the IR spectra of nearby RSGs, many works have invoked the presence of a MOLsphere: a non-photospheric layer of optically thick molecular material at 1--2$R_*$ with $T{\sim}1500$\,K \citep{Tsuji2000_molSPHERE,Montarges2014_alfOriMOLsphere,Ogorman2020} traced primarily by vibrational bands of CO and H$_2$O. A dense shell of CO could in theory help explain the compact rotational line emission; however, in this scenario, it is not trivial to explain the observed ``standard'' emission line profiles and their high \vexp.

\subsection{Mass-loss rates}

\subsubsection{Simple estimates}
\label{sect:gas_comp}
Using the intensities of the presented ALMA spectra as input to the \mdot-estimator from \citet{ramstedt2008}, we find \mdot-values in the range $(0.8-7.7)\times10^{-6}$\,\msunyr for the RSGC2 sources and in the range $(1.4-3.3)\times10^{-6}$\,\msunyr for the RSGC1 sources. These rates are systematically significantly lower than those presented in Sect.~\ref{sect:COmodeling} and App.~\ref{app:COmodels}. It is essential to note here that the \mdot-estimator was derived based on a grid of models representative of standard AGB CSEs, where the CO half-abundance radius $\rhalf$, and hence the extent of the molecular envelope, is set by the interstellar radiation field. Since this assumption is inherent to the estimator and does not match our observational constraints of more compact molecular CSEs, the prescription ends up underestimating \mdot compared to our modeling procedure by up to three orders of magnitude. This significant mismatch underscores the vital importance of spatial constraints on the emitting gas and a careful usage of (necessarily simplifying) \mdot-estimators.

\subsubsection{Comparing CO and SED models}
\label{sect:dustgas_comp}

Here, we compare the assumptions and results of our CO RT models with the dust models provided both in this work and in previous studies of the RSGC clusters. An example of this comparison is shown in Fig.~\ref{fig:RSGC1_3_Mdot_dens_profiles} for RSGC1-3, which displays the modeled gas and dust densities as a function of radius from our ALMA \twco RT-analysis, the \texttt{DUSTY} SED models of \citet{humphreys2020} and \citet{beasor2020}, and our own \texttt{RADMC-3D} models for this object. In our SED models, we investigate the impact of adopted optical constants on the retrieved wind properties, adopting the standard empirical grain opacities of \citet{Draine2003_dustopacs} as well as the laboratory-measured amorphous silicate opacities of \citet{Demyk2022_lab_silicates} (more detailed information on these dust models can be found in App.~\ref{app:SEDs}). Due to the wide range of acceptable CO rotational line models (see Sect.~\ref{sect:COmodeling}), the gas density constraints are displayed as polygonal regions in the $\rho-r$ space.

It is important to note with regard to Figure \ref{fig:RSGC1_3_Mdot_dens_profiles} that while the simulated region of our best-fit dust models (solid lines) extends a far beyond the CO outer radius, regions exterior to the CO extent contribute very little information to the SED fit. To test this, we additionally ran RADMC3D models with small outer boundaries of $r=10^{16}$\,cm, comparable to our retrieved $\rhalf$ for CO, and found only marginal changes in the resulting SED. Because the observed IR-excess is almost entirely contributed by dust interior to our CO constraints, the SEDs alone do not rule out an interpretation of the ALMA data where the overall CSM density falls off at large radii (i.e.\ a recent increase in the mass-loss rate).

Figure \ref{fig:RSGC1_3_Mdot_dens_profiles} illustrates some clear disagreements between our gas-phase constraints and the variable-$\dot{M}$ models from \citet{humphreys2020}. A simple comparison of the lower limit total $\dot{M}$ from our CO models ($3\times10^{-5}$\,\msunyr) with the $\dot{M}_{\rm dust}$ reported in that study ($7.5\times10^{-8}$\,\msunyr) implies a gas-to-dust mass ratio of 400. The gas-to-dust ratio is a highly uncertain quantity in RSG environments, with reported values spanning ${\sim}200$ all the way up to $\gtrsim2000$ measured for sources like Antares \citep{Cannon2021_Antares_sphere}, so a value of 400 is well within this range. However, Fig.\ \ref{fig:RSGC1_3_Mdot_dens_profiles} illustrates that the $\rho\propto r^{-1.5}$ density profile adopted in that \texttt{DUSTY} fit spans over an order of magnitude in instantaneous mass-loss rate, and the $\dot{M}$ that is reported (the average over the shell) represents the mass loss at $r{\sim}3\times10^{17}$\,cm, which is far outside the allowed maximum radii for CO. At small radii where the densities of the CO RT models and the variable $\dot{M}$ \texttt{DUSTY} model are both constrained, the implied gas-to-dust ratios are extremely high (${\sim}10^4$). Likewise, if CO were present at large radii ($>10^{17}$\,cm) with gas densities consistent with the model of \citet{humphreys2020}, we would expect to have detected significant spatially resolved emission in the ALMA data. These model discrepancies are ubiquitous throughout both cluster samples, and therefore favor an alternate interpretation of the SED shape rather than a decreasing-$\dot{M}$ scenario.

Comparing with the \texttt{DUSTY} fit from \citet{beasor2020}, we obtain a gas-to-dust ratio of ${\sim}1500$ for RSGC1-3 and the other two sources in common with that work; however, we also note that the inner radii of the dust shells are notably larger in those models, since the dust condensation temperature ($T_{\rm in}$) was treated as a free parameter in that study (with best-fit values of 400--500\,K). This is an important distinction from the other SED models displayed in Fig.\ \ref{fig:RSGC1_3_Mdot_dens_profiles}, which all use a fixed value of $T_{\rm in}=1000$\,K. If one adopts a smaller $T_{\rm in}$, the region where the gas and dust measurements overlap narrows to the very small range of $40{\sim}100$\,\rstar. This presents an additional discrepancy between our results and \citet{beasor2020}, as a substantial fraction of the CO flux in our accepted models is contributed from regions inside the implied inner dust radius of that work. Given that our CO RT models adopt the dust condensation radius as an inner boundary, extending this $R_{\rm in}$ to the location where $T{\sim}500$\,K would require an even higher total $\dot{M}$ and gas-to-dust ratio, as the CO-radiating region would shrink to a smaller volume (see Fig.\ \ref{fig:CO-RT-results-selected}, right panels).  

An alternative interpretation of the \twco observations and SED measurements is reached by invoking different optical properties of the dust grains. We observe in Fig.~\ref{fig:RSGC1_3_Mdot_dens_profiles} that the SED fit utilizing constants from \citet{Demyk2022_lab_silicates} (65\% MgO and 35\% SiO$_2$) requires a factor of ${\sim}$10 higher dust densities and mass-loss rates than the empirical opacities of \citet{Draine2003_dustopacs}. Two related but distinct effects drive this. First, the Fe-poor lab-measured grains have a much lower absorption efficiency in the near-IR, where the RSG radiation field peaks, so more dust mass is needed to absorb and re-radiate the same luminosity at the inner dust boundary even at fixed temperature. Second, beyond the main silicate resonance complex the Demyk grains have a shallower opacity spectral index $\beta$ ($\kappa\left(\lambda\right)\propto \lambda^{-\beta}$) which causes them to lose absorbed energy more efficiently as they move outward, producing a steeper radial temperature profile. Notably, this wavelength range is one where the \citet{Demyk2022_lab_silicates} optical constants are anchored directly to laboratory mass absorption coefficient measurements \citep{Demyk2017_abscoeffs}, whereas the \citet{Draine2003_dustopacs} constants are extrapolated beyond ${\sim}$20\,$\mu$m. The efficient cooling compresses the region of warm (${\sim}$few-hundred-K) grains that dominate the ${\sim}$20\,$\mu$m excess into a smaller volume, requiring higher densities to reproduce the same observed optical depth. Together, these effects push the SED fits toward substantially higher inferred densities and dust mass-loss rates for the lab-measured composition, which would instead imply gas-to-dust ratios ${\sim}500$--$1000$ for the full sample, which is closer to the standard assumptions used in dust models of galactic RSGs \citep{mauron_josselin_2011_masslossrsgs}.

The optical constants we adopt from \citet{Demyk2022_lab_silicates} represent a Mg-rich, Fe-free olivine solid mixture similar to forsterite. The Fe-content of silicates has a dramatic impact on their NIR absorbing efficiencies \citep{gail_2020_rsgs}, and is likely one of the main delimiting factors from the Draine astrosilicates, since those are derived to reproduce the extinction curve of interstellar grains which incorporate most of the iron in the ISM \citep{Dwek2016_iron}. Whether RSG silicates are formed sufficiently Fe-poor for laboratory forsterite-like optical constants to be appropriate is still uncertain, as Fe incorporation is expected to depend on the grain condensation sequence and local wind conditions \citep{Speck2000_Orich_dustcomp,Verhoelst2009_RSG_dustcond}. Nevertheless, these results demonstrate that plausible variations in silicate composition alone can shift SED-inferred dust-mass-loss rates by approximately an order of magnitude. 

\subsubsection{Comparing to \mdot-prescriptions}

\begin{figure*}[t]
    \centering
    \includegraphics[width=0.85\linewidth]{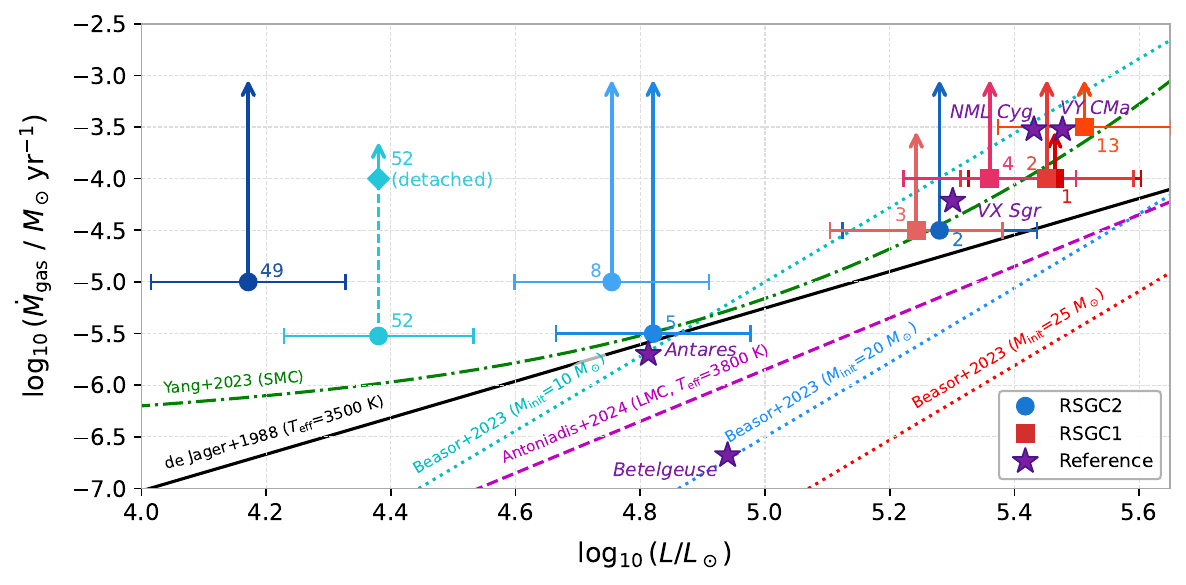}
    \caption{\twco-derived $\dot{M}$ constraints plotted against luminosity for detected RSGC2 (blue circles) and RSGC1 (red squares) sources, compared with various RSG mass-loss prescriptions.
    % (dashed lines) of \citet{deJager1988_Mdots} and the more recent relation from \citet{Beasor2023_RSG_Mdot_err}. 
    We show the prescription from \citet{Beasor2023_RSG_Mdot_err} for $M_{\rm init}=$10\,\msun (\textit{cyan}), 20\,\msun (\textit{blue}), and 30\,\msun (\textit{red}), a de Jager wind for $T_{\rm eff}=$3500\,K \citep[][\textit{black}]{deJager1988_Mdots}, and the prescriptions constrained with SMC and LMC RSGs by \citet[][\textit{green}]{Yang2023_SMC_RSGwinds} and \citet[][\textit{magenta}]{Antoniadis2024_LMC_RSG_precr}, respectively. Our RSGC mass-loss rates are lower limits, where higher $\dot{M}$-values for a given source correspond to smaller envelope sizes also consistent with the data due (Sect.\ \ref{sect:COmodeling}). For DFK\,52, two $\dot{M}$ values are shown with the larger describing an average over its complex detached wind, and the smaller reflecting the ``present-day'' component \citep{Siebert2025_DFK52}. Mass-loss rates for nearby RSGs (\textit{purple stars}) are adopted from \citet{Braun2012_alphaSco_hydro} and \citet{debeck2010_comdot,debeck2025_nmlcyg}. }
    \label{fig:mdot_lum}
\end{figure*}

A comparison of the CO-modeling results for RSGC1 and RSGC2 cluster members with various RSG-mass loss prescriptions is shown in Fig.~\ref{fig:mdot_lum}. Due to the degeneracy between envelope size and $\dot{M}$ discussed in the previous section, mass-loss rates are displayed as vertical ranges with the lower limit corresponding to the largest $\rhalf$ model that still produces an unresolved line profile and flux distribution. Because no suitable CO model was found for DFK\,1, it is omitted from this figure and discussion.

In general, the mass-loss rates allowed by our CO models are quite high in comparison to the prescriptions listed here, with many of our lowest acceptable $\dot{M}$ appearing ${\sim}$0.5--1 dex higher than the traditional de Jager wind (for $T_{\rm eff}=$3500\,K). For RSGC2 targets, which have wider spread in luminosity, we do find smaller allowed $\dot{M}$ for lower-$L$ stars DFK\,5 and DFK\,8, but all sources still require substantial mass-loss rates above 10$^{-6}$\,\msunyr. 

The mass-loss prescriptions of \citet{Yang2023_SMC_RSGwinds} and \citet{Antoniadis2024_LMC_RSG_precr}, which are based on the largest samples of dust-modeled RSGs in the SMC and LMC respectively, predict a flattening (or a ``kink'') in the $\dot{M}$ curve at low luminosity ($\log L<4.4$). The two relations differ by an order of magnitude; however, \citet{Antoniadis2024_LMC_RSG_precr} re-fit the same SMC sample as \citet{Yang2023_SMC_RSGwinds}, finding that applying the assumption of a steady-state wind in DUSTY reduces $\dot{M}$ measurements to agreement with LMC RSGs. This result suggests that metallicity has minimal impact on mass-loss rate (apart from the position of the kink in luminosity), so comparison to the galactic clusters here is appropriate. \citet{Antoniadis2024_LMC_RSG_precr} state that the gas constraints of \citet{decin2024_rsgc1} favor dust mass-loss rates obtained via the steady-state DUSTY velocity field assumption over that of a radiatively driven wind (RDW); however, our re-analysis of RSGC1 sources here yields considerably higher mass-loss rates than \citet{decin2024_rsgc1}, and is therefore more consistent with $\dot{M}-L$ relations constrained using the RDW assumption, such as \citet{Yang2023_SMC_RSGwinds}, \citet{vanLoon2005}, and \citet{Goldman2017_windspeeds}. Our additional results for RSGC2 members, as well as gas phase $\dot{M}$-constraints for nearby RSGs from the literature also support these stronger RSG wind prescriptions (Fig.\ \ref{fig:mdot_lum}).

For high-luminosity ($L>10^5 L_{\odot}$) sources, which include all detected RSGC1 stars and DFK\,2, we find mass-loss rates comparable to the high-$\dot{M}$ objects VX Sgr, NML Cyg, and VY CMa, also shown in Fig.\ \ref{fig:mdot_lum}. It is important to note that while this clustering of RSGs, both in $\dot{M}$ and wind velocity (Fig.\ \ref{fig:vexp_lum}), hints at a similarity in their wind properties, the observed molecular-emission morphologies stand in stark contrast to one another: VX Sgr, NML Cyg, and VY CMa are all known to show \twco emission on scales of ${\sim}$several thousand au \citep{Gottlieb2022_ATOMIUM,debeck2025_nmlcyg,singh2023}, whereas the cluster RSGs show very confined CO emission.  

The location of RSGC1 members in Fig.\ \ref{fig:mdot_lum} is particularly inconsistent with the prescription of \citet{Beasor2023_RSG_Mdot_err}, as the initial mass of RSGs in RSGC1 has been measured as 25$\pm$2\,\msun (red dash-dotted line). This mismatch results from a more fundamental incongruency between our result and that work, as \citet{Beasor2023_RSG_Mdot_err} use SED fits to RSGC1 members to constrain the $M_{\rm init}$-dependence of the $\dot{M}-L$ relation, and we obtain mass-loss rates 1--2 orders of magnitude higher using molecular gas. As discussed in Section \ref{sect:dustgas_comp}, this requires very high gas-to-dust ratios ($>$2000) for the three sources in common between our sample and theirs, or alternative assumptions on the dust models. While our results for RSGC1 tentatively point toward a more relaxed dependence on initial mass than in prescriptions derived by \citet{Beasor2023_RSG_Mdot_err} and \citet{decin2024_rsgc1}, we refrain from putting forward any updated $\dot{M}-L$ relation from these observations alone, as it is critical to first place spatial constraints on the molecular component to reduce the uncertainties and better understand the unique phase structure of CSM in these cluster environments.

It is important to caveat Fig.\ \ref{fig:mdot_lum} with the selection bias inherent to our sample: the plot excludes 17 ALMA-observed cluster RSGs not detected in \twco\ (8 in RSGC2 and 9 in RSGC1), biasing it toward objects with higher \mdot and denser molecular CSM. Several non-detections (e.g.\ DFK\,3, 6, and 10; RSGC1-5) have luminosities comparable to or exceeding those of detected targets \citep{humphreys2020}, suggesting that the true cluster-wide \mdot--$L$ relation may lie systematically lower, with our detections representing outliers. Including non-detections as upper limits would help assess this possibility, but requires assuming an envelope size, which remains poorly constrained and anomalously small among detected sources. Consequently, absent \twco\ emission may indicate either genuinely lower \mdot or suppressed molecular emission from comparably massive outflows, for example through reduced atmospheric CO abundance or enhanced dissociating radiation. The non-detections therefore provide limited additional constraints on mass-loss prescriptions from CO observations alone.

Similar to the results of our $v_{\rm}-L$ analysis (Sect.~\ref{sect:linefits}, Fig.~\ref{fig:vexp_lum}), DFK\,49 and DFK\,52 appear as clear outliers in Fig.\ \ref{fig:mdot_lum}, as they are the lowest luminosity RSGs in the sample and at the same time exhibit the largest disagreements with empirical mass-loss prescriptions. The case of DFK\,49 is discussed further in App.~\ref{app:SEDs}, as we revise its luminosity by over an order of magnitude, down to ${\sim}$15,000\,\lsun, based on new constraints to its SED. This places it on (single star) evolutionary tracks for a ${\sim}$12\,\msun star \citep{Ekstrom2012_evomodels_rot} rather than a massive post-RSG as it was classified previously \citep{humphreys2020}. Likewise, DFK\,52 is thought to be less massive than the cluster average. The surprisingly fast and dense winds derived for these two RSGC2 members, and the detached nature of DFK\,52's complex CSM \citep{Siebert2025_DFK52}, therefore suggest avenues for efficient and potentially episodic mass loss for lower-mass RSGs that are not currently accounted for in evolutionary models.

An alternative explanation for the peculiar objects DFK\,49 and DFK\,52 in our sample could relate to multiplicity effects and companion interactions that have impacted the evolution of RSGC2. Binary population synthesis models have shown that mass-transfer and mergers during the main sequence in young open clusters can significantly extend the RSG luminosity distribution \citep{Eldridge2020_RSGbinary_populations,Wang2025_binaryRSGevo}. For a cluster  with age $t{\sim}20$\,Myr like RSGC2, \citet{Eldridge2020_RSGbinary_populations} predict that this effect would introduce a spread of $\Delta\log L{\sim}0.6$ for member RSGs, whereas a younger cluster like RSGC1 with $t{\sim}10$\,Myr would be much smaller ($<0.3$\,dex). These luminosity ranges are broadly consistent with the observed targets in RSGC1 and RSGC2. However, DFK\,49 and DFK\,52 remain notably underluminous relative to the most luminous RSG in RSGC2, by approximately 1 and 0.8\,dex, respectively.

Binary evolution models therefore provide a possible explanation for the high-luminosity end of the RSG population, with the highest-$L$ RSGs like DFK\,2 representing products of earlier mergers, making them appear younger in the sample, while the low-$L$ stars in the cluster would correspond to single star or non-interacted systems. DFK\,49 and DFK\,52 could be special cases of cluster members that have mostly undergone uninterrupted single star evolution up to this point and only very recently experienced a companion interaction or merger, enhancing their mass-loss rates. This scenario is supported by observational similarities, namely in the shell-like density structure (for DFK\,52), with other known recently-interacting RSG systems such as V838\,Mon \citep{Kaminski2021_V838Mon} and AFGL\,4106 \citep{Tomassini2026_AFGL4106}. Stellar evolution models have also shown that rapid envelope stripping resulting from common envelope evolution is met with a dramatic drop in the luminosity \citep{Lohev2019_TypeIIb_CEE}, which could explain why these sources appear underluminous relative to the rest of cluster. Thus, a recent binary interaction provides a plausible explanation for the observed properties of DFK\,49 and DFK\,52, though significant follow-up observations constraining the current stellar types are necessary to confirm this.

\section{Conclusions}\label{sect:conclusion}
We have presented a millimeter study of the circumstellar environments of RSGs in Galactic open clusters, including new ALMA observations of RSGC2 and a re-analysis of the RSGC1 data from \citet{decin2024_rsgc1}. With 28 sources and 11 detections in \twco\ $J=2-1$, this is the largest sample of its kind at these wavelengths and enables a systematic comparison of gas and dust CSM properties across a wide range of RSGs.

We examined wind expansion and systemic velocities using empirical fits to the observed line profiles and a $\chi^2$-minimization procedure that accounts for ISM-contaminated channels despite the resulting non-Gaussian parameter space. Our $v_{\rm exp}$-values for RSGC1 members are systematically larger by a factor of 2--3 than those reported by \citet{decin2024_rsgc1}, and are comparable to those of the most luminous RSGC2 members. Across both clusters, $v_{\rm exp}$ follows a clear luminosity trend consistent with the established $v_{\rm exp}\propto ZL^{0.4}$ relation. This relation should be incorporated into SED-based mass-loss prescriptions, as assuming constant $v_{\rm exp}$ can bias the inferred slope of the \mdot curve. Complementary molecular tracers such as thermal SiO lines can also provide cleaner constraints on expansion and systemic velocities along low-Galactic-latitude sight lines where \twco\ is heavily contaminated.

We investigated the spatial size of observed circumstellar emission in both clusters, and uncover surprisingly small envelopes, with nearly all observed \twco emission coming from regions interior to $r{\sim}2\times10^{16}$\,cm. This result suggests a physical distinction between cluster RSG winds and closer sources like VY CMa, NML Cyg, and VX Sgr. We find that this cannot be attributed solely to the increased flux of interstellar dissociating radiation expected closer to the Galactic center, so additional mechanisms are required to explain the envelope sizes. The unresolved CO emission presents a critical degeneracy in modeling the physical properties of the envelopes, as models with higher densities (and $\dot{M}$) yet smaller CO photodissociation radii cannot be ruled out by the observations, so our derived mass-loss rates represent lower limits on the total gas phase contribution.

Our radiative transfer analysis yielded substantial mass loss ($\dot{M}\gtrsim10^{-5}$\,\msunyr) for detected RSGs, implying gas-to-dust ratios of ${\sim}$several thousand when compared with IR studies of RSGC1 and RSGC2 by \citet{beasor2020} and \citet{humphreys2020}. We performed our own modeling of the SEDs and found that alternative assumptions on the dust-grain opacities could imply substantially lower gas-to-dust ratios, but this quantity remains highly uncertain. Our results currently favor high-$\dot{M}$ prescriptions like the winds of \citet{Yang2023_SMC_RSGwinds} or the traditional de Jager wind \citep{deJager1998}.

The density profiles we measure from circumstellar CO are comparable to the CSM densities measured for strongly interacting Type IIp SNe \citep[e.g.\ SN2023ixf;][]{Nayana2025_SN2023ixf_late} via multi-wavelength follow-up monitoring. It is important to note that the material probed by our observations spans regions which would be impacted by the shock front $1{\sim}3$\,yr after explosion, as opposed to the very close-in environment ($r<10^{15}$\,cm) traced by flash-spectroscopy \citep{Jacobson-Galan2023_sn2023ixf_CSM}. While the unresolved nature of the observed CO envelopes does not rule out its interpretation as a dense confined superwind like SN2023ixf, we find it statistically unlikely that such a phase of pre-SN mass-loss would be occurring simultaneously for all detected RSGs in the sample. 

The RSGC2 members DFK\,52 and DFK\,49 appeared as consistent outliers in our analysis. A separate analysis for DFK\,52 was presented in \citet{Siebert2025_DFK52}, where it was shown to have a massive detached wind component despite its relatively low luminosity, signaling a prior period of rapid mass loss. DFK\,49 was previously classified as a luminous post-RSG; however, we have revised its luminosity using new long-wavelength constraints by over an order of magnitude to $L{\sim}15,000$\,\lsun. This revision makes its expansion velocity unusually high for its luminosity, and its gas-derived mass-loss rate is $1-2$ orders of magnitude higher than any empirical prescriptions predict. Together, DFK\,52 and DFK\,49 may suggest that efficient mass-loss channels may operate at earlier evolutionary stages than predicted by current models, or that these stars have undergone very recent interaction or a merger with binary companions. 

Our results highlighted some outstanding discrepancies between observations of cluster RSGs and nearby archetypal sources. In order to reduce the uncertainties and degeneracies that remain for mass-loss determinations in environments like RSGC1 and RSGC2, we emphasize the need for a more complete view of the phase structure and composition of material in the close-in circumstellar regions. In particular, observations of spatially resolved CO lines covering multiple rotational-energy states, in addition to atomic tracers like CI, would provide better constraints on the gas densities and temperatures throughout the sample. At the same time, future spectral characterizations in the mid-IR, for example with JWST, would offer useful insights into the grain composition and thereby directly inform optical constants used in SED models. Together, these measurements would provide an empirically grounded, source-specific physical framework for comparing circumstellar properties within the clusters, which is a critically necessary step toward understanding mass loss broadly in RSGs.

\begin{acknowledgements}
This paper makes use of the following ALMA data: ADS/JAO.ALMA\#2023.1.01519.S, ADS/JAO.ALMA\#2013.1.01200.S, ADS/JAO.ALMA\#2024.A.00018.S. ALMA is a partnership of ESO (representing its member states), NSF (USA) and NINS (Japan), together with NRC (Canada), NSTC and ASIAA (Taiwan), and KASI (Republic of Korea), in cooperation with the Republic of Chile. The Joint ALMA Observatory is operated by ESO, AUI/NRAO and NAOJ. MS acknowledges support from the Norwegian Research Council through ESGC project (project No. 335497). 
GQL acknowledges funding support from the Spanish Ministerio de Ciencia, Innovación y Universidades through grant PID2023-147545NB-I00.
\end{acknowledgements}

\bibliographystyle{aa} 
\bibliography{ref.bib}

\begin{appendix}

\section{SED modeling}
\label{app:SEDs}

\begin{figure}[h]
\includegraphics[trim=0.2cm 1.6cm 4.7cm 0.36cm,clip,height=0.347\linewidth]{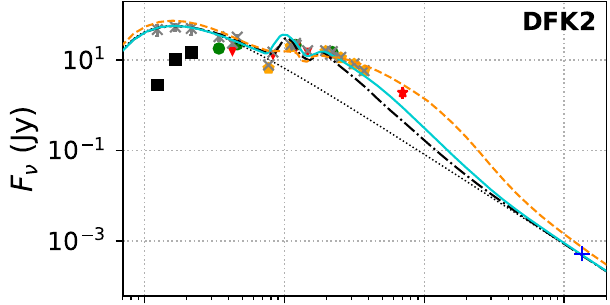}\\
\includegraphics[trim=0.2cm 1.6cm 4.7cm 0.36cm,clip,height=0.347\linewidth]{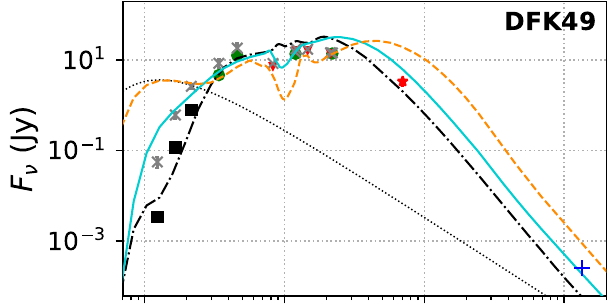}\hfill\\
\includegraphics[trim=0.2cm 1.6cm 4.7cm 0.36cm,clip,height=0.347\linewidth]{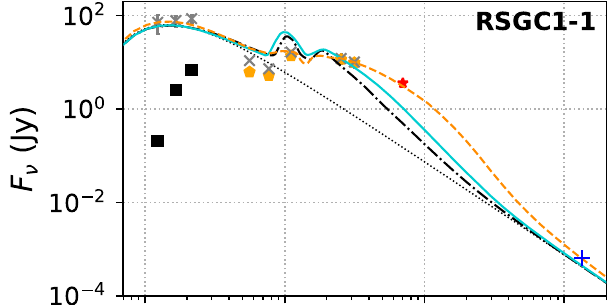} \hfill\\
\includegraphics[trim=0.2cm 0cm 0.2cm 0.0cm,clip,width=\linewidth]{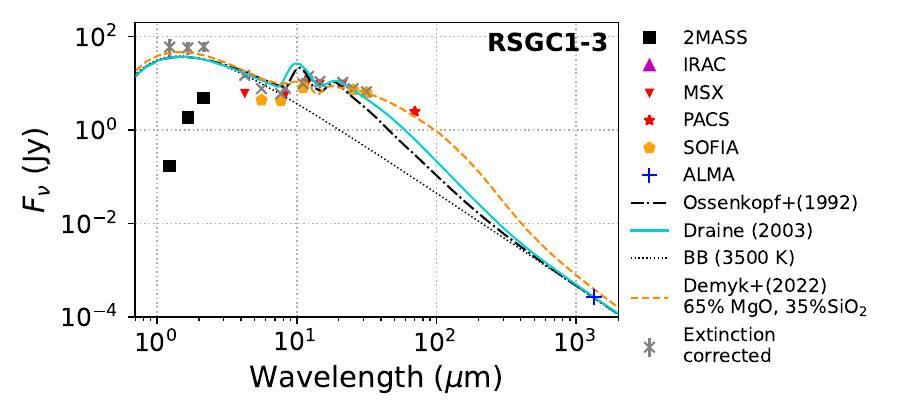}\hfill
\caption{Example extinction-corrected SEDs for RSGC2 and RSGC1 targets. 
% with added ALMA continuum and Herschel/PACS photometry measurements. 
Best-fit \texttt{RADMC-3D} models are shown under three different dust-grain-opacity assumptions: \citet{Oseenkopf1992_silicates} silicates, \citet{Draine2003_dustopacs} silicates, and laboratory-derived constants of \citet{Demyk2022_lab_silicates} for an amorphous silicate mixture of 65\% MgO and 35\% SiO$_2$. The luminosity and $\dot{M}$ reported in Table~\ref{tab:sed_results} correspond to the \citet{Draine2003_dustopacs} silicates model. The \citet{Oseenkopf1992_silicates} models are shown for equivalent optical depth at 0.55$\,\mu$m.}
\label{fig:SED_fits}
\end{figure}

Here, we discuss the methodology of the dust models described \ref{sect:DUSTmodeling}, and present their detailed results. Though \texttt{RADMC-3D} is a three-dimensional code, we model all sources as spherically symmetric, steady-state winds. In general, we follow a similar procedure to \citet{humphreys2020}: adopting the same spectral types, interstellar extinction corrections, and fixed inner radius dust temperature (1000\,K). The main difference in our approach is that instead of varying the slope of the density profile, we fix it to the standard $\rho\propto r^{-2}$ and 
% alternatively 
adopt the empirically measured silicate optical constants from \citet{Draine2003_dustopacs} instead of those of \citet{Oseenkopf1992_silicates} (Figure \ref{fig:SED_fits}). The \citeauthor{Draine2003_dustopacs} grains have comparatively larger far-IR emissivities, shifting the IR-excess to longer wavelengths. This has a similar effect to adopting a shallower density profile, as it lowers the temperature of the dust column and thereby reduces the spectral index in the 20--100\,$\mu$m range, which is often needed to reproduce the shallow shape of the observed SEDs (Fig.\ \ref{fig:SED_fits}). To demonstrate the dependence of the SED on input dust properties, we additionally compute the same set of models using grains from \citet{Oseenkopf1992_silicates} as well as the laboratory-measured opacities from \citet{Demyk2022_lab_silicates} (65\% MgO, 35\% SiO$_2$)\footnote{\href{https://github.com/Thiebauts/DustOpacityGenerator}{https://github.com/Thiebauts/DustOpacityGenerator}}. Optical constants were generated using Optool \citep{Dominik2021_Optool} using a $a^{-3.5}$ power-law grain size distribution with $a_{\rm min}=0.01$\,$\mu$m and $a_{\rm max}=1.0$\,$\mu$m \citep{Mathis1977_grainsize}.

For each SED and grain type, we run a grid of \texttt{RADMC-3D} models with total $\dot{M}$ ranging from $10^{-8}-5\times10^{-4}$\,\msunyr and determine the best-fit model using the modified $\chi^2$-metric of \citet{Yang2023_SMC_RSGwinds}, based on the ratio of model and observed fluxes:
\begin{equation}
\chi_{\rm mod}^2 = \frac{1}{N-p-1}
\sum_i w_i 
\frac{\left[1-F_i\left({\rm Model, }\lambda\right)/F_i\left({\rm Obs, }\lambda\right)\right]^2}
{F_i\left({\rm Model, }\lambda\right)/F_i\left({\rm Obs, }\lambda\right)}
\end{equation}
where we include added weights based on the measurement uncertainties:
\begin{equation}
w_i = \frac{\tilde{\sigma}_i^{-2}}{\sum_j \tilde{\sigma}_j^{-2}}, \qquad
\tilde{\sigma}_i = \max\left(\frac{\sigma_i}{F_i},\, \epsilon\right).
\end{equation}

\begin{table*}[htb!]
\caption{%
\texttt{RADMC-3D} SED-fitting results compared to the analysis of \citet[][H20]{humphreys2020}. }\label{tab:sed_results}
\centering
\begin{tabular}{l c|c c c c c c c c}
\hline\hline
& ID & $T_\star$ $^{a}$ & $L_{\rm H20}$ & $L_{\rm fit}$ $^{b}$& $\dot{M}_{{\rm H20, }n\neq2}$ $^{c}$ & $\dot{M}_{{\rm H20, }n=2}$ & $\tau_{\rm{fit}}$ $^{d}$ & $\dot{M}_{\rm fit}$ & $R_{\rm in}$ \\
& & [K] & [$10^5\,L_\odot$] & [$10^5\,L_\odot$] & [$10^{-6}\,M_\odot\,\mathrm{yr}^{-1}$] & [$10^{-6}\,M_\odot\,\mathrm{yr}^{-1}$] & & [$10^{-6}\,M_\odot\,\mathrm{yr}^{-1}$] & [au] \\
\hline
% \rule{0pt}{2.5ex}\textbf{RSGC2} &  &  &  &  &  &  &  & \\
\textbf{RSGC2}
& \rule{0pt}{2.5ex}1 & 3200 & \nodata & $3.21 \pm 1.15$ & \nodata & \nodata & $0.8$ & $5.94^{+1}_{-1}$ & $61.1$ \\
& \rule{0pt}{2.5ex}2 & 3200 & $1.60 \pm 0.70$ & $1.91 \pm 0.68$ & $13.00$ & $1.23$ & $0.42$ & $2.29^{+0.6}_{-0.4}$ & $44.9$ \\
& \rule{0pt}{2.5ex}3 & 3400 & $0.88 \pm 0.38$ & $1.00 \pm 0.36$ & \nodata & $0.51$ & $0.1$ & $0.40^{+0.08}_{-0.10}$ & $32.5$ \\
& \rule{0pt}{2.5ex}5 & 3400 & $1.00 \pm 0.40$ & $0.66 \pm 0.24$ & $5.80$ & $0.55$ & $0.21$ & $0.67^{+0.1}_{-0.2}$ & $26.9$ \\
& \rule{0pt}{2.5ex}6 & 3600 & $0.53 \pm 0.23$ & $0.58 \pm 0.21$ & \nodata & $0.20$ & \nodata & $<0.05$ & $25.5$ \\
& \rule{0pt}{2.5ex}8 & 3900 & $0.84 \pm 0.34$ & $0.57 \pm 0.20$ & \nodata & $0.26$ & $0.05$ & $0.15^{+0.03}_{-0.04}$ & $26.5$ \\
& \rule{0pt}{2.5ex}49 & 4000 & $3.90 \pm 1.70$ & $0.15 \pm 0.05$ & $770.00$ & $130.00$ & $28$ & $81.52^{+15}_{-17}$ & $24.2$ \\
\hline
% \rule{0pt}{2.5ex}\textbf{RSGC1} &  &  &  &  &  &  &  & \\
\textbf{RSGC1}
& \rule{0pt}{2.5ex}1 & 3500 & $3.35 \pm 1.60$ & $2.92 \pm 0.93$ & $17.00$ & $2.50$ & $0.39$ & $2.83^{+0.5}_{-0.5}$ & $59.2$ \\
& \rule{0pt}{2.5ex}2 & 3700 & $2.15 \pm 1.00$ & $2.83 \pm 0.90$ & $9.00$ & $2.10$ & $0.37$ & $2.79^{+0.5}_{-0.5}$ & $60.9$ \\
& \rule{0pt}{2.5ex}3 & 3500 & $1.20 \pm 0.50$ & $1.75 \pm 0.56$ & $15.00$ & $1.42$ & $0.39$ & $2.19^{+0.4}_{-0.4}$ & $45.8$ \\
& \rule{0pt}{2.5ex}4 & 3800 & $3.80 \pm 1.80$ & $2.29 \pm 0.73$ & $9.70$ & $2.27$ & $0.19$ & $1.25^{+0.2}_{-0.2}$ & $54.0$ \\
& \rule{0pt}{2.5ex}13 & 4200 & $2.90 \pm 1.40$ & $3.25 \pm 1.04$ & $27.00$ & $6.31$ & $0.34$ & $2.99^{+0.5}_{-0.7}$ & $72.1$ \\
\hline\hline
\end{tabular}
\tablefoot{ All mass-loss rates are cumulative (gas+dust) assuming a wind expansion velocity of 25\,\kms and gas-to-dust ratio of 200. (a) Spectral types are from \citet{davies2007_rsgc2,davies2008}, or the updated values from \citet{negueruela2012} for RSGC2 stars 2, 3, 5, and 6. (b) Errors in reported luminosities incorporate the cluster distance uncertainty ($\pm1$\,kpc for both) as well as the error in the adopted interstellar extinction $A_K$. (c) $\dot{M}_{{\rm H20, }n\neq2}$ is the published mass-loss rate from \citet{humphreys2020} averaged over a variable $\dot{M}$ profile, while $\dot{M}_{{\rm H20, }n=2}$ is the value yielding equivalent optical depth assuming a constant mass-loss rate. (c) Optical depth of the best-fit model at 0.55\,$\mu$m.}
\end{table*}

Here, $F_i\left({\rm Obs, }\lambda\right)$ and $F_i\left({\rm Model, }\lambda\right)$ are the observed and modeled fluxes, $N$ is the number of data points, $p$ is the number of free parameters (1), $\sigma_i$ is the photometric uncertainty (including both catalog-reported errors and extinction-correction uncertainties), and $\epsilon$ is a fractional error floor introduced to prevent a small number of high-precision photometric points from dominating the fit; we adopt $\epsilon = 0.2$ for all fits. Adopting a ratio-based minimization as opposed to a traditional $\chi^2$ additionally ensures that long-wavelength points are adequately considered in the fit as opposed to only the very bright NIR fluxes which dominate a traditional $\chi^2$ \citep{Yang2023_SMC_RSGwinds}.

Since all SEDs are initially computed for a fixed luminosity of $10^5\,L_\odot$, the above minimization is used to determine a multiplicative scaling factor applied to each model, and the scaled luminosity is then adopted as our final reported value.
The stellar radius, dust inner radius, and mass-loss rate are subsequently rescaled as well to preserve the stellar effective temperature and optical depth of the optimal model.

Figure \ref{fig:SED_fits} shows the SED-fitting results for two RSGs in each cluster, including newly-measured ALMA fluxes as well as archival Herschel/PACS photometry measurements that have not yet been published. We note that the ALMA continuum fluxes measured in this work are consistent (within a factor of 2) with purely stellar emission extrapolated from the extinction-corrected 2MASS fluxes. In contrast, the Herschel/PACS 70\,$\mu$m fluxes are often underestimated by the dust models using grain opacities from \citet{Oseenkopf1992_silicates} and \citet{Draine2003_dustopacs}, especially for RSGC1 sources. While this could be evidence for a variable mass-loss rate interpretation like that adopted by \citet{humphreys2020}, we stress again that it may be a shortcoming of the assumed optical properties of the dust. The laboratory-measured constants from \citet{Demyk2022_lab_silicates} produce much colder grains and can thus accurately reproduce the mid-to-far-IR shape for RSGC1 sources without changing the slope of the density profile (Fig.\ \ref{fig:SED_fits}). A full exploration of the dust grain composition and size distribution parameter space is outside the scope of this work, so we emphasize the need for future constraints on these properties and on the conditions in the dust-formation regions to physically interpret this behavior.

Our derived luminosities and dust mass-loss rates, obtained from the scaled best-fit model using \citet{Draine2003_dustopacs} opacities, are presented in Table \ref{tab:sed_results} along with the results of \citet{humphreys2020}. For comparison with the results of \citet{beasor2020}, for which we have three sources in common, we refer to the Tables in that work. We find that our measured dust optical depths are similar to those found by \citet{beasor2020} and \citet{humphreys2020}. Because the same gas-to-dust ratio and expansion velocity were applied in computing the total $\dot{M}$ listed in Table \ref{tab:sed_results}, the discrepancies between mass-loss rates can be mainly attributed to the different assumptions in dust properties and density/temperature structure. We note that our reported $\dot{M}$ is consistently smaller than the quoted average over the variable mass loss models of \citet{humphreys2020}, but comparable or slightly larger than their equivalent $\tau$ model with $n=2$ (calculated from Eqs.\ A8--A9 of that work). For sources in common with \citet{beasor2020} (RSGC1-1, 2, 3), our retrieved $\dot{M}$ is smaller by a factor of ${\sim}2$.

The target with the largest discrepancy in luminosity relative to previous work is DFK\,49. Both \citet{davies2007_rsgc2} and \citet{humphreys2020} note its extremely reddened SED and suggest that the observed $K$-band flux suffers an additional ${\sim}3$ magnitudes of circumstellar extinction, implying a luminosity of $3.9\times10^5$\,\lsun and a tentative post-RSG classification. However, if this interpretation were correct, we would expect the attenuated near-IR flux to be reradiated at wavelengths beyond 20\,$\mu$m, which are now constrained by the far-IR and mm observations. Instead, the SED model for DFK\,49 (Fig.\ \ref{fig:SED_fits}) yields a bolometric luminosity of $L=1.5\times10^4$\,\lsun, while direct integration of the photometry gives $L=1.9\times10^4$\,\lsun. These results disfavor a post-RSG interpretation and instead highlight this as a peculiar object showing unusually high $\dot{M}$ relative to its comparatively low luminosity.

% \clearpage

\section{All detections and shell fits}\label{app:linefits}

\begin{figure}[!htbp]
    % --------- Row 1 ---------
    \centering
    \includegraphics[width=0.65\linewidth]{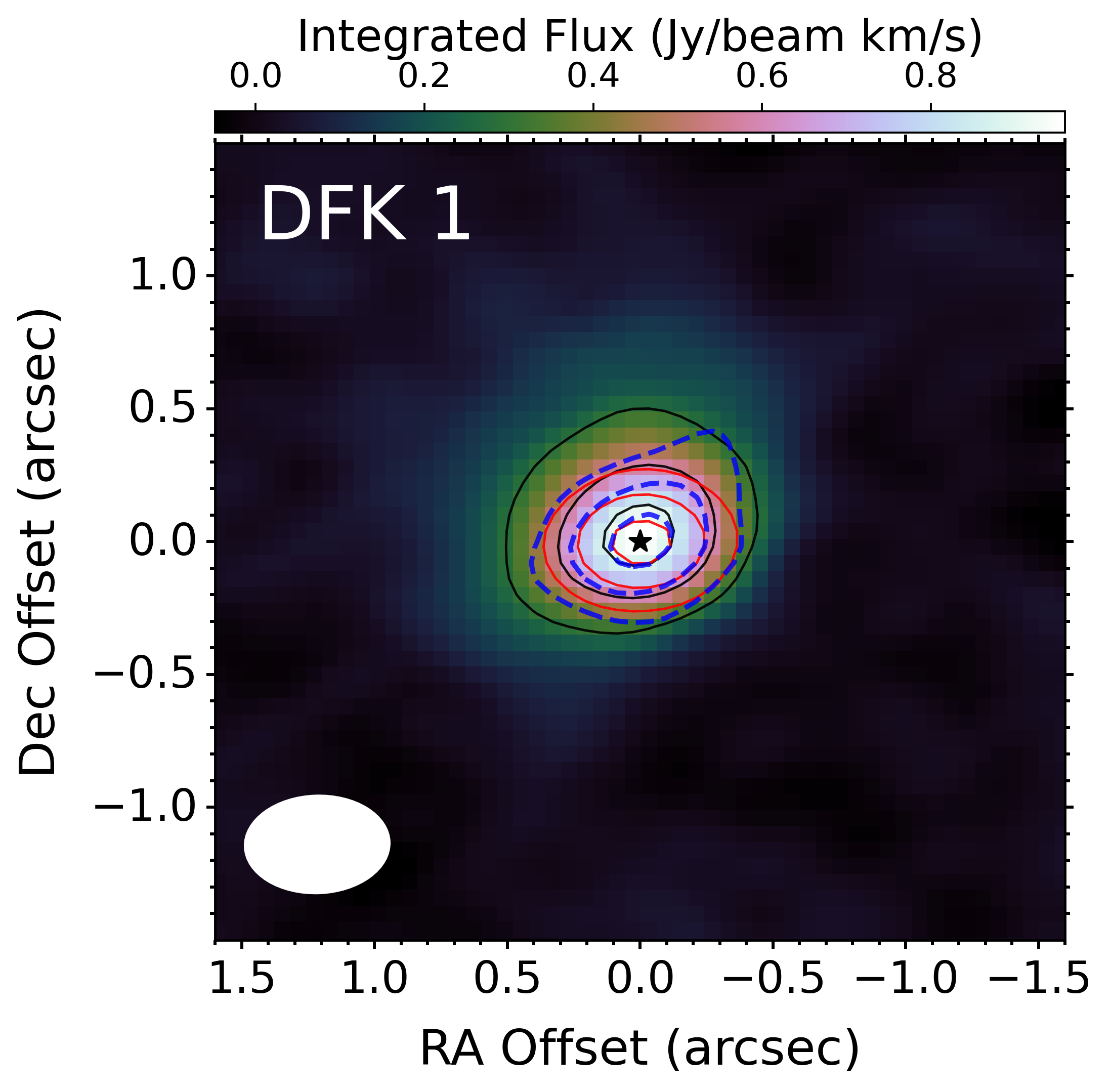} 
    \includegraphics[width=0.65\linewidth]{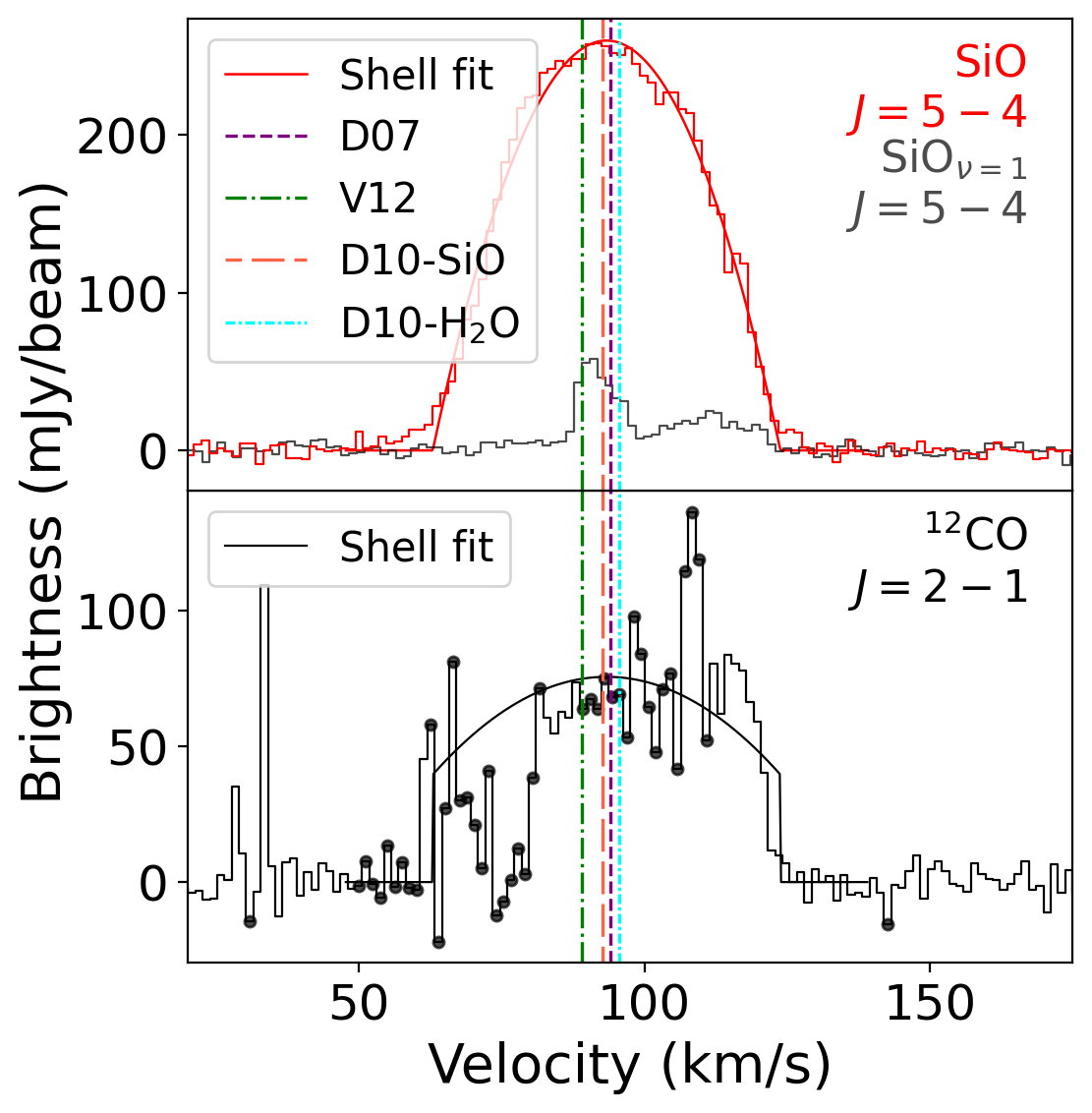} 
    
    % --------- Row 3 ---------
    \includegraphics[width=0.65\linewidth]{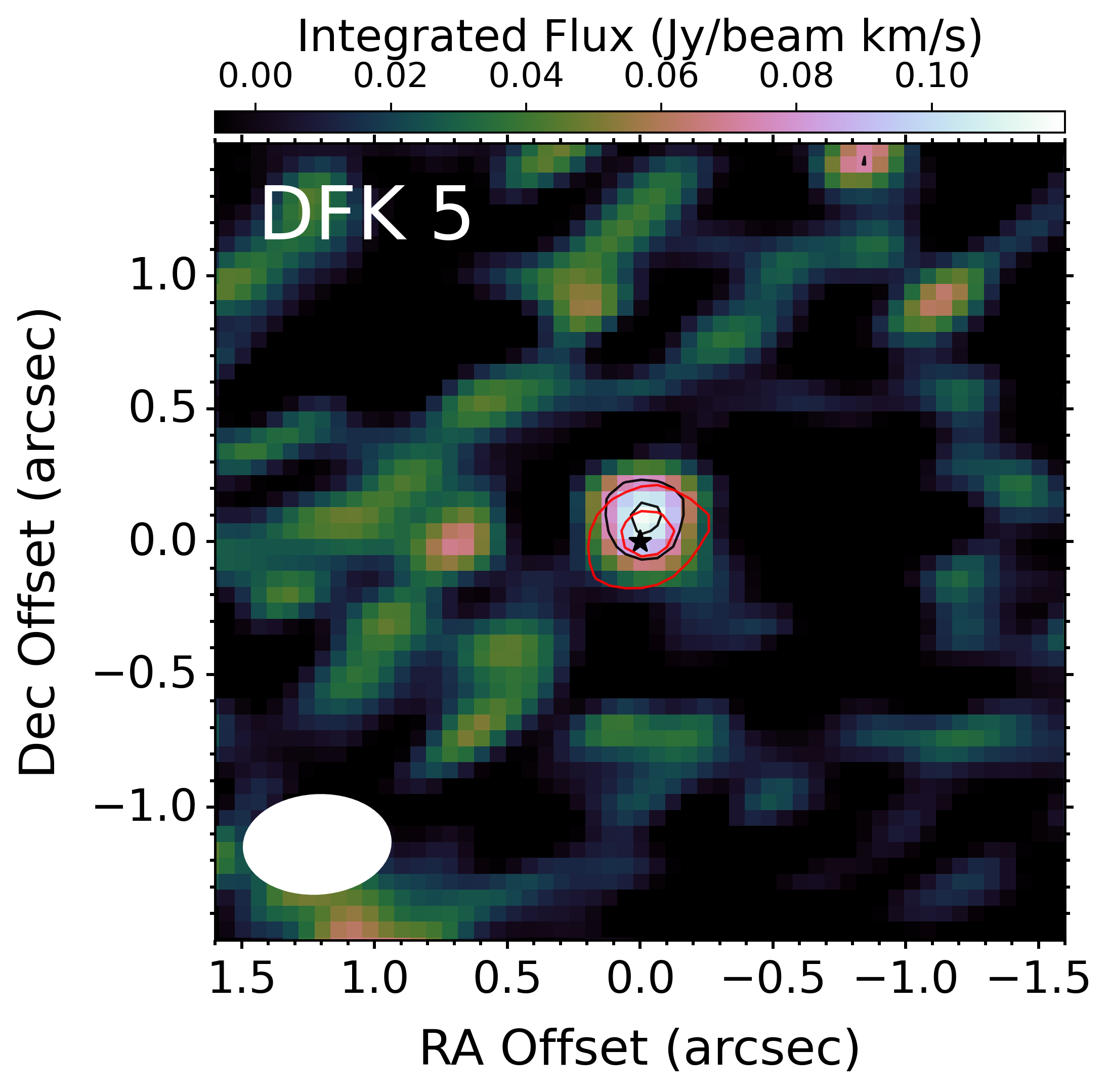} 
    \includegraphics[width=0.65\linewidth]{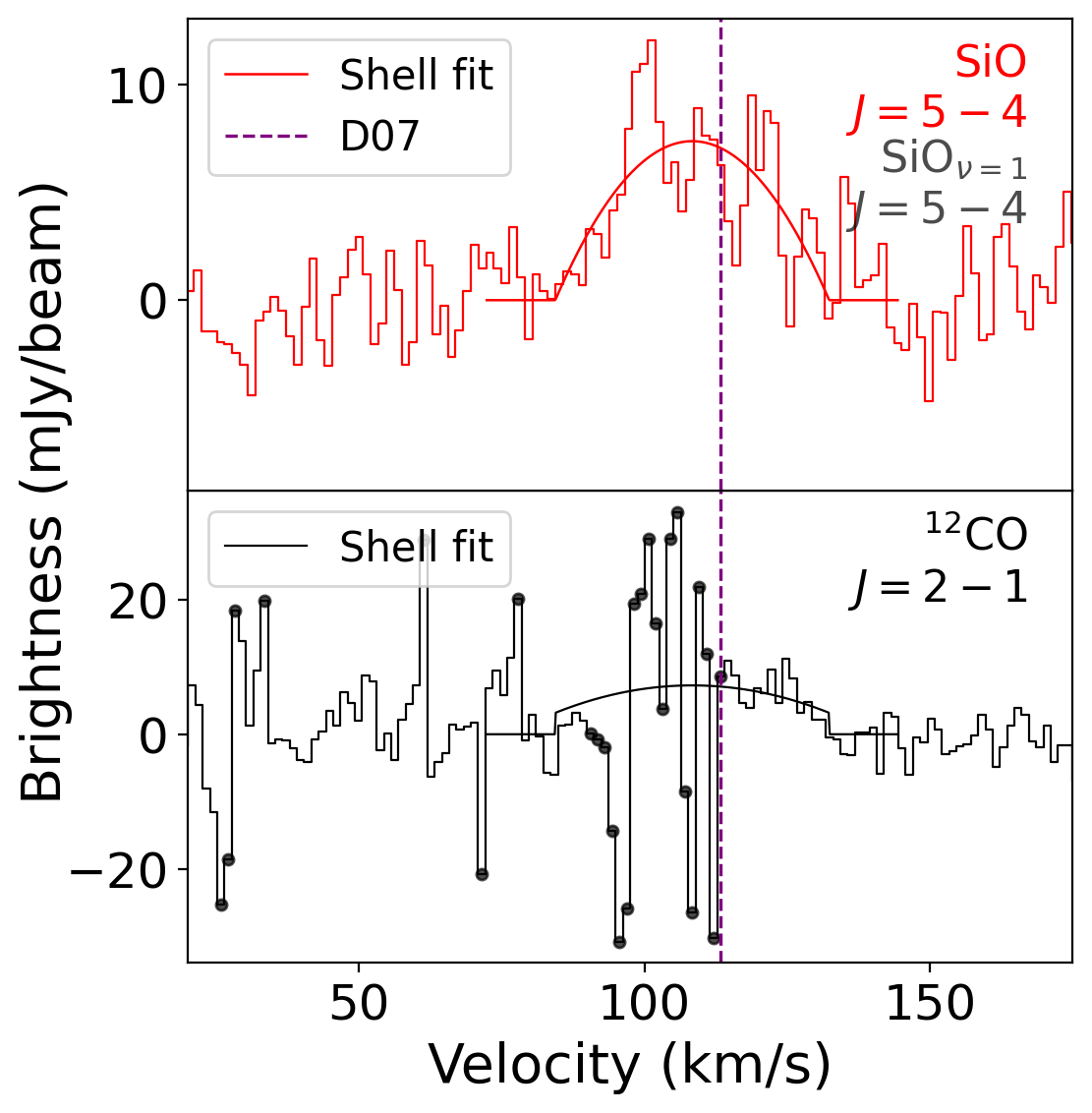} 

    \caption{Same as Fig.\ \ref{fig:obs_summary} for all remaining detected RSGs in both clusters.}\vspace{-1em}
    \label{fig:obs_summary_full}
\end{figure}

\begin{figure*}[p]
    % --------- Row 1 ---------
    \ContinuedFloat
    \centering
    \includegraphics[width=0.3\linewidth]{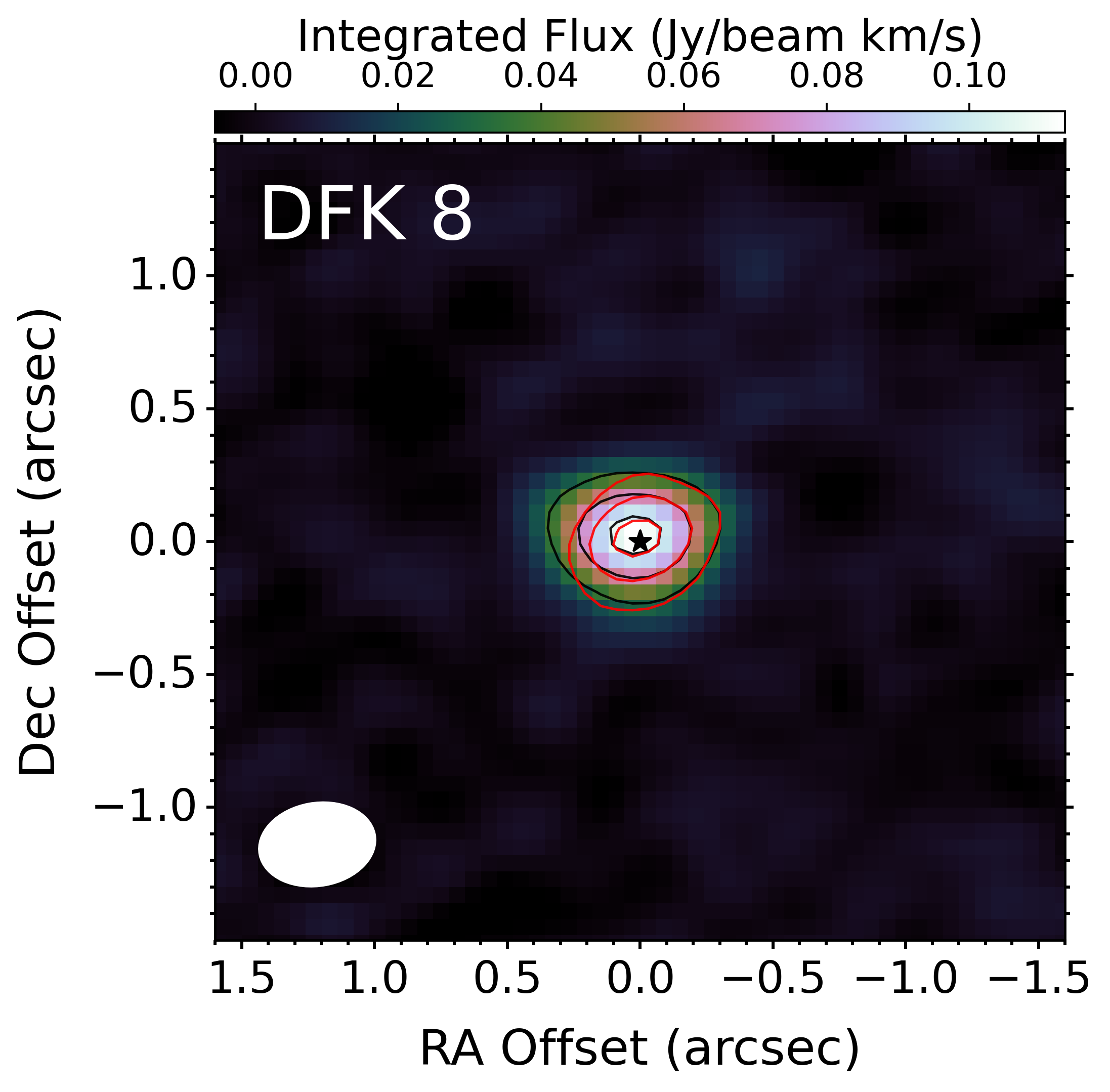} 
    \includegraphics[width=0.3\linewidth]{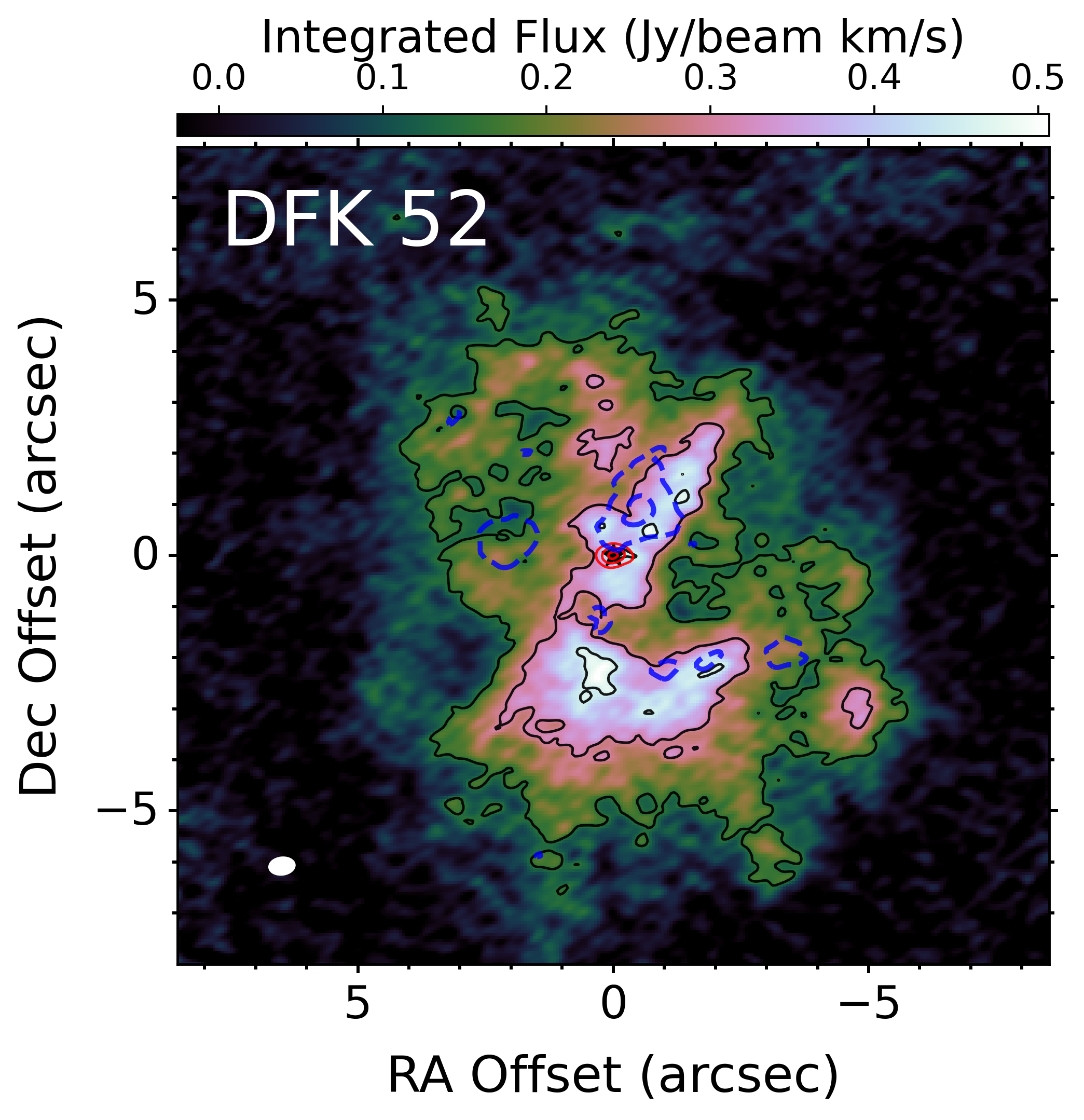} 
    \includegraphics[width=0.3\linewidth]{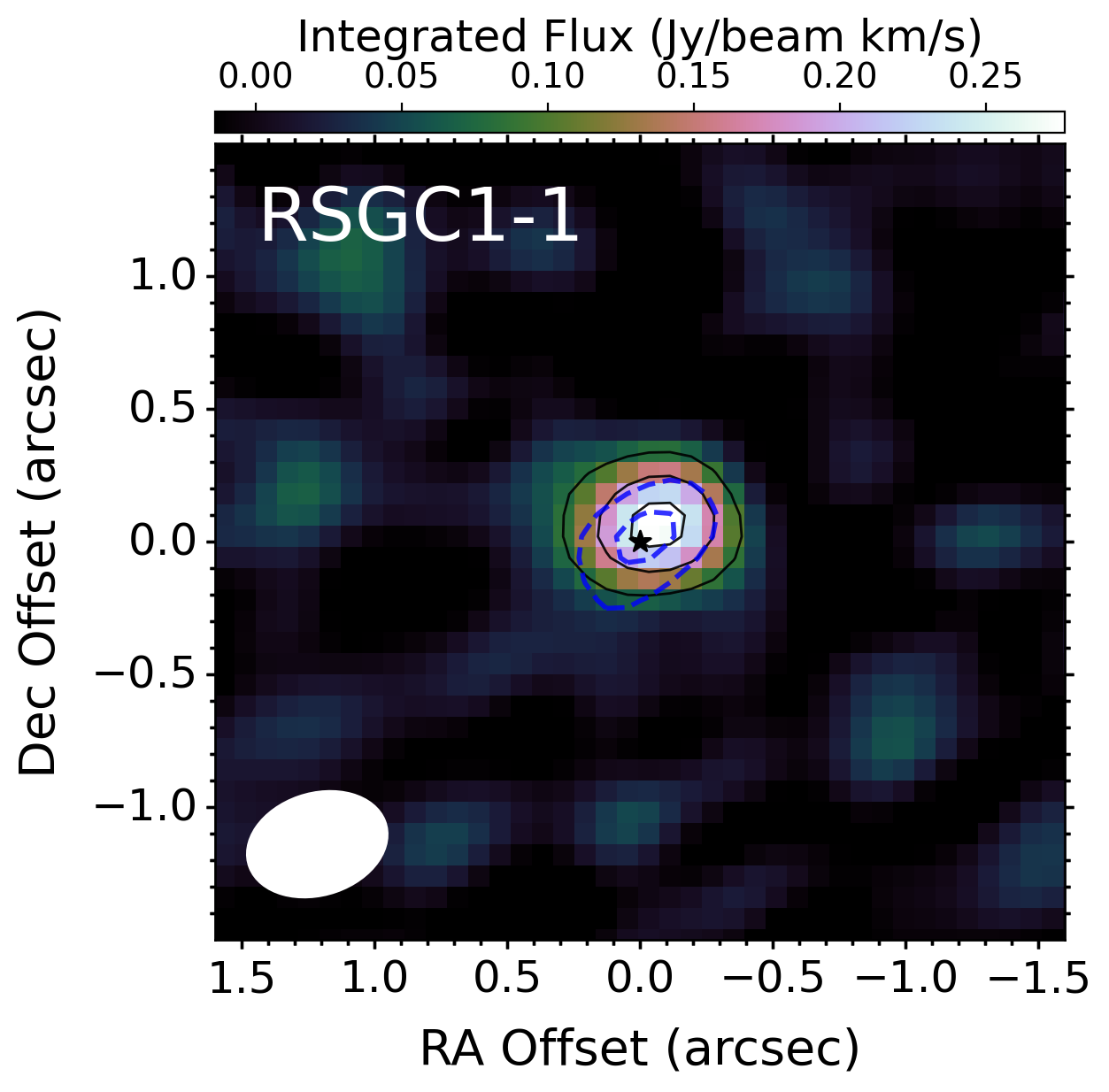} 
 
    % --------- Row 2 ---------
    \includegraphics[width=0.3\linewidth]{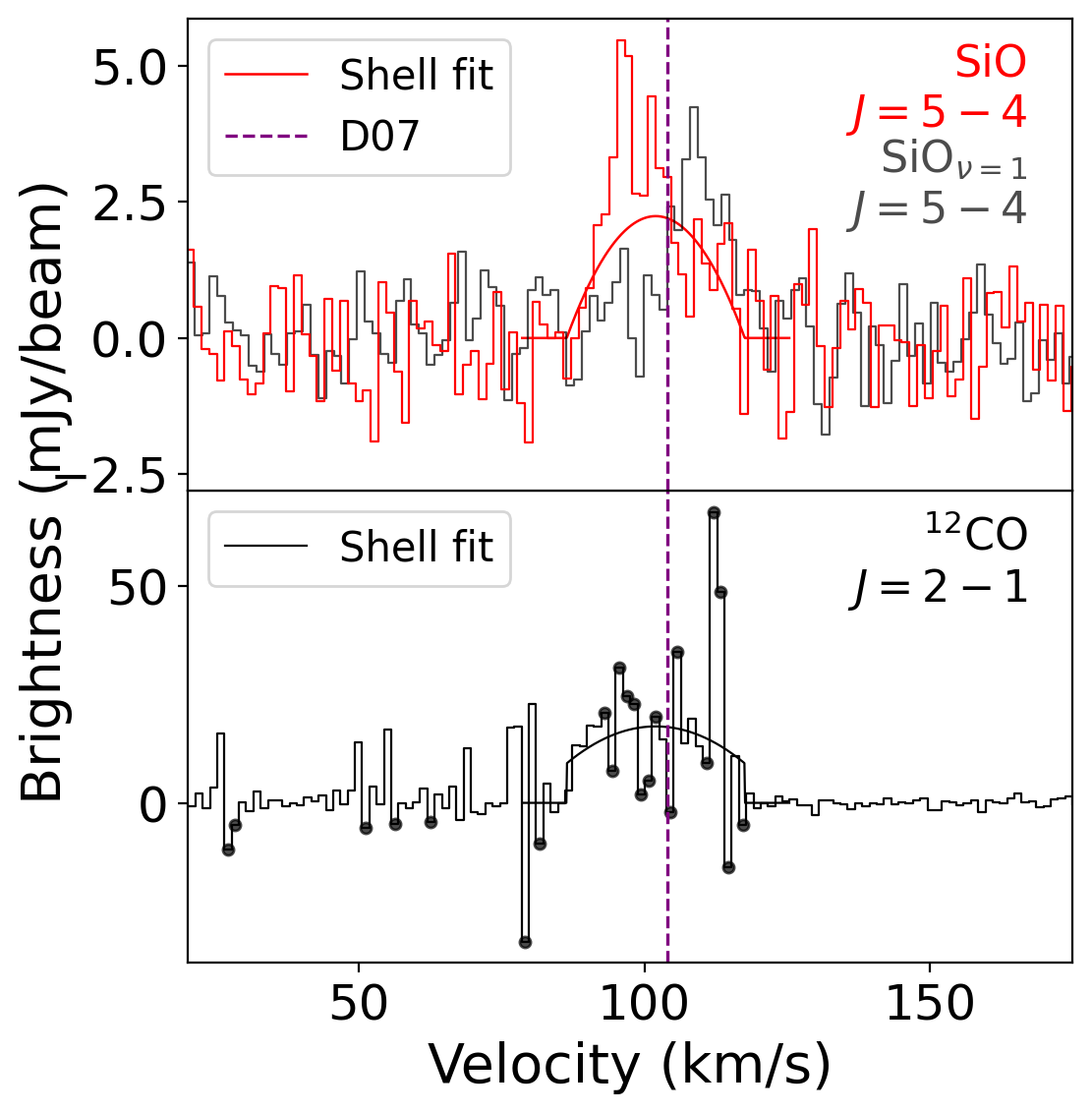} 
    \includegraphics[width=0.3\linewidth]{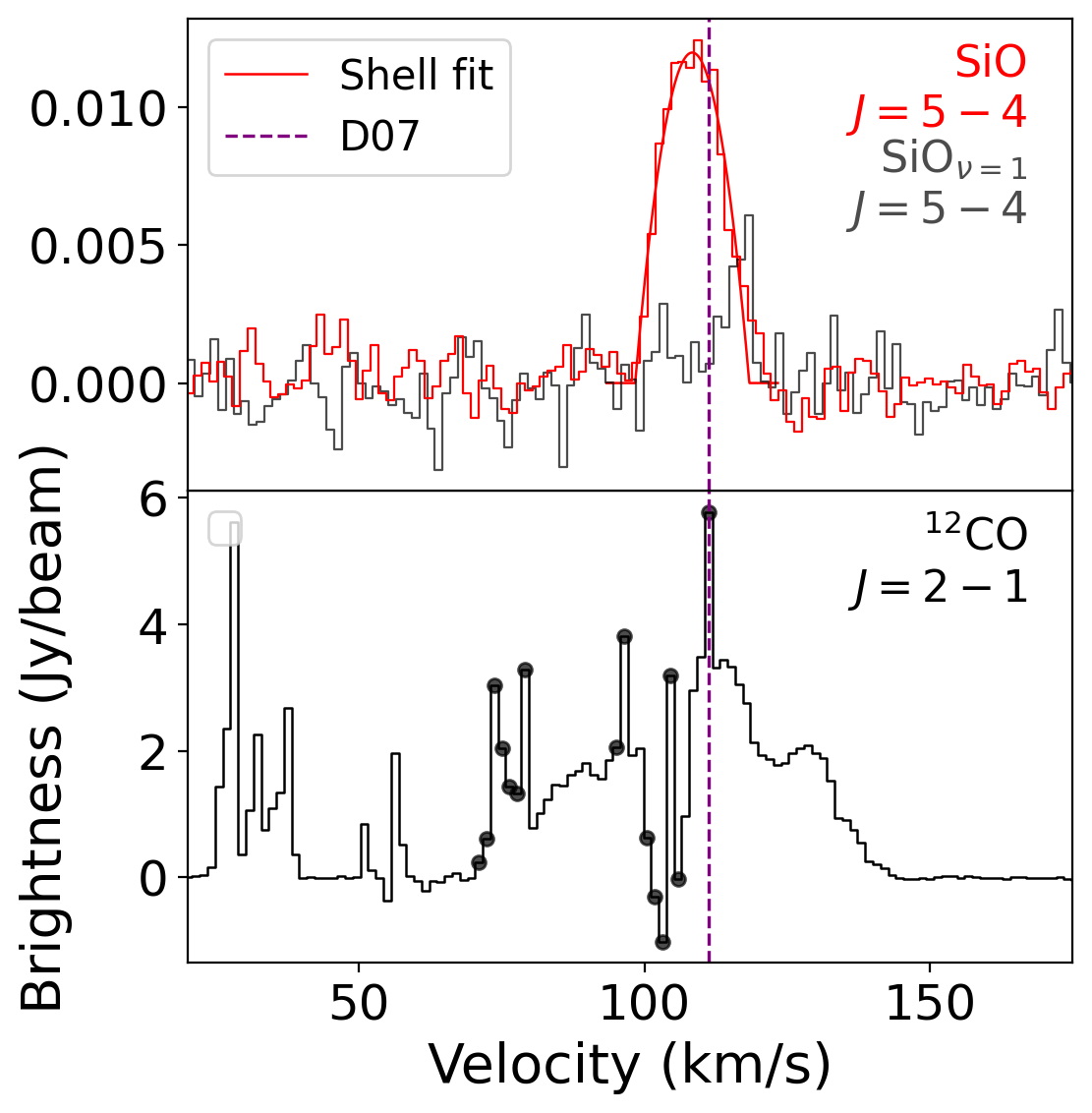} 
    \includegraphics[width=0.3\linewidth]{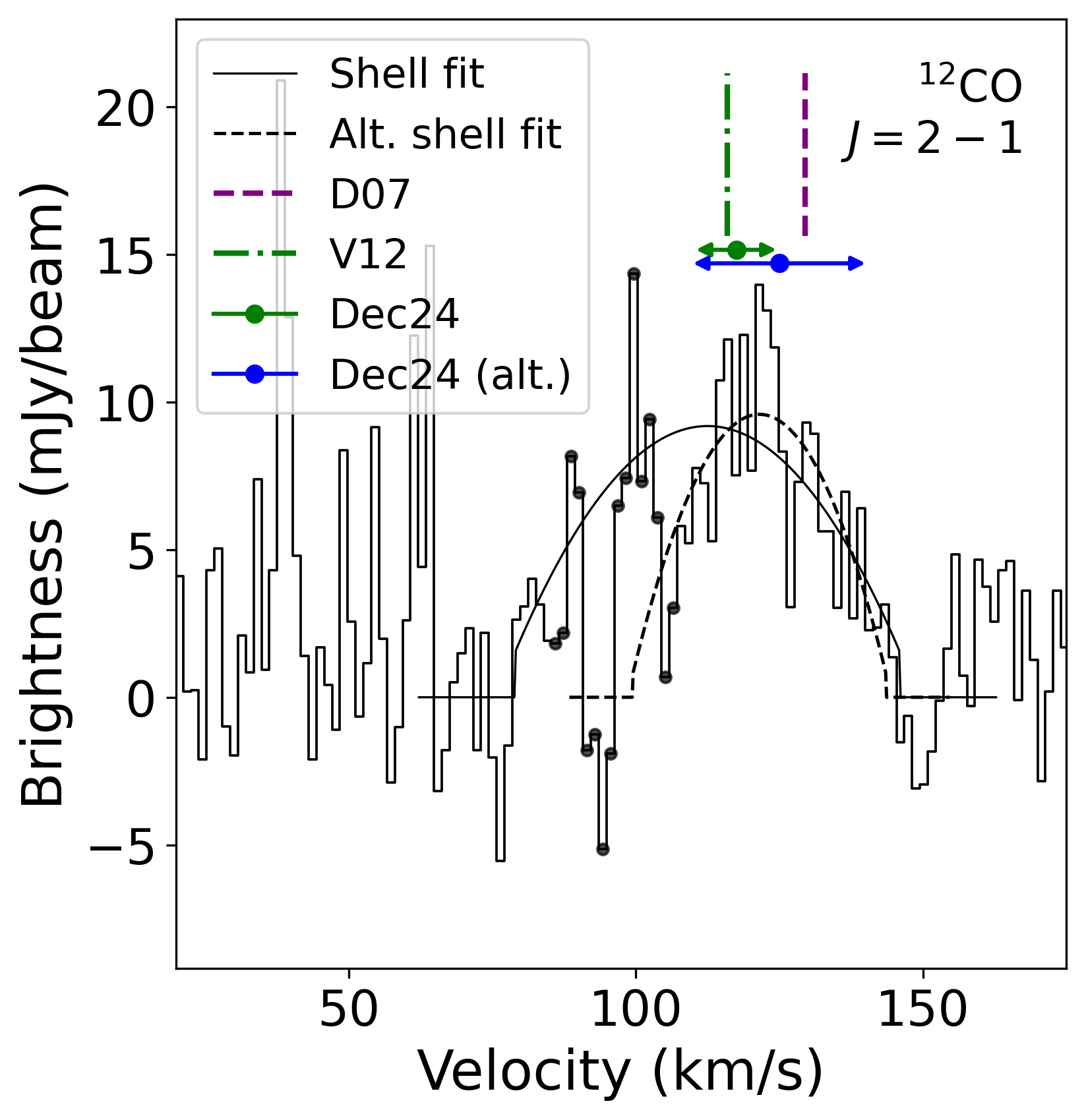} 

    \includegraphics[width=0.3\linewidth]{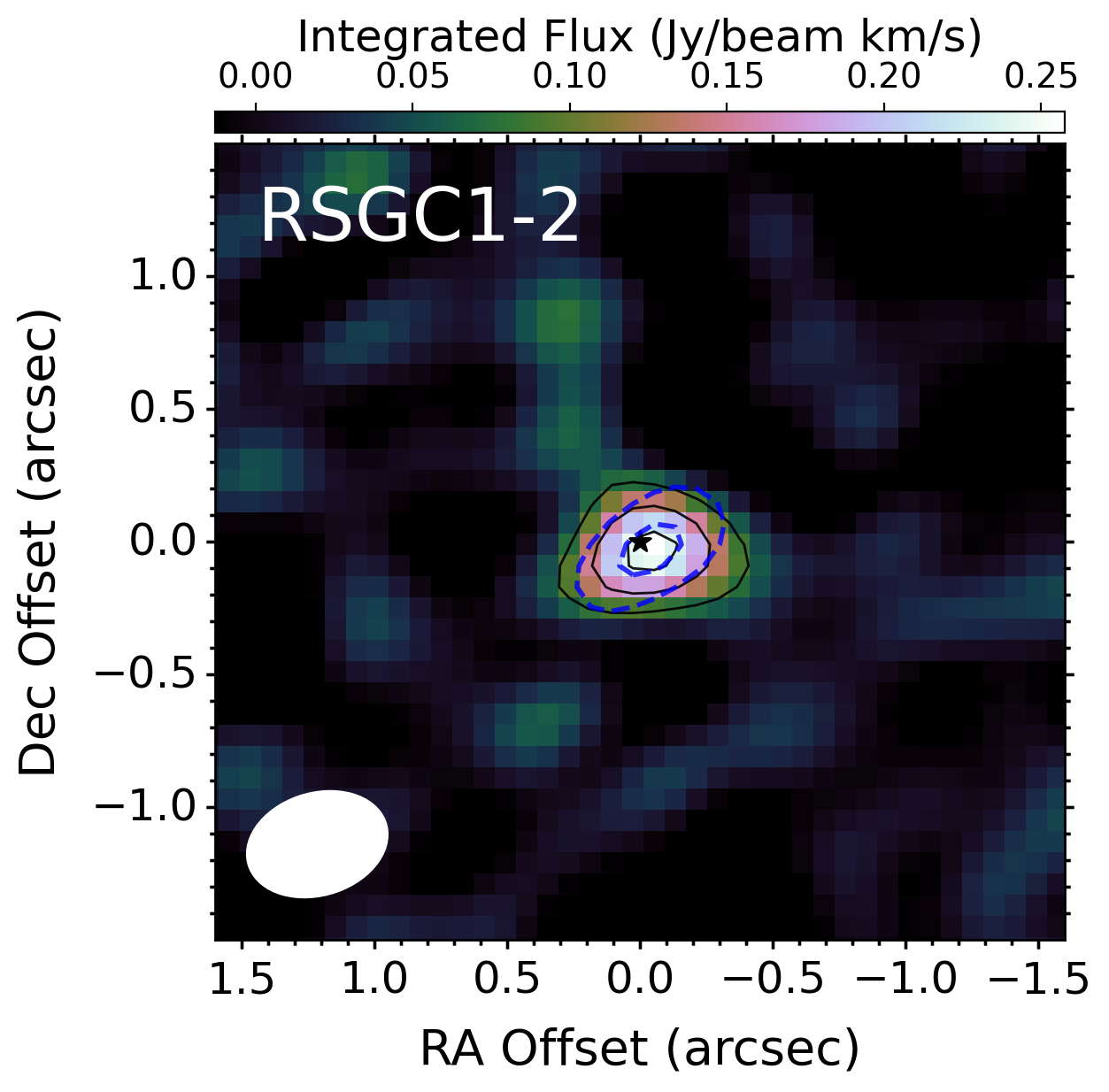} 
    \includegraphics[width=0.3\linewidth]{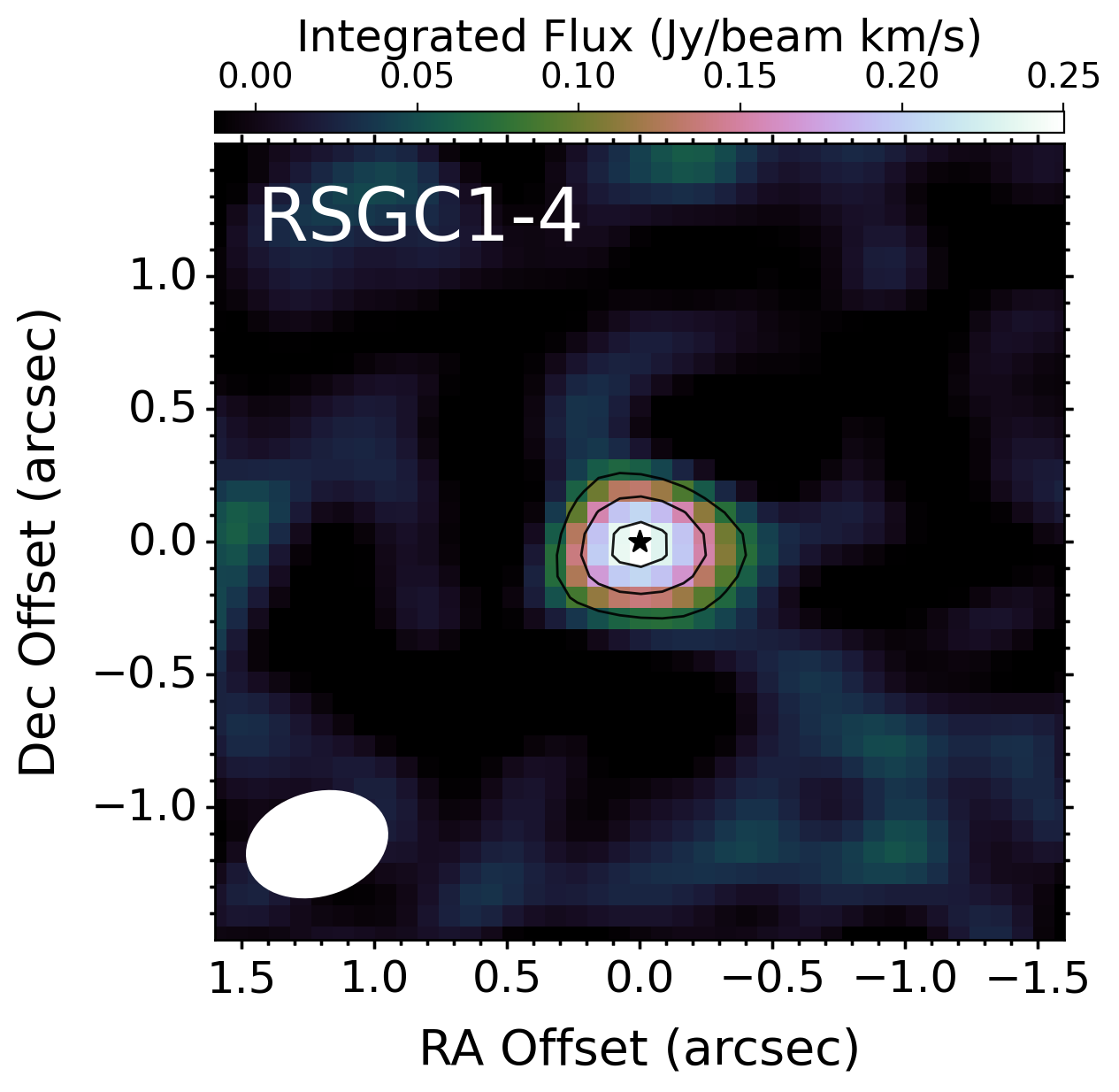} 
    \includegraphics[width=0.3\linewidth]{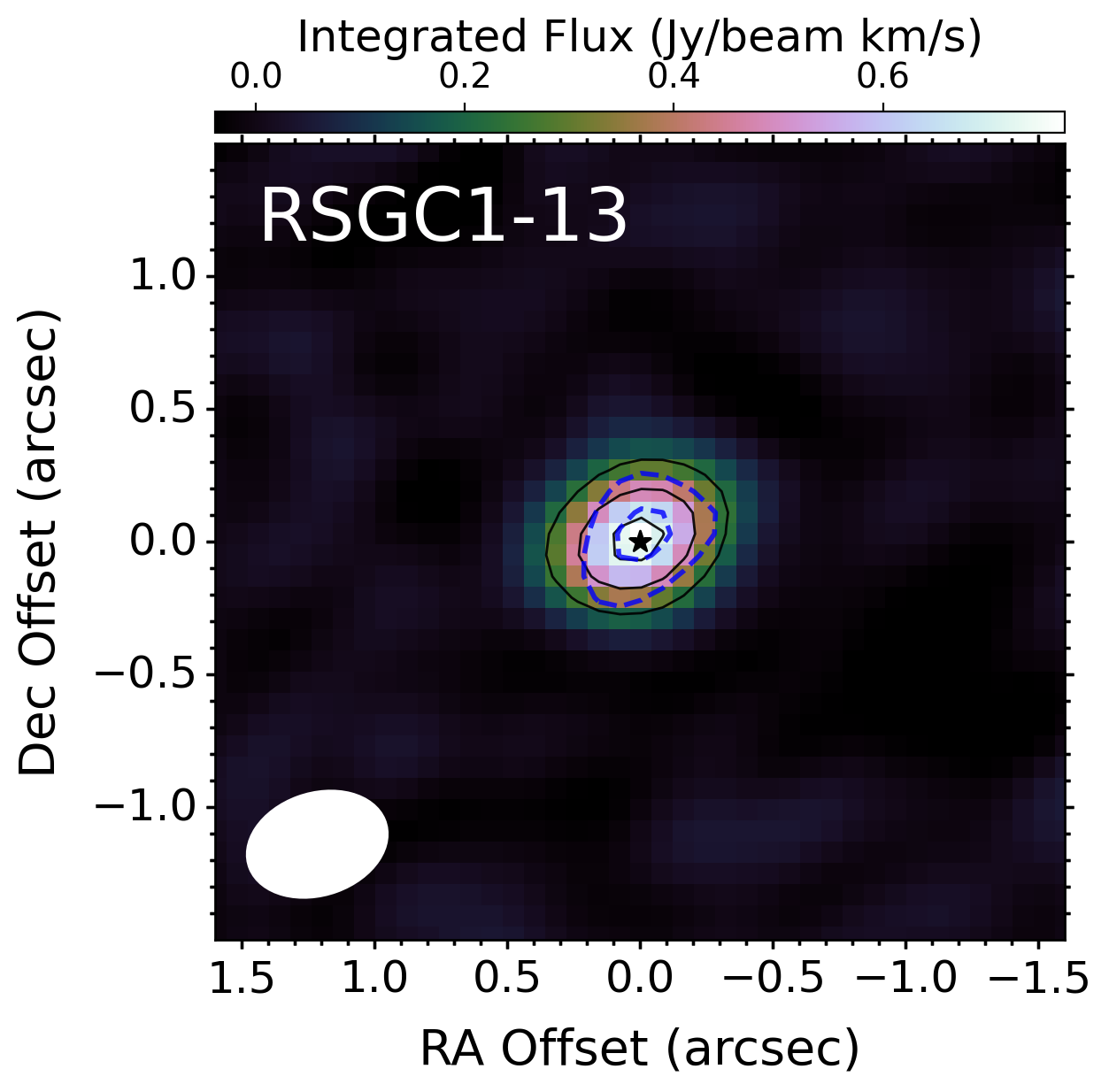} 
 
    % --------- Row 3 ---------
    \includegraphics[width=0.3\linewidth]{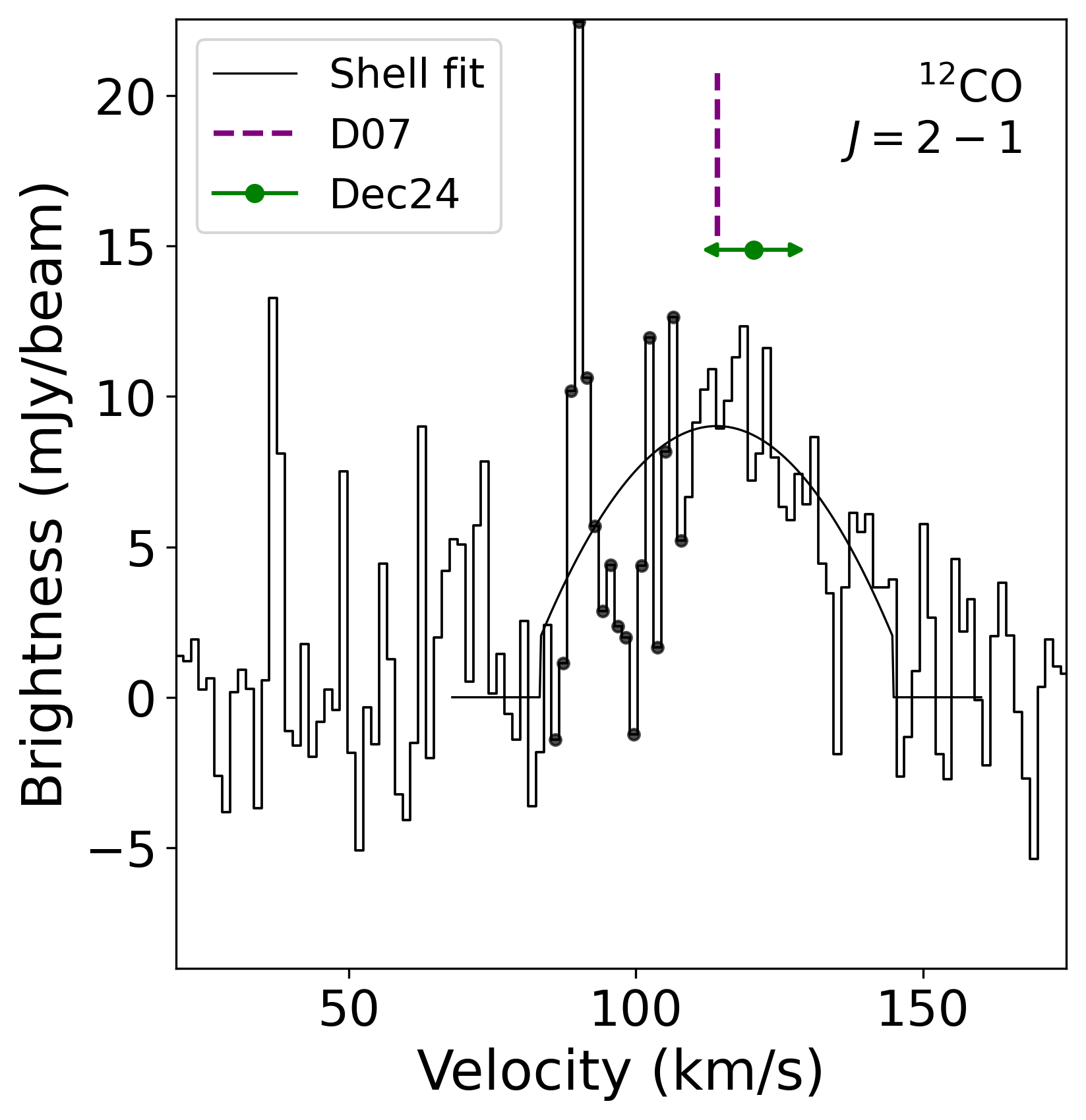} 
    \includegraphics[width=0.3\linewidth]{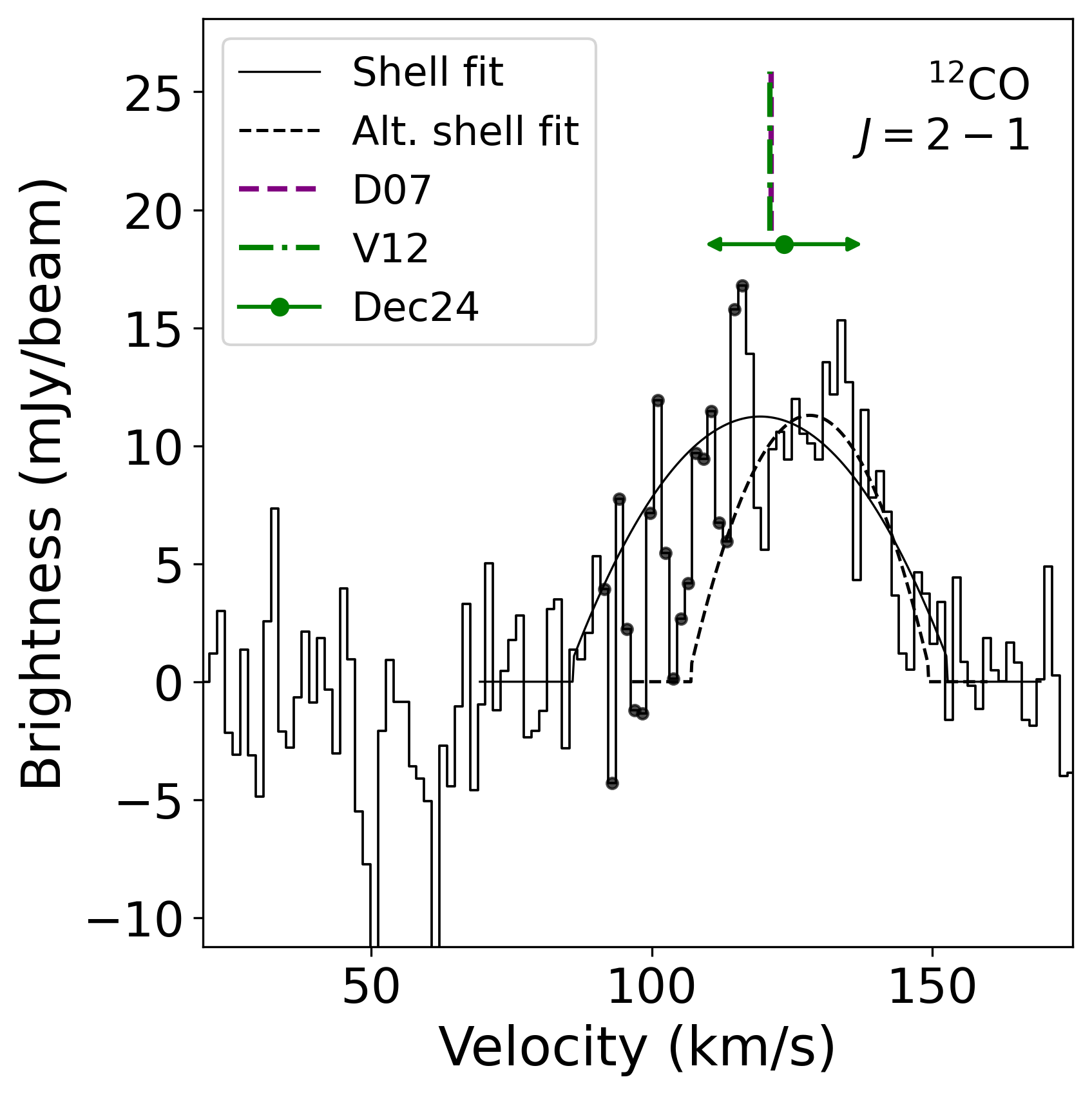} 
    \includegraphics[width=0.3\linewidth]{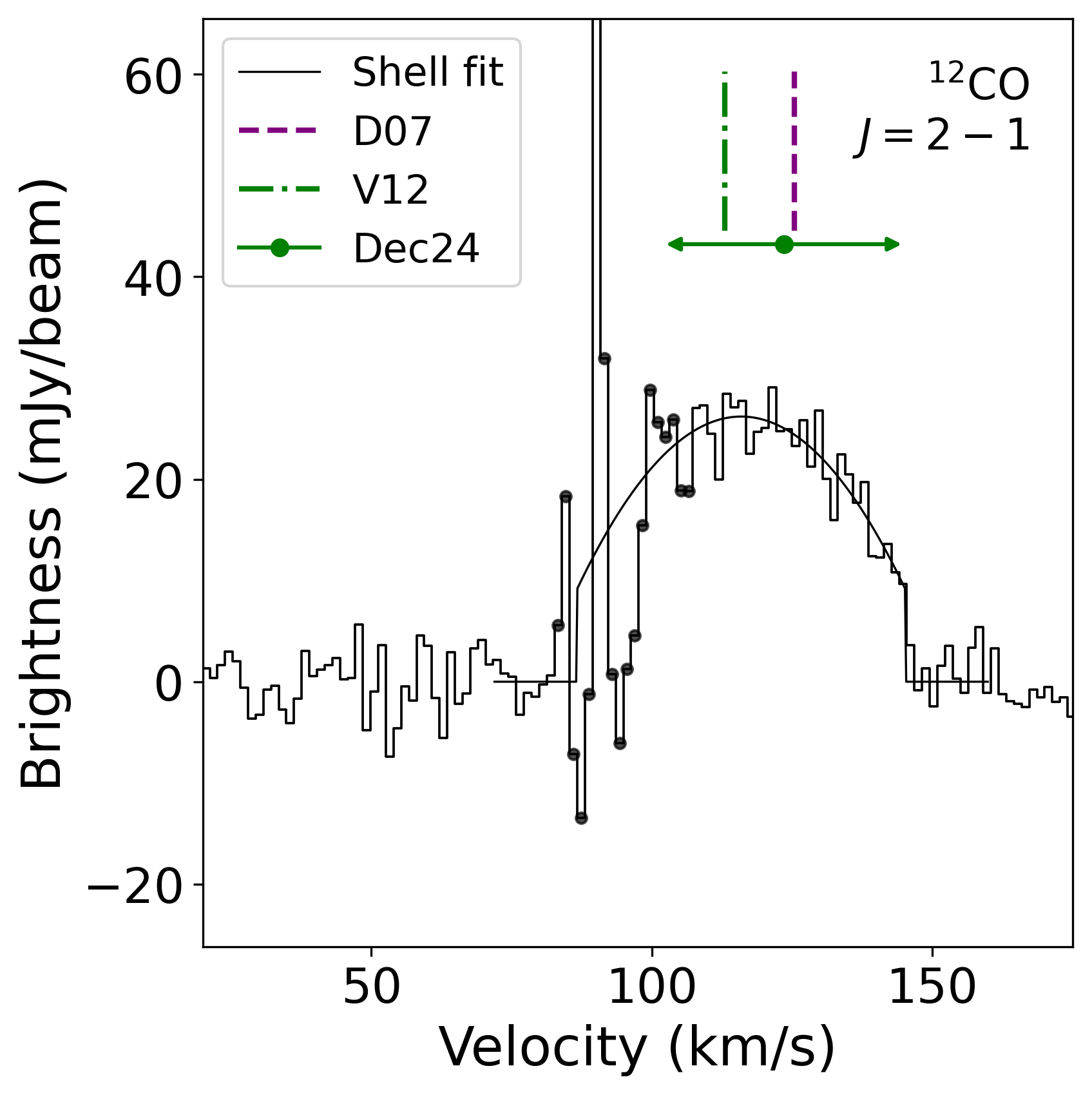} 

    \caption{continued}\vspace{-1em}
\end{figure*}

Figure \ref{fig:obs_summary_full} shows the additional ALMA molecular line and continuum detections utilized in this work, as well as our shell profile fits to each CO and SiO$_{\nu=0}$ line described in Section \ref{sect:linefits}. Here, we describe the line profile fitting procedure, and the resulting source-specific uncertainties in detail. Due to the non-uniform sensitivity, line coverage, and ISM-affected velocity ranges in the data sets, the degree to which the line width and shape can be constrained varies from source to source.

Under the shell line profile assumption (Eq.~\ref{eq:shell}), apart from the rigid amplitude scaling parameter $A$, the centroid and line width parameters ($v_{\mathrm{LSR}}$ and $v_{\mathrm{exp}}$, respectively) have the greatest influence on the geometric structure of the modeled emission. Variations in these velocity parameters shift the locations of sharp spectral features and thus produce strongly non-linear behavior in the residuals, particularly in the presence of masked velocity channels. We therefore construct a two-dimensional grid in $v_{\mathrm{LSR}}$ and $\,v_{\mathrm{exp}}$ for each source. For each grid point, the remaining parameters $A$ and $H$ are optimized via local minimization. This allows us to determine the best-fit velocity parameters and their associated confidence regions through minimization and inspection of $\chi^2(v_{\mathrm{LSR}}, v_{\mathrm{exp}})$:
\begin{equation}
\chi^2(v_{\mathrm{lsr}}, v_{\mathrm{exp}}) = \sum_{i=1}^{N}\left[\frac{S_i - M_i\!\left(v_i \mid A, v_{\mathrm{lsr}}, v_{\mathrm{exp}}, H\right)}{\sigma}\right]^2
\end{equation}
where $S_i$ is the observed flux density in spectral channel $i$, $M_i(v_i \mid A, v_{\rm lsr}, v_{\mathrm{exp}}, H)$ is the shell model evaluated at velocity $v_i$, $\sigma$ is the rms noise per spectral channel, and $N$ is the number of unmasked channels. 
The ISM-masked \twco $J=2-1$ line was used for $\chi^2$-minimization for RSGC1 targets and DFK\,49. For the remaining RSGC2 targets, SiO$_{\nu=0}\,J=5-4$ emission was used with $H=-1$, representing a parabolic line profile.

The results of this $\chi^2$-analysis are shown in Fig.~\ref{fig:chi2_maps} for nine of the detected RSGs in both clusters. We find that for sources fit using SiO emission, the $\chi^2$-surface is nearly Gaussian and mainly dependent on the S/N of the line. DFK\,5 and DFK\,8 have relatively weak SiO lines (Fig.~\ref{fig:obs_summary_full}) and, therefore, have very poorly constrained velocities by this method in comparison to DFK\,1, DFK\,2, and DFK\,52. 

For sources where only \twco was available, we find strong degeneracies between \vLSR and \vexp,
% the two velocity parameters, 
as the missing channels on the blue-shifted edge of the line allow for a wide range of line center/width combinations. This effect appears as diagonal stripes in the $\chi^2$-surface with discretized steps corresponding to locations where the sharp edge of the shell function passes over channel boundaries. These effects lead to larger uncertainties of $2{-}6$\,\kms for sources where only \twco is available.

To quantify the spatial dependence of the line shapes and intensities for each target, the shell function was fit multiple times using spectra extracted from image cubes convolved to increasing circular beam sizes (1.1\arcsec, 1.6\arcsec, 2.1\arcsec, etc.). For all of these fits, the $v_{\rm exp}$ and $v_{\rm LSR}$ were fixed to their best-fit values for the smallest beam, as the S/N decreases when smoothing to larger beams. These yield the same result as discussed in Sect.~\ref{sect:spatial} and Fig.~\ref{fig:radial_profs}, where no statistically significant extended emission is seen for stars apart from DFK\,1, DFK\,52, and tentatively DFK\,49.

\begin{figure*}[t]
\centering
\subfigure[DFK\,1, SiO$_{\nu=0}\,J=5-4$]{%
\includegraphics[width=.325\linewidth]{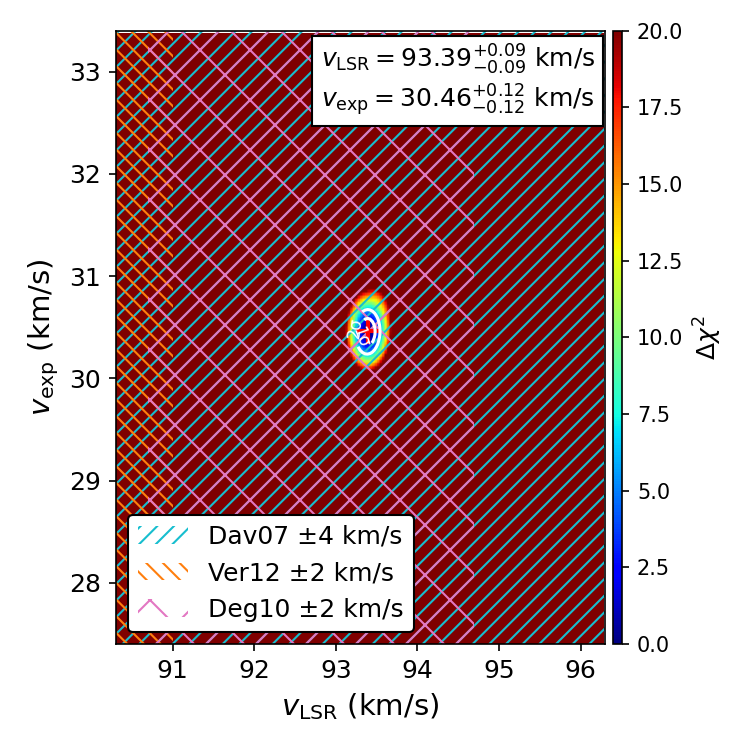}
\label{fig:DFK_1_chi2}}
\subfigure[DFK\,2, SiO$_{\nu=0}\,J=5-4$]{%
\includegraphics[width=.325\linewidth]{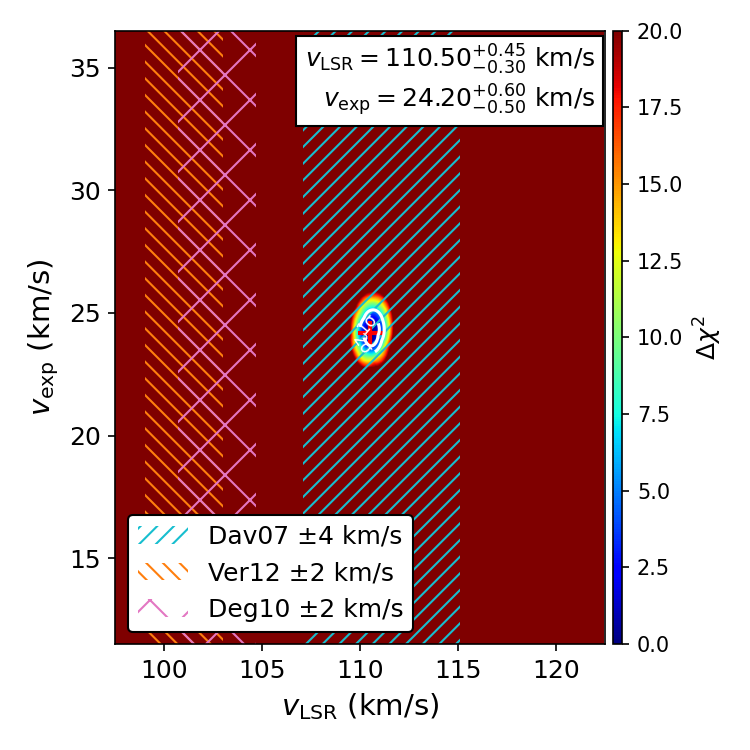}
\label{fig:DFK_2_chi2}}
\subfigure[DFK\,5, SiO$_{\nu=0}\,J=5-4$]{%
\includegraphics[width=.325\linewidth]{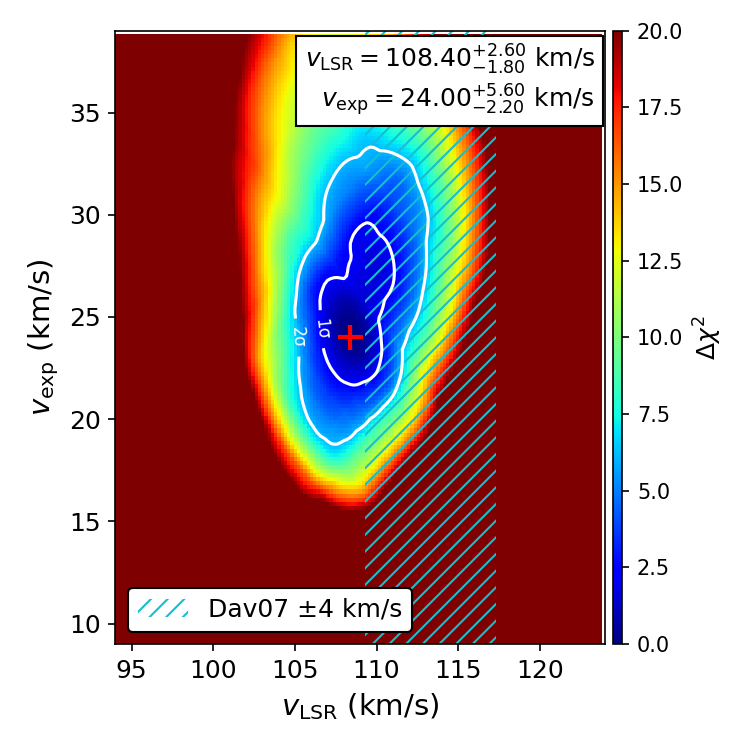}
\label{fig:DFK_5_chi2}}
\subfigure[DFK\,8, SiO$_{\nu=0}\,J=5-4$]{%
\includegraphics[width=0.325\linewidth]{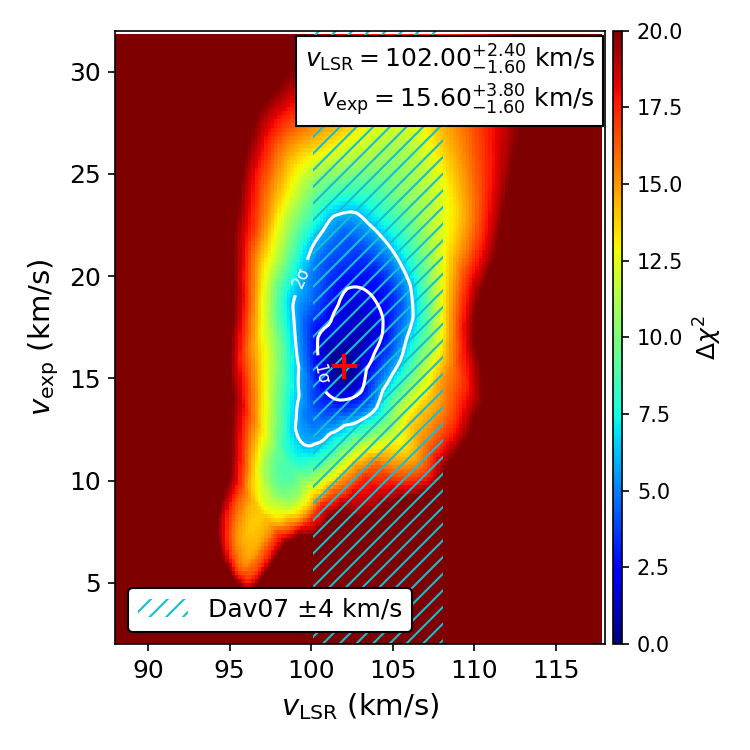}
\label{fig:DFK_8_chi2}}
\subfigure[DFK\,49, \twco $J=2-1$]{%
\includegraphics[width=0.325\linewidth]{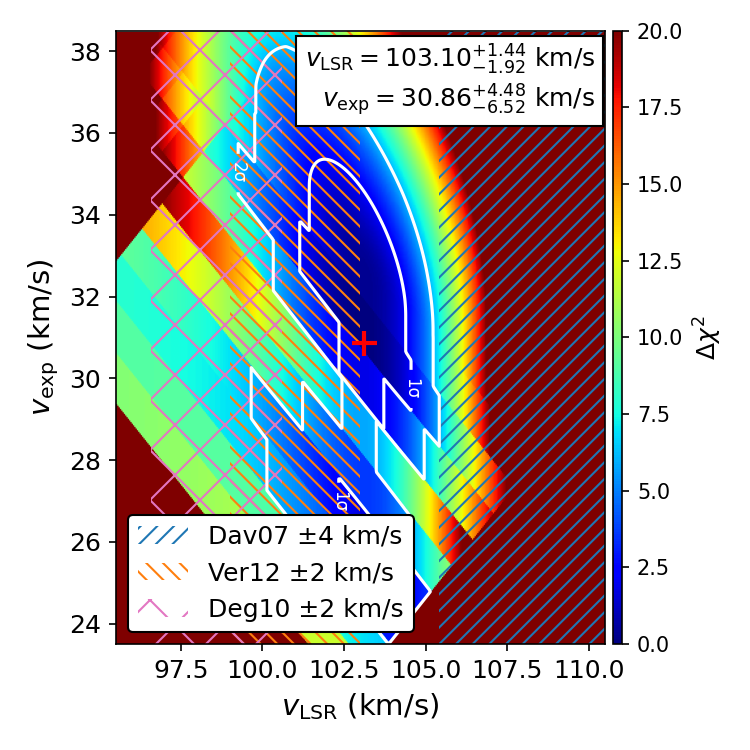}
\label{fig:DFK_49_chi2}}
\subfigure[DFK\,52, SiO$_{\nu=0}\,J=5-4$]{%
\includegraphics[width=0.325\linewidth]{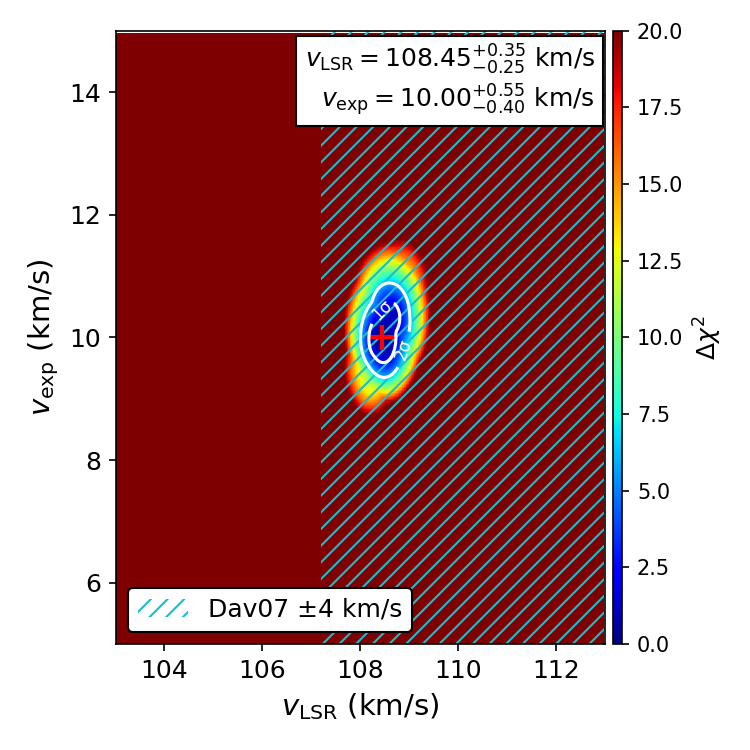}
\label{fig:DFK_52_chi2}}
\subfigure[RSGC1--1, \twco $J=2-1$]{%
\includegraphics[width=0.325\linewidth]{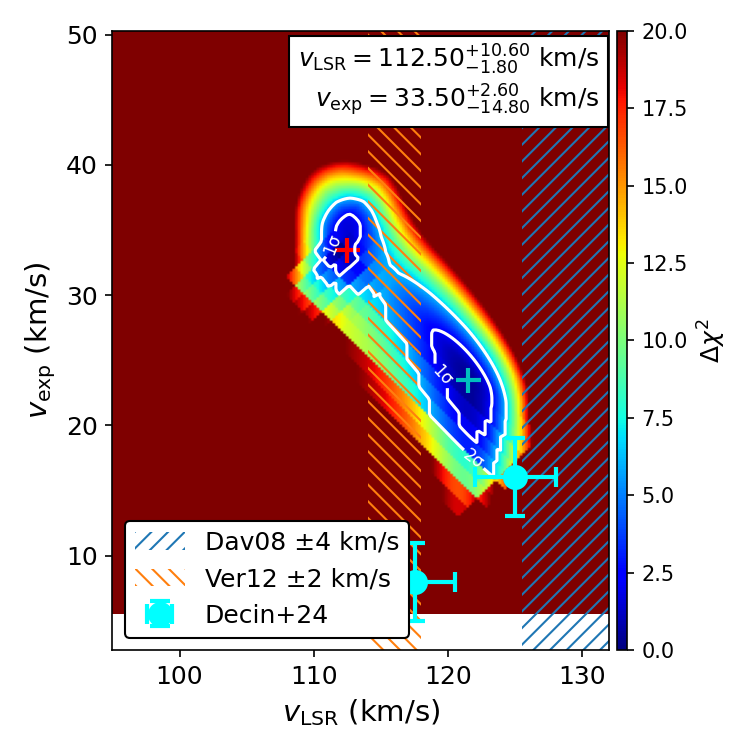}
\label{fig:RSGC1_1_chi2}}
\subfigure[RSGC1--2, \twco $J=2-1$]{%
\includegraphics[width=0.325\linewidth]{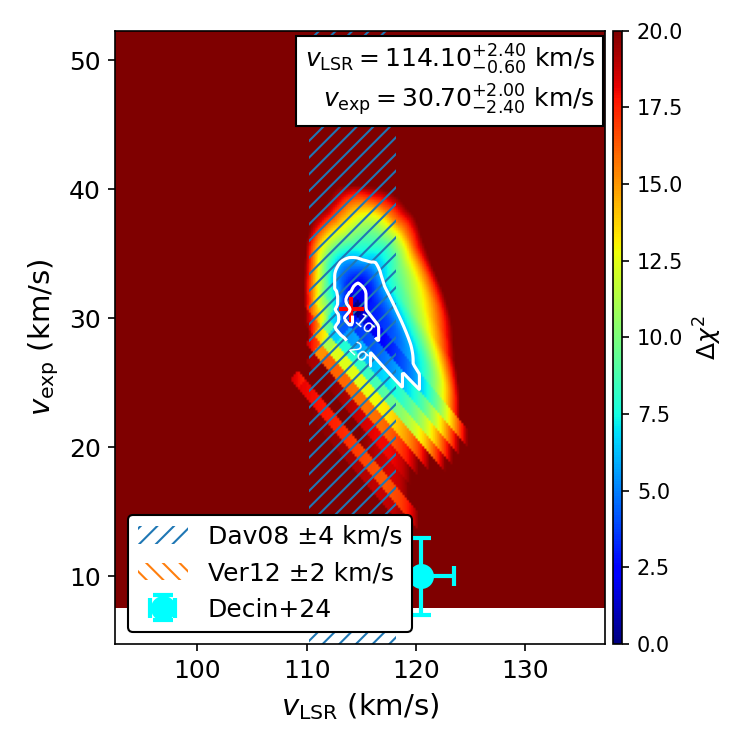}
\label{fig:RSGC1_2_chi2}}
\subfigure[RSGC1--3, \twco $J=2-1$]{%
\includegraphics[width=0.325\linewidth]{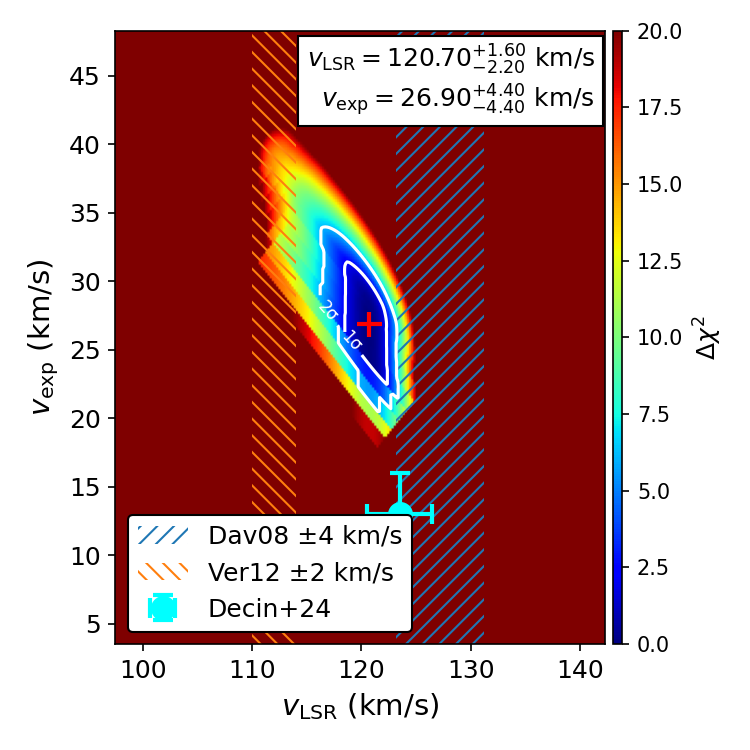}
\label{fig:RSGC1_3_chi2}}
\caption{$\chi^2$-maps for the systemic velocity $v_{\mathrm{LSR}}$ and expansion velocity $\,v_{\mathrm{exp}}$ calculated during fitting of molecular line profiles. 
The minimum (or local minima) is marked with a cross (\textit{red} or \textit{cyan} for secondary), and contours represent the $1\sigma$ and $2\sigma$ surfaces. Previously measured velocities from the 2.3\,$\mu$m bandhead \citep[][for RSGC2, RGSC1, respectively]{davies2007_rsgc2,davies2008}, SiO$_{\nu=1}$ $J=2-1$ \citep{verheyen2012}, and SiO$_{\nu=1}$ $J=1-0$ \citep{deguchi2010} masers are drawn with hatched regions (\textit{blue, orange, pink}, respectively) spanning their quoted uncertainties. For RSGC1 sources, we also show the $v_{\mathrm{LSR}}$ and $\,v_{\mathrm{exp}}$ deduced from the same ALMA observations by \citet[][\textit{cyan points}]{decin2024_rsgc1}.
\vspace{2cm}
}
\label{fig:chi2_maps}
\end{figure*}

\begin{figure*}[t]
\ContinuedFloat
% \sidecaption
\centering
\subfigure[RSGC1--4, \twco $J=2-1$]{%
\includegraphics[width=0.32\linewidth]{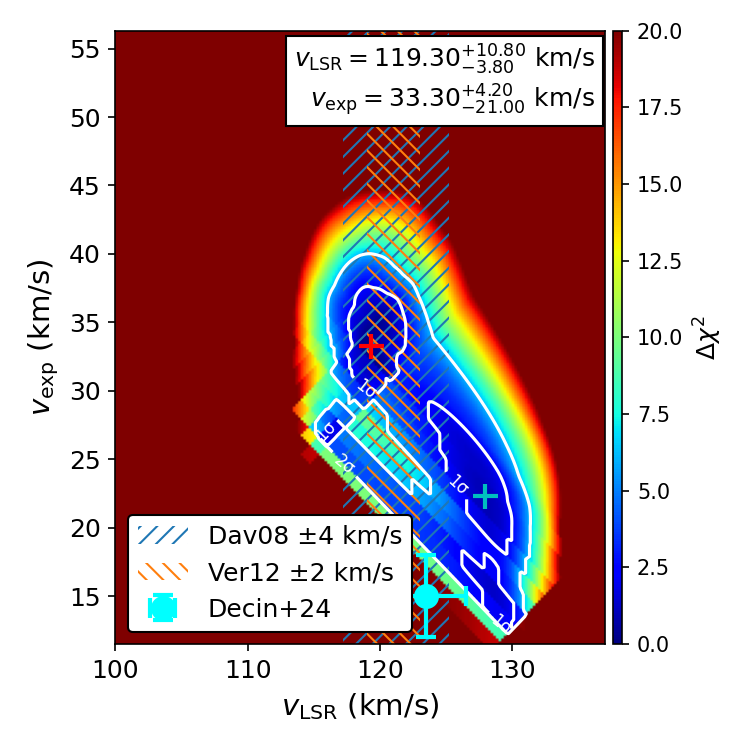}
\label{fig:RSGC1_4_chi2}}
\subfigure[RSGC1--13, \twco $J=2-1$]{%
\includegraphics[width=0.32\linewidth]{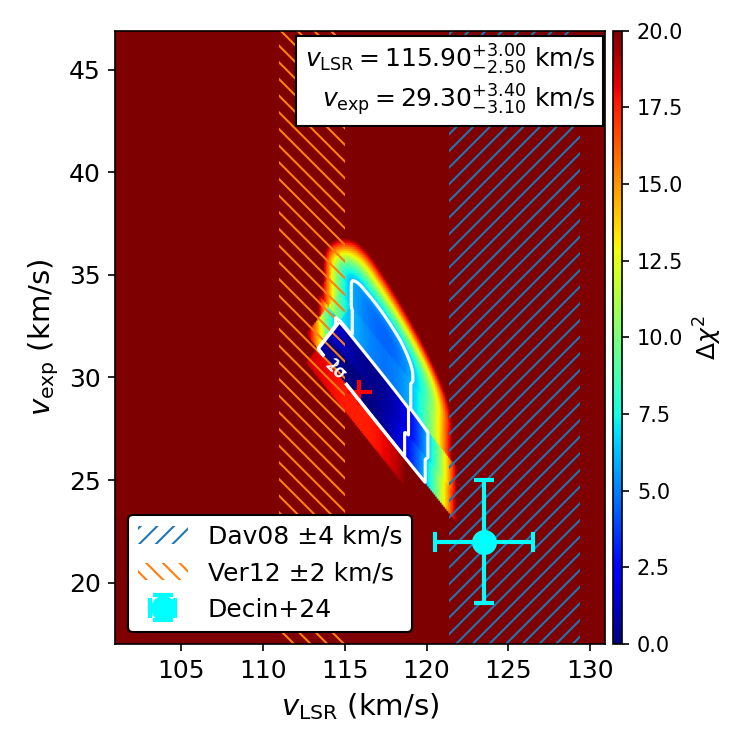}
\label{fig:RSGC1_13_chi2}}
\caption{Continued.}
\end{figure*}

\FloatBarrier

\section{CO models}\label{app:COmodels}
\subsection{CO photodissociation models}
\begin{figure}[!h]
    \centering
    \includegraphics[width=\linewidth]{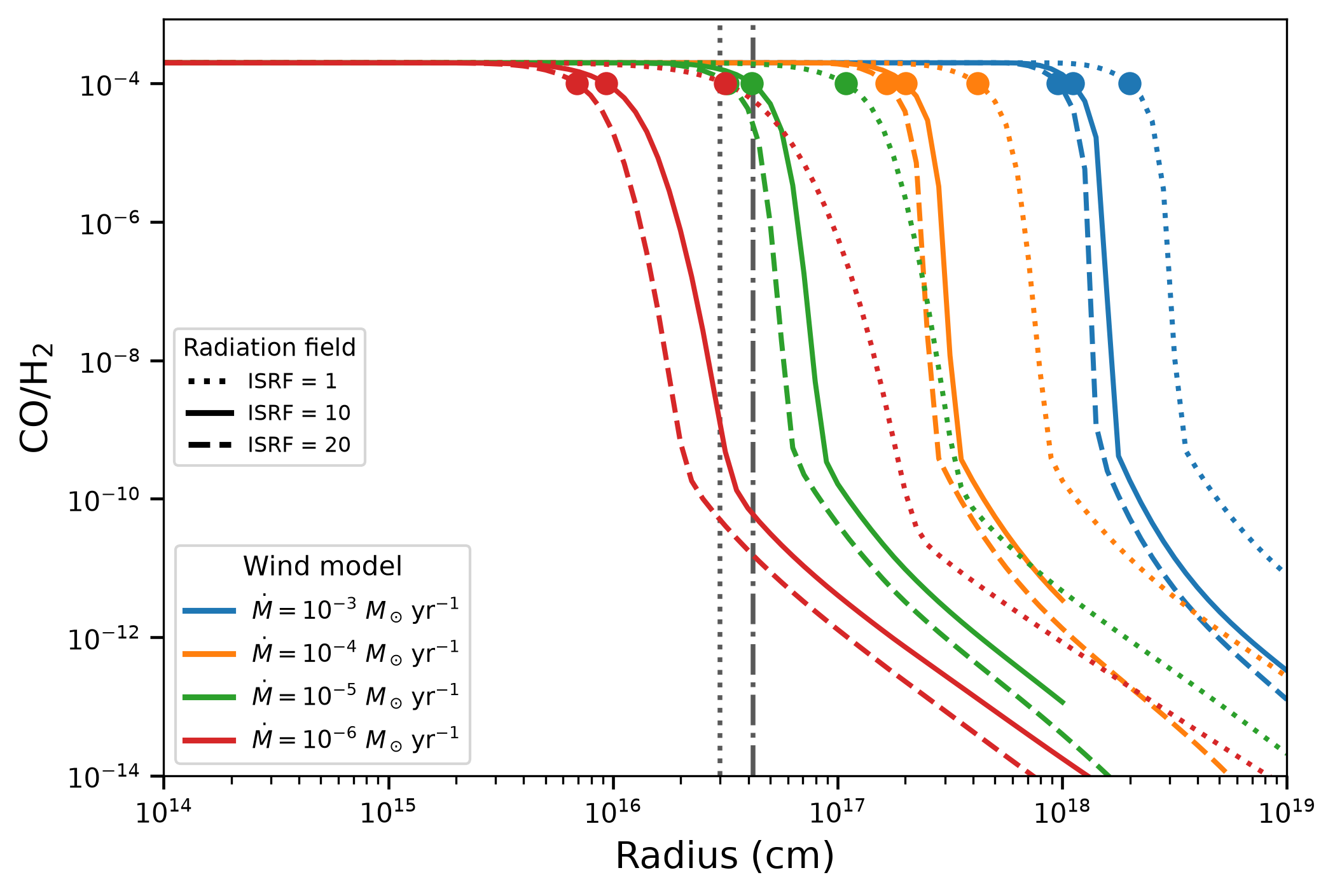}
    \caption{CO abundance profiles calculated using photodissociation models according to \citet{saberi2019} for outflows with \vexp = 25\,\kms, \mdot in the range $10^{-6}-10^{-3}$\,\msunyr, and assumed CO peak abundance of $2\times10^{-4}$, embedded in differently strong local interstellar radiation fields, ranging over 1, 10, and 20  (\emph{dotted, full, dashed curves}) times the average Galactic radiation field. The half-abundance radius is indicated on each curve with a filled circle. The vertical lines correspond to half the synthesized-beam-size for the ALMA observations of RSGC2 (\emph{dotted}) and RSGC1 (\emph{dash-dotted}). }
    \label{fig:co_efold}
\end{figure}

Figure~\ref{fig:co_efold} shows CO abundance profiles calculated by the photodissociation model presented by \cite{saberi2019} for circumstellar envelopes with a representative expansion velocity (25\,\kms), a range of mass-loss rates ($10^{-6}-10^{-3}$\,\msunyr), which are embedded in differently strong local interstellar radiation fields (ISRF), ranging over 1, 10, and 20 times the average Galactic radiation field. The models were calculated for an assumed CO peak abundance of $2\times10^{-4}$, very close to the values assumed in our radiative-transfer models (RSGC1: $1.8\times10^{-4}$, RSGC2: $1.6\times10^{-4}$). The results from \citet{saberi2019} show that such a small difference in abundance incurs only very small changes in the calculated CO half-abundance radii. 

The increase in ISRF intensity has a different impact across the relevant \mdot-range: the 20-fold increase leads to a reduction of $\rhalf$ by a factor $\approx2$ at $\dot{M}=10^{-3}$\,\msunyr, whereas this becomes a factor $\approx5$ at $\dot{M}=10^{-6}$\,\msunyr. 

The half-abundance radii in the dissociation models (Fig.~\ref{fig:co_efold}) are generally significantly beyond the half-beam size of the ALMA observations. Add to that the fact that there is significant contribution to the CO $J=2-1$ emission beyond $\rhalf$ (e.g. Fig.~\ref{fig:CO-RT-results-selected}) and we have to conclude the abundance profiles in Fig.~\ref{fig:co_efold} would give rise to extended emission in CO $J=2-1$ in all cases, except for perhaps the lowest mass-loss rate combined with the highest ISRF intensities. 

\subsection{CO radiative transfer modeling results}
\begin{figure*}[b]
\centering
    \subfigure[DFK\,1 \label{fig:comodels-dfk1}]{\includegraphics[trim=0.0cm 0.2cm 0.0cm 0.3cm,width=0.9\linewidth]{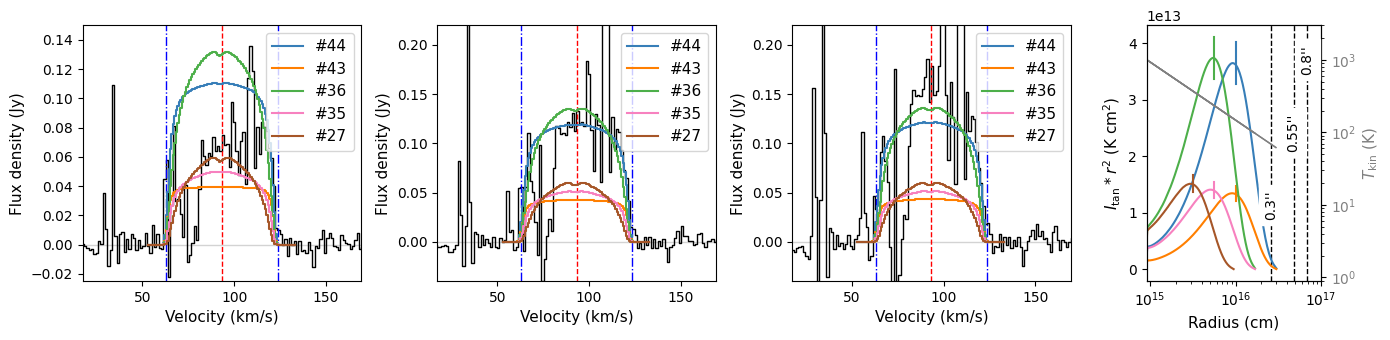}}
    \\
    \subfigure[DFK\,5 \label{fig:comodels-dfk5}]{\includegraphics[trim=0.0cm 0.2cm 0.0cm 0.3cm,width=.45\linewidth]{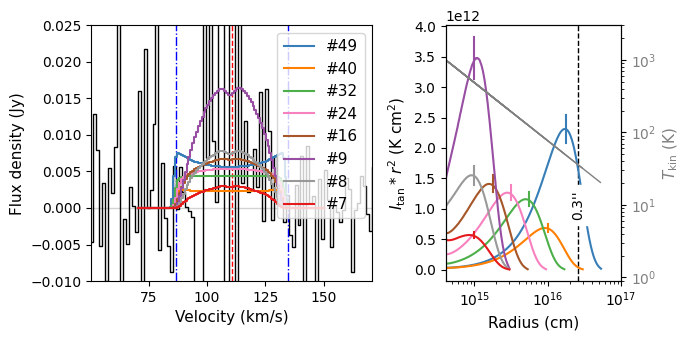}}
    % \hfill
    \subfigure[DFK\,8 \label{fig:comodels-dfk8}]{\includegraphics[trim=0.0cm 0.2cm 0.0cm 0.3cm,width=.45\linewidth]{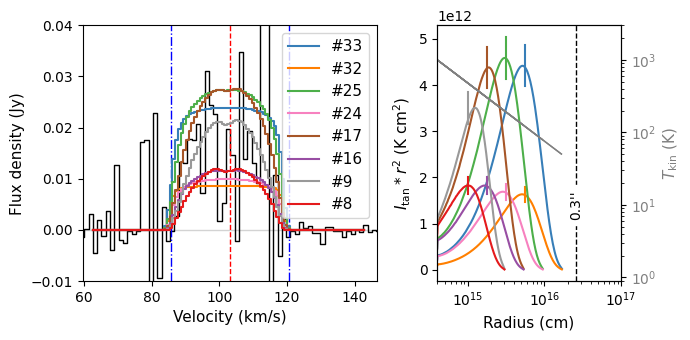}}
    \\
    \subfigure[DFK\,49 \label{fig:comodels-dfk49}]{\includegraphics[trim=0.0cm 0.2cm 0.0cm 0.3cm,width=0.9\linewidth]{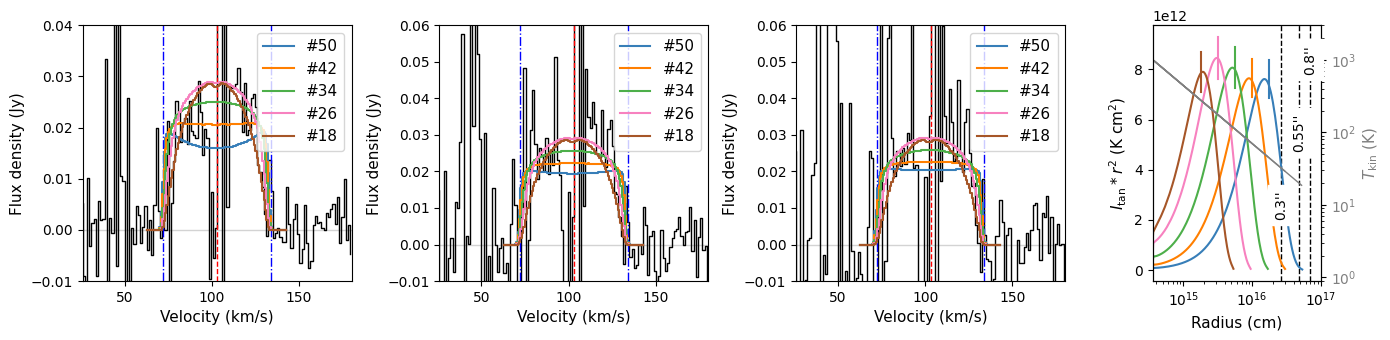}}
    
    \subfigure[RSGC1-1 \label{fig:comodels-rsgc1-1}]{\includegraphics[trim=0.0cm 0.2cm 0.0cm 0.3cm,width=.45\linewidth]{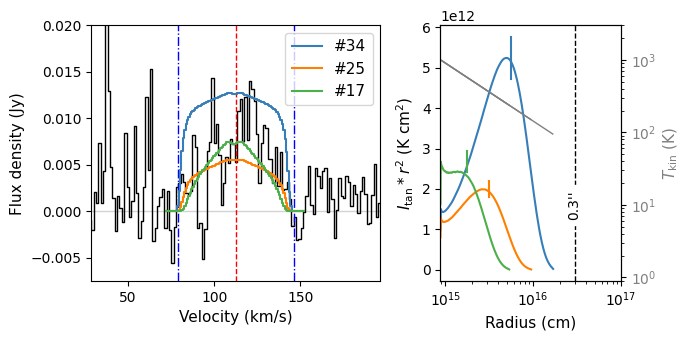}}
    \subfigure[RSGC1-2 \label{fig:comodels-rsgc1-2}]{\includegraphics[trim=0.0cm 0.2cm 0.0cm 0.3cm,width=.45\linewidth]{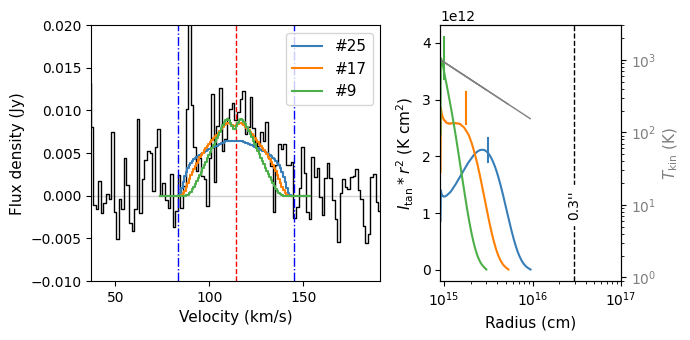}}
    \\
    \subfigure[RSGC1-4 \label{fig:comodels-rsgc1-4}]{\includegraphics[trim=0.0cm 0.2cm 0.0cm 0.3cm,width=.45\linewidth]{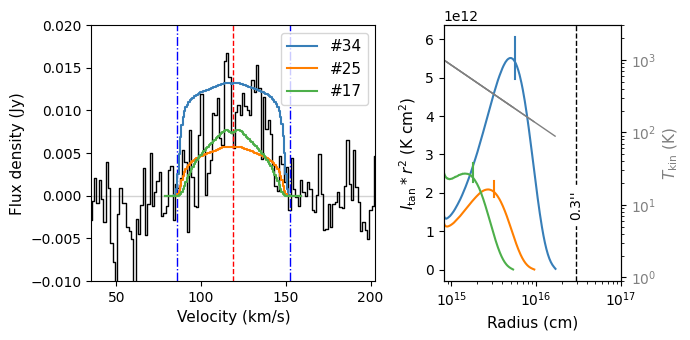}}
    \subfigure[RSGC1-13 \label{fig:comodels-rsgc1-13}]{\includegraphics[trim=0.0cm 0.2cm 0.0cm 0.3cm,width=.45\linewidth]{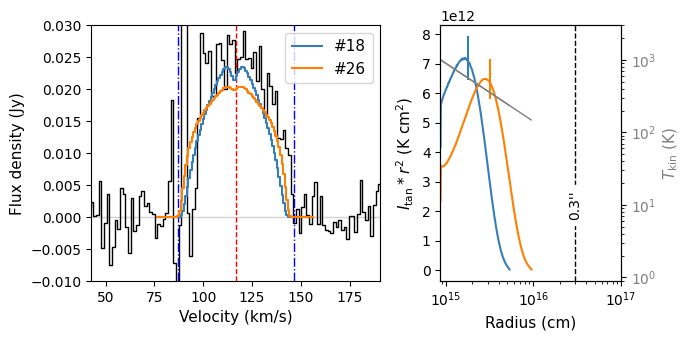}}

    \caption{    Same as Fig.~\ref{fig:CO-RT-results-selected} for the remainder of the cluster stars.     
   In the case of DFK\,1 and DFK\,49, we show additional panels for the 1.1\arcsec\/ and 1.6\arcsec\/ beams (increasing beam size from left to right). 
   }
   \label{fig:comodels-rsgc-appendix}
\end{figure*}

Table~\ref{tab:CO_RT_inputs} lists all accepted radiative transfer models for the remaining sources. Figure~\ref{fig:comodels-rsgc-appendix} 
shows the observed CO spectral lines and the lines predicted by our radiative transfer models (see Sect.~\ref{sect:COmodeling}) for all sources, except DFK\,2 and RSGC1-3 (see Fig.~\ref{fig:CO-RT-results-selected}).

\begin{table}
\caption{Accepted CO models and their input. }
    \label{tab:CO_RT_inputs}
    \centering
    \begin{tabular}{c | cccc }% | l}
    \hline\hline\\[-2ex]
    Star    
    & Model& log(\mdot     & log($\rhalf$ & $\rhalf$ \\%\\
            & \#   & /(\msunyr))  &/cm)& ($R_{\star}$)\\% & \\
    \hline\\[-2ex]
    DFK\,1\tablefootmark{a,b}  
&27&$-$3.0	&	15.50	&	25	\\%&	\\
&35,36&$-$3.5,$-$3.0	&	15.75	&	44	\\%&	\\
&43,44&$-$4.0,$-$3.5	&	16.00	&	78	\\%&	\\

    \hline\\[-2ex]        
    DFK\,2   
&9&$-$3.0	&	15.00	&	10	\\%&	\\
&17&$-$3.5	&	15.25	&	18	\\%&		\\
&25&$-$4.0	&	15.50	&	32	\\%&		\\
&33&$-$4.5	&	15.75	&	57	\\%&		\\
                
    \hline\\[-2ex]
    DFK\,5   
&7, 8, 9&$-$4.0,$-$3.5, $-$3.0	 &	15.00	&	19	\\%&		\\    
&16&$-$4.0	&	15.25	&	34	\\%&		\\
&24&$-$4.5	&	15.50	&	61	\\%&		\\
&32&$-$5.0	&	15.75	&	108	\\%&		\\
&40&$-$5.5	&	16.00	&	193	\\%&		\\
&49&$-$5.5	&	16.25	&	343	\\%&		\\
    
    \hline\\[-2ex]            
    DFK\,8   
&8, 9&$-$3.5, $-$3.0	&	15.00	&	27	\\%&&	\\
&16, 17 &$-$4.0, $-$3.5	&	15.25	&	49	\\%&	\\
&24, 25&$-$4.5, $-$4.0	&	15.50	&	87	\\%&	\\
&32, 33&$-$5.0, $-$4.5	&	15.75	&	155	\\%&	\\
    
    \hline\\[-2ex]        
    DFK\,49\tablefootmark{a}  
&18&$-$3.0	&	15.25	&	100	\\%&		\\
&26&$-$3.5	&	15.50	&	178	\\%&		\\
&34&$-$4.0	&	15.75	&	317	\\%&		\\
&42&$-$4.5	&	16.00	&	564	\\%&		\\
&50&$-$5.0	&	16.25	&	1002	\\%&	\\

    \hline\hline\\[-2ex]
    RSGC1-1 
&17&$-$3.5	&	15.25	&	18	\\%&		\\
&25&$-$4.0	&	15.50	&	31	\\%&		\\ 
&34&$-$4.0	&	15.75	&	55  \\

    \hline\\[-2ex]
    RSGC1-2 
&9&$-$3.0	&	15.00	&	11	\\%&		\\
&17&$-$3.5	&	15.25	&	20	\\%&		\\
&25&$-$4.0	&	15.50	&	35	\\%&		\\
    
    \hline\\[-2ex]
    RSGC1-3 
&17&$-$3.5	&	15.25	&	23	\\%&		\\
&25&$-$4.0	&	15.50	&	40	\\%&		\\
&33&$-$4.5	&	15.75	&	71	\\%&		\\
    
    \hline\\[-2ex]
    RSGC1-4 
&17&$-$3.5	&	15.25	&	23	\\%&		\\
&25&$-$4.0	&	15.50	&	41	\\%&		\\
&34&$-$4.0	&	15.75	&	73  \\

    \hline\\[-2ex]
    RSGC1-13 
&18&$-$3.0	&	15.25	&	29	\\%&		\\
&26&$-$3.5	&	15.50	&	52	\\%&		\\
    \hline
    \end{tabular}
   \tablefoot{
The columns list the stellar identifier, the model identifiers that are used in Figs.~\ref{fig:CO-RT-results-selected} and \ref{fig:comodels-rsgc-appendix}, % , and \ref{fig:comodels-rsgc1-appendix}, 
the mass-loss rate (\mdot), and CO half-abundance radius ($\rhalf$) in the models, the latter presented both in absolute scale and relative to the stellar radius \rstar. The assumed stellar temperatures and luminosities are taken as $T_{\star}$ and $L_{\mathrm{fit}}$ from Table~\ref{tab:sed_results} and wind-expansion velocities \vexp are those from Table~\ref{tab:linefits}.
\tablefoottext{a}{The CO emission from DFK\,1 and DFK\,49 extends beyond the synthesised beam (see Sect.~\ref{sect:spatial} and Figs.~\ref{fig:radial_profs} and \ref{fig:comodels-rsgc-appendix}).}   
\tablefoottext{b}{No model manages to reproduce the first two apertures simultaneously (see Fig.~\ref{fig:comodels-rsgc-appendix}).}
   
   } 
\end{table}

\end{appendix}

\end{document}